\documentclass[aps,onecolumn,showpacs,nofootinbib]{revtex4-2}
\usepackage{graphicx}
\usepackage{graphics}
\usepackage{textcomp}
\usepackage{psfrag}
\usepackage{epstopdf}
\usepackage{amssymb}
\usepackage{amsmath}
\usepackage{xcolor}
\usepackage{float}
\usepackage{lineno}
\usepackage{makecell}
\usepackage{setspace}
\usepackage{lipsum}%
\usepackage{inputenc}
\usepackage{tabularx}
\usepackage{soul,collcell,datatool}
\usepackage[colorlinks=true,linkcolor=blue,citecolor=blue,urlcolor=blue]{hyperref}
\usepackage[normalem]{ulem}
\graphicspath{{./plots/}}

\begin{document}
\setstretch{1.0}
\renewcommand\linenumberfont{\normalfont\scriptsize\sffamily\color{red}}

\title{Optoelectronics and Magnetic properties calculation of RE$_2$MnNiO$_6$ (Re=
La–Lu,Y) using Density Functional Theory} 

\author{Debidutta Pradhan}
\email{debi444@utkaluniversity.ac.in}
\affiliation{Center of Excellence in High Energy and Condensed Matter Physics, Department of Physics, Utkal University, Bhubaneswar 751004, India }
\begin{abstract}
RE$_{2}$NiMnO$_6$ (RE = La-Lu) family of ordered double-perovskite oxides hosts a corner-sharing network of alternating NiO$_6$ and MnO$_6$ octahedra whose electronic and magnetic ground states are systematically governed by the A-site ionic radius through the lanthanide contraction. The strong localisation of RE $4f$ electrons poses a fundamental challenge to density-functional treatments, yet the hybridisation between $4f$ and neighbouring $5d$ (RE) or $3d$ (Ni, Mn) states is central to the origin of exchange interactions and optoelectronic response across the series. We present a comprehensive first-principles study of the electronic structure, lattice dynamics, and optical properties of representative RE$_2$NiMnO$_6$ compounds within the DFT+U framework. To disentangle the role of Kondo-type $4f-d$ hybridisation, calculations are performed with the RE $4f$ electrons treated both as frozen core states and explicitly in the valence manifold, enabling a direct assessment of their contribution to the band structure, dielectric function, and phonon dispersion. Spin-polarised calculations reveal significant spin-channel asymmetry, with magnetic moments reaching up to 30 $\mu_{B}$ per formula unit for select members of the series. The results establish a unified picture of how $4f$ occupancy and octahedral distortion collectively determine the magnetic and optoelectronic potential of this double-perovskite family. 
\end{abstract}
\maketitle
\section{Introduction}
Rare earth elements play a vital role in modern technology because of their unusual 4f electron-derived magnetic, optical, and catalytic properties. They are used extensively in high-tech commercial and domestic devices, such as; glass polishing, catalytic converter, coloured phosphors, water treatment, alloying agent, high resolution imaging, in small electronics gadgets, nuclear reactor control rods and also for energy and data storage devices \cite{Ormerod23}. Lutetium (Lu) ions are used to color and polish optical glasses, and lanthanum is a key component (about 50$\%$) of camera lenses where as, Cerium and lanthanum catalysts are essential in petroleum refining and automotive exhaust converters. Neodymium, praseodymium, and dysprosium are critical for producing the strongest permanent magnets (N$d$-Pr-Fe-B) used in computer drives, electric motors, and wind turbines, while Gadolinium is used in MRI and other magnetic imaging purposes. In industry, rare earths enable high-strength magnets in wind turbines and electric vehicle motors; efficient rechargeable batteries (e.g., nickel-metal hydride cells use lanthanum-based alloys); and catalysts (lanthanum and cerium in refineries and converters), and corrosion-resistant steels. Domestically, REEs are present in virtually every electronic device: smartphones, computers, TVs, LED bulbs, lasers, and clean-energy products. For example, hybrid car batteries can contain 10-15 kg of lanthanum, and (N$d$-Fe-B) motors in new cars use praseodymium and neodymium magnets \cite{Binnemans18}. Importantly, these elements are not extremely rare, but their ores are diffuse and difficult to mine and separate, requiring specialized processing \cite{Stephen06,Walters10}. This combination of unique properties and complex supply makes lanthanides rare and expensive minerals for both industry and research.

In double perovskite, the B-site ordering offers extra flexibility to design novel properties that can yield tunable bandgaps and magnetic moments, better than single perovskites\cite{Kim19,Xudong25}.
 
This work focuses on RE$_2$MnNiO$_6$, where RE represents rare-earth elements consisting of the lanthanide series (La--Lu) along with Y and Sc, as the later two share similar physical and chemical properties and are often extracted from the same ore deposits. The lanthanides belong to Group 3 and Period 6, where their atomic and ionic radii decrease with increasing atomic number due to the lanthanide contraction, a phenomenon arising from the poor shielding effect of the 4$f$ electrons. Other Group 3 elements, such as Sc and Y, with electronic configurations [Ar]4$s^2$3$d^1$ and [Kr]5$s^2$3$d^1$ respectively, are also sometimes classified as rare-earth elements. However, in the present work, Sc is excluded as it exhibits more pronounced transition-metal characteristics compared to the lanthanides. 

\begin{figure}[H]
\begin{center}
\includegraphics[height=5.5cm,width=6.3cm]{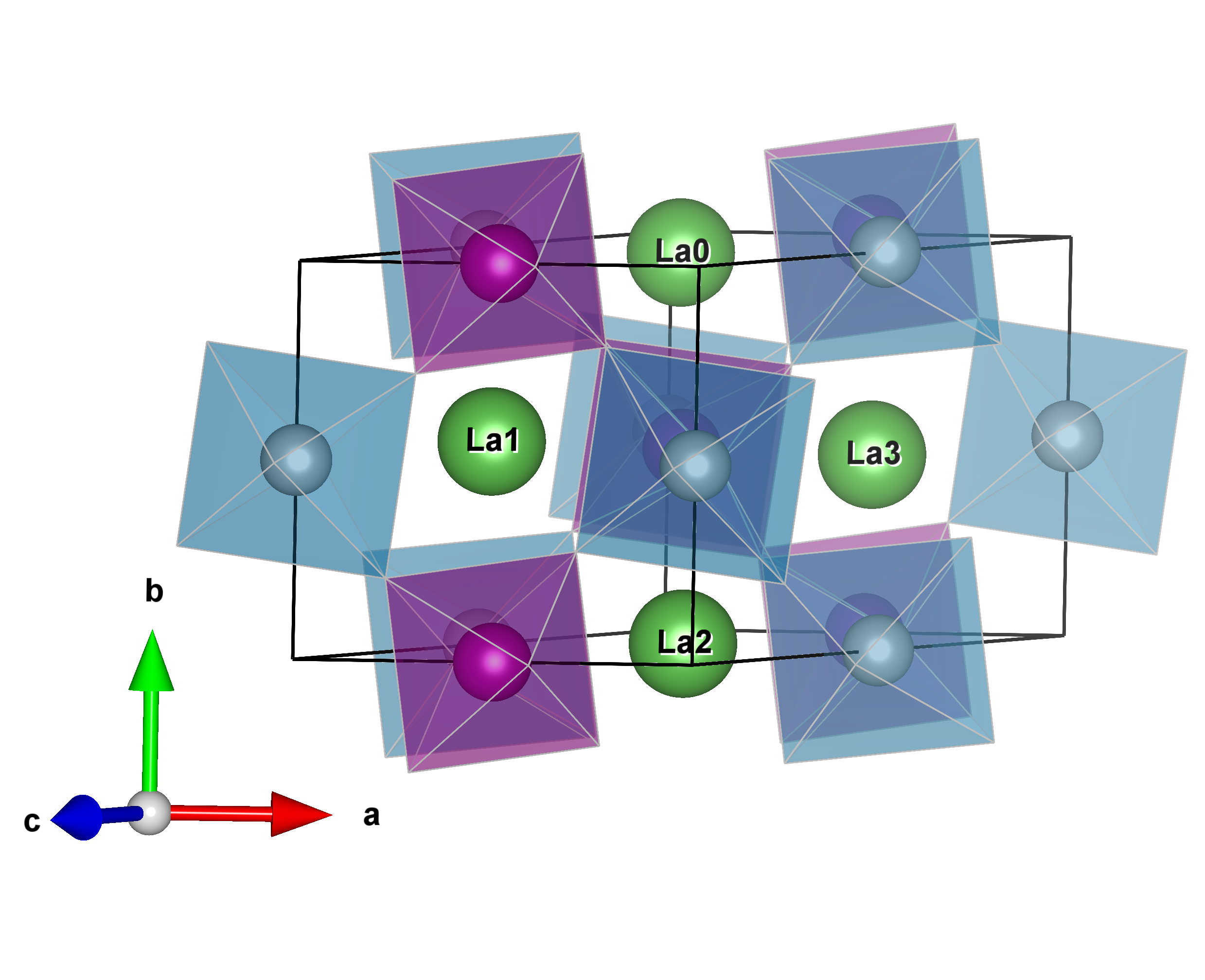} \hspace{1cm}
\includegraphics[height=5.5cm,width=6.3cm]{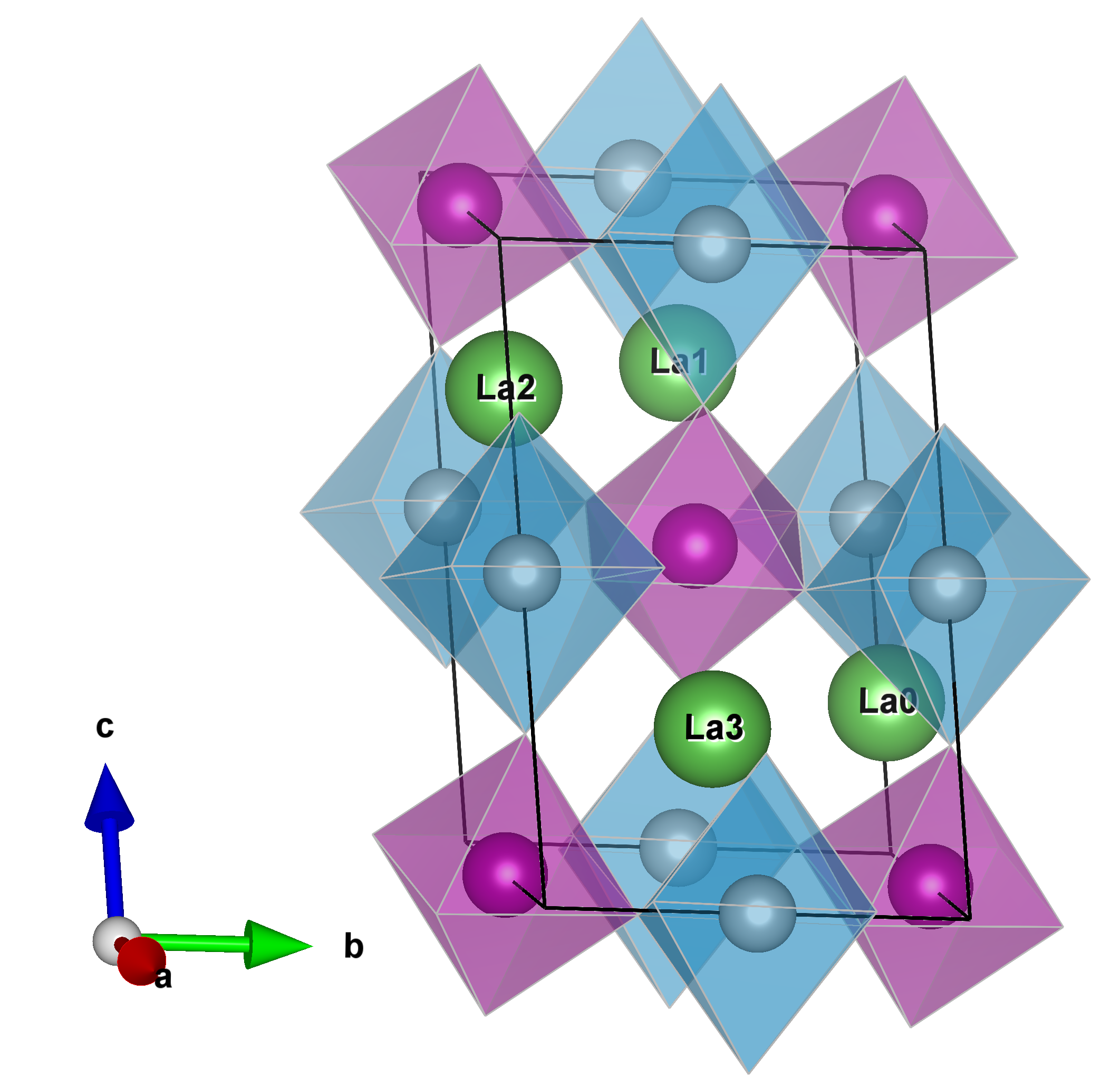} 
\caption{LMNO crystal structure.}
\label{fig:structure6}
\end{center} 
\end{figure}

Unlike $p$- or $d$-orbitals, where the orbital lobes are oriented in a two-dimensional plane, $f$-orbitals exhibit a complex three-dimensional spatial distribution. This results in a high degree of $f$-electron localization in three-dimensional space, enhancing the probability of hybridization in multiple directions. In contrast to $d$-orbitals, whose energy levels split significantly under crystal field effects, the energy levels of $f$-orbitals experience minimal splitting, thereby retaining their degeneracy. This property further facilitates extensive hybridization. Additionally, $f$-orbitals interact weakly with ligands regardless of their strength or number, enabling $f$-electrons to remain in excited states for longer durations, a property that makes these elements highly suitable for laser applications. Apart from configurations with completely empty, hal$f$-filled, or fully filled $f$-orbitals, lanthanides often form colored crystals, which makes them widely used in phosphor materials \cite{Hosoya14,King20}. 
For convenience, RE$_2$NiMnO$_6$ will hereafter be denoted as RMNO. RMNO can crystallize in multiple symmetry configurations, among which the monoclinic $P2_{1}/c$ phase have greater occurrence and exhibits better thermodynamic stability. Therefore, in this work, all investigated perovskite structures are considered in monoclinic symmetry (see Fig. \ref{fig:structure6}).
 
La$_2$NiMnO$_6$ (LMNO) is a prototype of the RE$_2$NiMnO$_6$ double perovskite family and has been widely studied. It crystallizes in an ordered monoclinic (P2$_1$/c) structure, where Ni$^{2+}$ and Mn$^{4+}$ octahedra alternate. This is a ferromagnetic semiconductor, showing strong ferromagnetism (with a Curie temperature around 270-280 K) due to Ni-O-Mn superexchange interactions \cite{Housni19}. This band gap combined with strong visible absorption makes LNMO interesting for UV-filter and optoelectronic applications. Importantly, LNMO is lead-free and non-toxic, unlike many halide perovskites. Recent studies have proposed LNMO as an absorber material in perovskite solar cells, predicting high efficiencies (around 20-25$\%$) when optimized. It can be integrated into thin-film devices and often shows good dielectric properties and charge transport. The coupling of magnetic and electric order also suggests uses in multiferroic and spin-based devices.

Replacing La$^{3+}$ with other lanthanides (Ce to Lu) in RE$_2$NiMnO$_6$ systematically tunes the material’s structure and optical band gap. As the rare-earth ionic radius decreases down the series (from La$^{3+}$ at 1.16 angstroms to Lu$^{3+}$ at approximately 0.977 angstroms), the lattice contracts and the Ni-O-Mn bond angles change. First-principles and experimental studies show that smaller rare-earth ions tend to alter the band structure. Materials with smaller band gap, make it nearly optimal for single-junction solar absorption and also more favorable for visible or near-infrared absorption. In general, as the size of Re$^{3+}$ decreases, the overlap between Ni/Mn 3d orbitals and O 2p orbitals changes, leading to either widening or narrowing of the band gap depending on specific structural configurations. By selecting the Re cation, it is possible to tailor the optical absorption edge and electronic properties of RE$_2$NiMnO$_6$. RMNO can be considered as a heavy-fermion system, where Kondo hybridisation is observed. Kondo hybridisation arise in rare earth and actinoid metals, where highly localised $f$-orbital along with d and p orbitals form hybrid orbitals. These hybrid orbitals form as a result of weak electronic interaction which reduce the overall energy of the system \cite{Frantzeskakis13,Kim2022}. 

By correlating A-site ionic radius with electronic bandwidth, phonon scattering and optical transition energies, our results demonstrate that RE$_2$NiMnO$_6$ double perovskites can be rationally engineered to meet the combined requirements of wide band gap, and strong UV absorption - key attributes for next-generation optoelectronic devices. These lanthanide-tunable double perovskites offer a versatile platform for designing new optoelectronic and magnetic materials.

\section{Computational Method}
In this work, first-principles calculations were performed within the framework of density functional theory (DFT) to obtain and analyse various properties of RMNO. All structures were obtained from Materials Project for RE = La, Pr, Nd, Sm, Gd, Tb, Dy and Ho and the remaining structures were constructed in VESTA by referencing the crystal structures of neighbouring lanthanide perovskites \cite{Matpro,VESTA}. The monoclinic symmetry with space group P2$_1$/c was chosen, as it is the dominant phase among all the structures. The structures were optimized using high-precision spin-polarised calculations with an increased planewave cutoff energy of 520 eV, corresponding to actual k-spacings of $0.5$ per angstrom. Electronic convergence threshold of 10$^{-5}$ $eV$ was used with blocked Davidson algorithm with reciprocal space projection operators, while GGA-PBEsol exchange-correlation functional was used to describe the interaction. For correction of strongly correlated localized electrons Simplified LSDA+U approach was implemented with on-site Coulomb terms of U-J value, for $d$ orbital of Mn and Ni atoms, as 4 and 5 $eV$ respectively. Optimised structures were then used to compute electronic band structures and the corresponding density of states, as well as for the optical properties. Gaussian smearing was used with a width of 0.05 eV, where the k-mesh was forced to be centered on gamma point where, Projector augmented-wave (PAW) method was also used. Two set of calculations were performed: in one set, the $f$-electrons were taken as core electrons to compare the results with those obtained from previous works, as valence $f$-electrons can lead to convergence problems. Valence electronic configurations were 5s$^{2}$5p$^{6}$6s$^{2}$5d$^{1}$ for La, Ce, Pr and Nd with 11 valence electrons, whereas rest of the elements had 9 valence electrons with 5p$^{6}$6s$^{2}$5d$^{1}$ configuration. The phonon band structure and phonon density of states were calculated for a $2 \times 2 \times 2$ supercell, and vibrational properties along with the corresponding thermodynamic parameters. Optical properties were also computed using this pseudopotential. In the second set, partially filled $f$ states were considered for valence electron configuration as 5s$^{2}$5p$^{6}$6s$^{2}$5d$^{1}$4f$^{1}$ and 5s$^{2}$5p$^{6}$6s$^{2}$5d$^{1}$4f$^{14}$ for Ce and Lu respectively. Other elements have fractional configuration of 5s$^{2}$5p$^{6}$6s$^{2}$5d$^{n}$4f$^{m+n}$, where n = 0.5 and m = number of $f$ electrons (2-13) of lanthanides, due to their higher tendency to attain mixed valency of RE$^3+$/RE$^4+$, resulting a better PAW construction. Electronic and magnetic charge densities were also analysed with initial magnetic moment based on spin-only magnetic moment. All the simulations were performed using the Vienna Ab initio Simulation Package (VASP) \cite{MedeA,vasp1,vasp2,vasp3}.

\section{Results and Discussion}
Detailed analysis of electronic, optical and thermal properties of the lanthanide based composites for possible potential applications has been performed in this work. The above properties are discussed in detail in the following subsections.

\subsection{Structural Analysis}
\begin{figure}[h]
\begin{center}
\includegraphics[height=5.5cm,width=7cm]{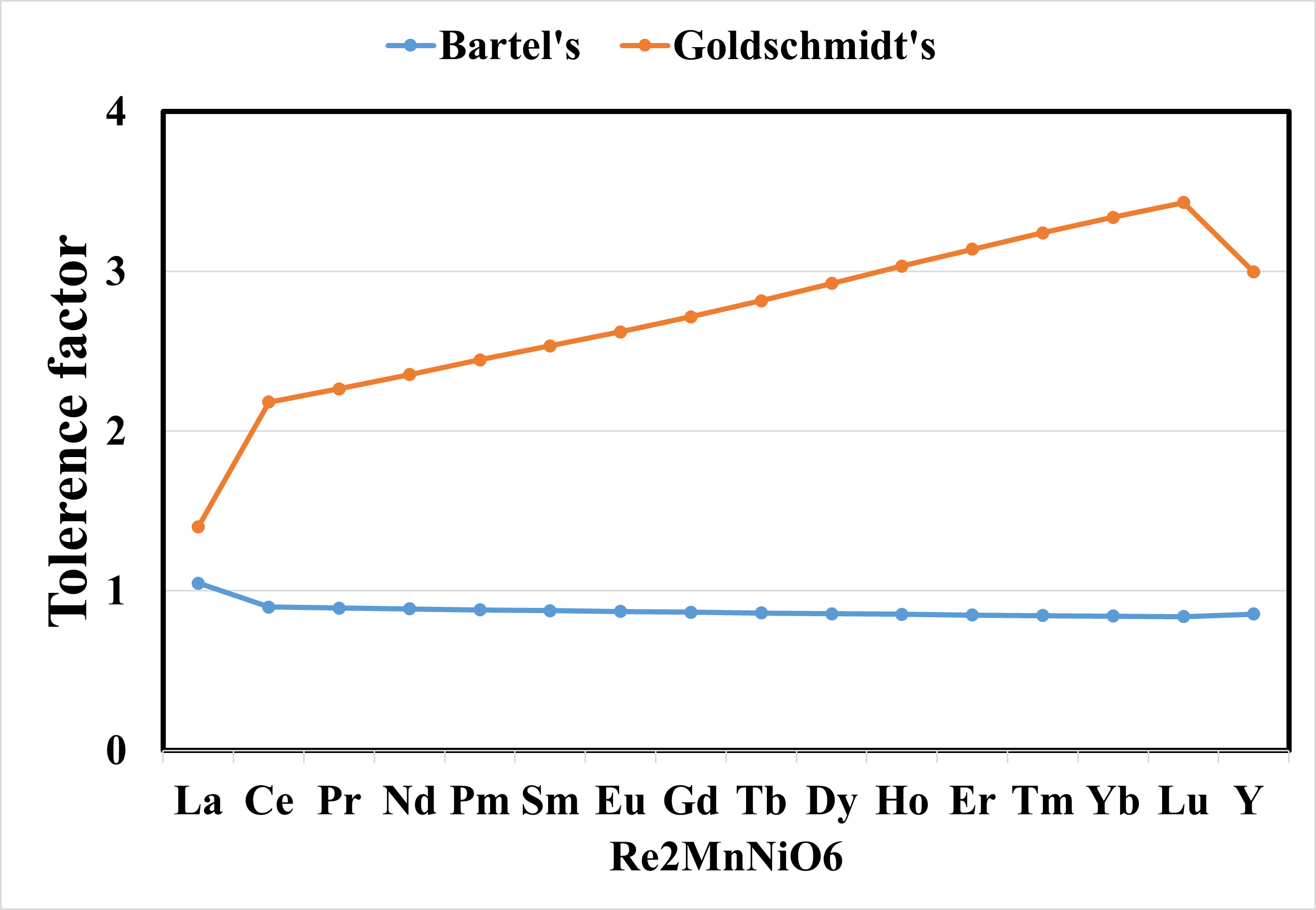}
\caption{Tolerance factor estimation RMNO.}
\label{fig:Brillouin_zone6}
\end{center}
\end{figure}

Rare earth elements in this work have +3 oxidation state and Mn, Ni have +4 and +2 states respectively. As the ionic radius decreases, the crystal lattice length also ($a$ and $c$) decreases, leading to a simultaneous reduction in volume. Cell angles remain constant (90$^o$), except $\beta$ which changes constantly. This leads to higher density of the lattice from La to Lu-perovskite (shown in Table. \ref{tab:structure_details6}). Because of the equal number of atoms, same symmetry and equal occupancies of electrons, energy values remain similar throughout all lanthanides. These values for the set-I, were in similar agreement with experimental results \cite{Nasir2019}. For the set-II, lattice constants remained the same but the energy values were increased due to involvement of extra $f$-electrons in the system. Since $f$-electrons do not interact strongly with surrounding orbitals, a slight increase of total energy is observed. The tolerance factor confirms the perovskite stability by Goldschmidt's and Bartel's formula \cite{Goldschmidt26,Bartel19}. Goldschmidt's tolerance factor should be within the range of 0.8 to 1.0, whereas Bartel's factor should be below 4.1; and in this case all compounds satisfy the same, as shown in Fig. \ref{fig:Brillouin_zone6}. Electronic and magnetic properties depend on both atomic and lattice contributions of the system. In the lattice the bond angle of Ni--O--Mn is considered to be one of the primary factors give rise to the intrinsing magnetic moment of the composite. The correlation of this factor along with the change in tilting angle of the perovskite is one of the important factors for static magnetisation. Ce$_2$MnNiO$_6$ have Ni--O--Mn bond angle of 157$^o$, Gd$_2$MnNiO$_6$ shows 147$^o$ and Lu$_2$MnNiO$_6$ with 140$^{o}$ shows the decreasing order along the period. This order is also responsible for the change in magnetic moment and magnetisation of the material along with other properties.

\begin{table}[htbp]
\centering
\caption{Lattice parameters, volume, and density for RMNO.}
\setlength{\tabcolsep}{6pt}
\begin{tabular}{|c|c|c|c|c|c|c|c|}
\hline 
Structures & $a$ & $b$ & $c$ & $\beta$ & Volume & Density & Energy \\ 
 & ( \AA ) & ( \AA ) & ( \AA ) & ($^o$) & ( \AA $^3$) & ($Mg m^{-3}$) & (eV) \\ 
\hline 
La$_2$MnNIO$_6$ & 5.466 & 5.447 & 9.451 & 54.567 & 229.265 & 7.061 & -156.617 \\ 
\hline 
Ce$_2$MnNIO$_6$ & 5.466 & 5.470 & 9.461 & 54.587 & 230.546 & 7.056 & -154.313 \\ 
\hline 
Pr$_2$MnNIO$_6$ & 5.435 & 5.443 & 9.405 & 54.591 & 226.172 & 7.197 & -154.786 \\ 
\hline 
Nd$_2$MnNIO$_6$ & 5.374 & 5.499 & 9.335 & 54.872 & 225.645 & 7.331 & -154.986 \\ 
\hline 
Pm$_2$MnNIO$_6$ & 5.325 & 5.521 & 9.264 & 55.057 & 223.236 & 7.490 & -155.265 \\ 
\hline 
Sm$_2$MnNIO$_6$ & 5.300 & 5.507 & 9.253 & 54.877 & 220.939 & 7.671 & -155.305 \\ 
\hline 
Eu$_2$MnNIO$_6$ & 5.264 & 5.520 & 9.210 & 54.873 & 218.892 & 7.792 & -155.543 \\ 
\hline 
Gd$_2$MnNIO$_6$ & 5.227 & 5.539 & 9.116 & 54.819 & 216.968 & 8.023 & -155.788 \\ 
\hline 
Tb$_2$MnNIO$_6$ & 5.183 & 5.535 & 9.088 & 54.848 & 213.219 & 8.216 & -155.870 \\ 
\hline 
Dy$_2$MnNIO$_6$ & 5.172 & 5.510 & 9.094 & 54.788 & 211.792 & 8.383 & -155.951 \\ 
\hline 
Ho$_2$MnNIO$_6$ & 5.157 & 5.528 & 9.080 & 54.781 & 211.532 & 8.470 & -155.980 \\ 
\hline 
Er$_2$MnNIO$_6$ & 5.146 & 5.504 & 9.069 & 54.789 & 209.912 & 8.609 & -156.042 \\ 
\hline 
Tm$_2$MnNIO$_6$ & 5.118 & 5.508 & 9.036 & 54.750 & 208.059 & 8.739 & -156.129 \\ 
\hline 
Yb$_2$MnNIO$_6$ & 5.111 & 5.486 & 9.026 & 54.795 & 206.828 & 8.923 & -156.184 \\ 
\hline 
Lu$_2$MnNIO$_6$ & 5.085 & 5.480 & 9.003 & 54.760 & 204.941 & 9.068 & -156.201 \\ 
\hline 
Y$_2$MnNIO$_6$ & 5.279 & 5.599 & 9.268 & 54.86 & 224.092 & 5.742 & -162.863 \\ 
\hline 
\end{tabular} 
\label{tab:structure_details6}
\end{table}

\subsection{Band structures and density of states}
The band structure analysis for the rare-earth elements was carried out using two different pseudopotentials in this work. The band structure and density of states were plotted in the energy range from -6 $eV$ to +6 $eV$ along the high-symmetry path $C-Y-\Gamma-B-A-E-Z$. This symmetry path corresponds to the monoclinic space group $P2_{1}/c$. The obtained band gaps for the ground-state and NSCF calculations, along with the energy gaps for both spin-up and spin-down channels in set I, are listed in Table. \ref{tab:bandgaps6}. The results show that the spin-up band gaps are consistently smaller than the spin-down band gaps. In band disperssion plots, blue lines represent the spin-up channel and red lines represent the spin-down channel. The lower energy of the spin-up bands compared to the spin-down bands indicates a greater contribution from the majority spin channel, supporting the ferromagnetic nature of the composites \cite{Jain18}.
\begin{table}[H]
\begin{center}
\setlength{\tabcolsep}{10pt}
\begin{tabular}{|c|c|c|c|c|}
\hline 
Re(in RMNO) & SCF & NSCF & Spin-down & Spin-up \\ 
\hline 
La & 1.537 & 1.329 & 3.15 & 1.32 \\ 
\hline 
Ce & 1.739 & 1.573 & 3.17 & 1.57 \\ 
\hline 
Pr & 1.646 & 1.457 & 3.19 & 1.46 \\ 
\hline 
Nd & 1.874 & 1.709 & 3.12 & 1.68 \\ 
\hline 
Pm & 1.914 & 1.750 & 3.10 & 1.73 \\ 
\hline 
Sm & 1.788 & 1.611 & 3.09 & 1.59 \\ 
\hline 
Eu & 1.828 & 1.634 & 3.08 & 1.61 \\ 
\hline 
Gd & 2.005 & 1.803 & 3.05 & 1.78 \\ 
\hline 
Tb & 1.997 & 1.803 & 3.05 & 1.78 \\ 
\hline 
Dy & 1.806 & 1.619 & 3.08 & 1.60 \\ 
\hline 
Ho & 2.023 & 1.852 & 3.07 & 1.83 \\ 
\hline 
Er & 1.918 & 1.742 & 3.09 & 1.72 \\ 
\hline 
Tm & 1.923 & 1.909 & 3.08 & 1.87 \\ 
\hline 
Yb & 1.812 & 1.802 & 3.10 & 1.76 \\ 
\hline 
Lu & 1.961 & 1.942 & 3.10 & 1.90 \\ 
\hline 
Y & 1.767 & 1.587 & 3.12 & 1.68 \\ 
\hline 
\end{tabular} 
\caption{Spin-up and spin-down bandgaps for various lanthanide elements.}
\label{tab:bandgaps6}
\end{center}
\end{table}
For all the materials, the Fermi level lies between the valence and conduction bands which indicates that the materials are semiconductors with moderate band gaps. Unlike increasing trend of band gaps along the series from La to Lu is not observed here \cite{Nasir2019,Zhao2014}. Overall the bandgap is increasing in nature along the period for more than half of the elements (La, Ce, Nd, Pm, Gd, Tb, Ho, Tm, Lu). Occurrence of splitting of bandgaps between up and down spin confirms the magnetic nature of the material.   

\begin{figure}[H]
\begin{center}
\includegraphics[height=5.5cm,width=7.5cm]{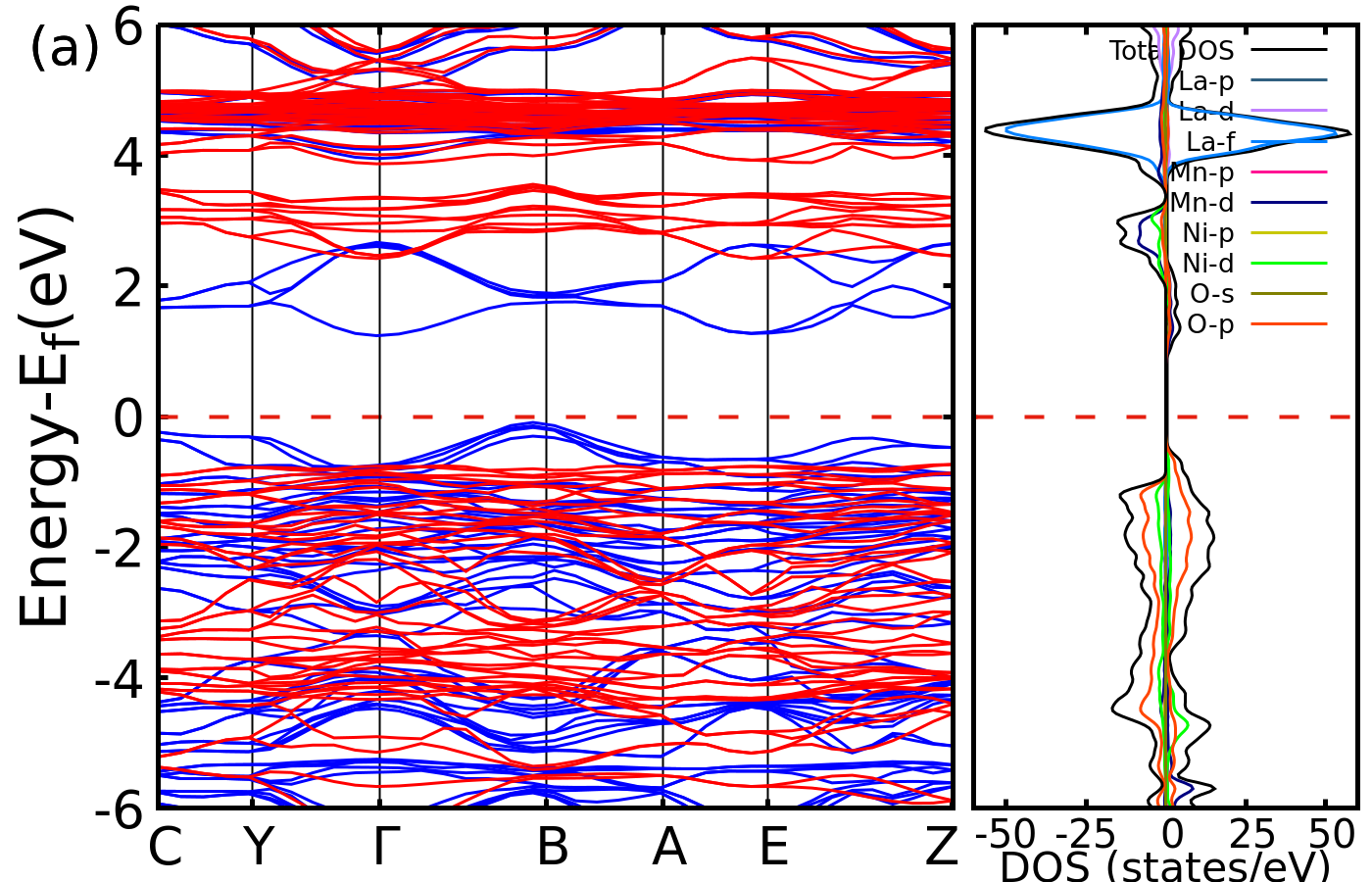}
\includegraphics[height=5.5cm,width=7.5cm]{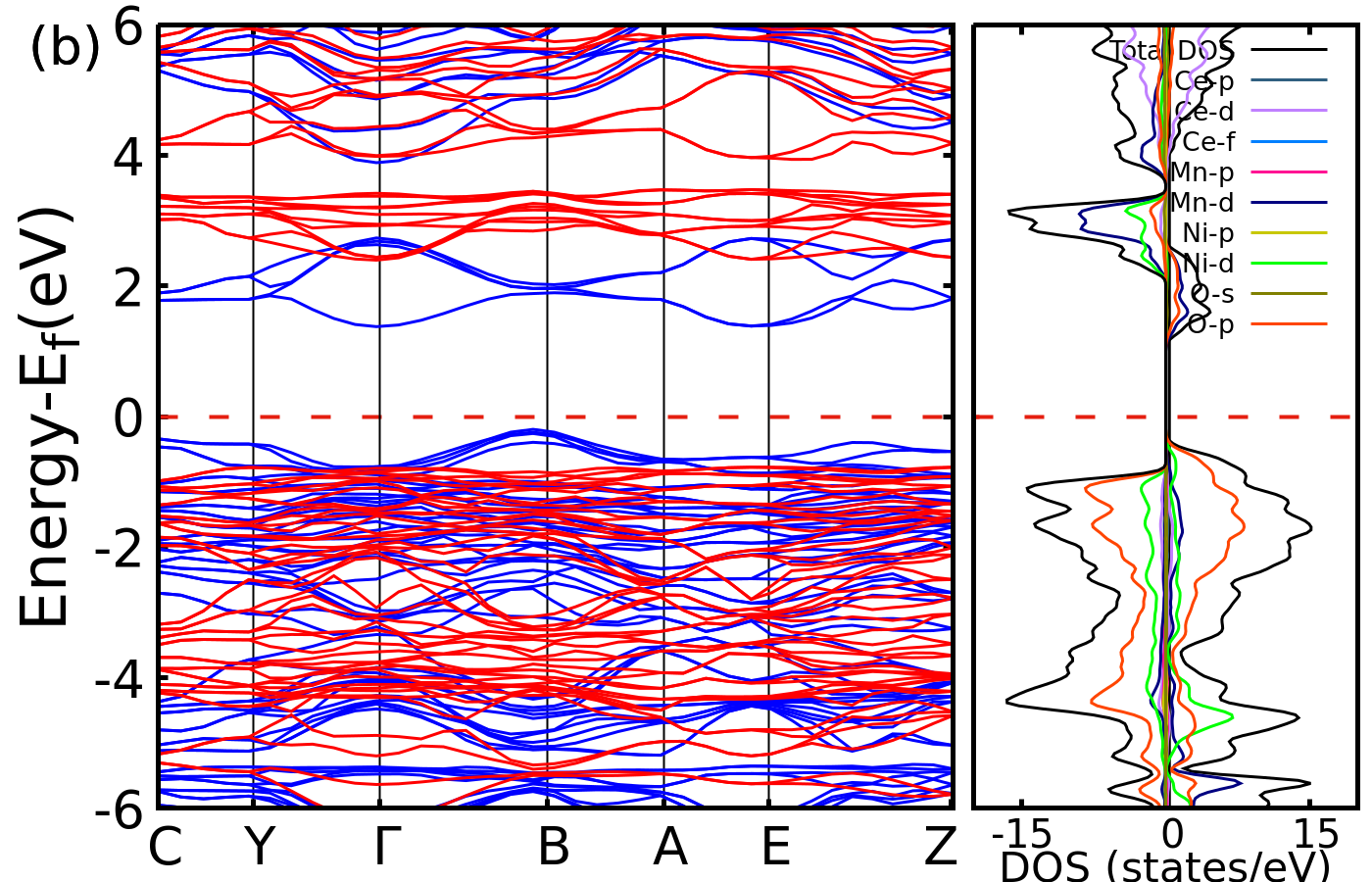}\\ \vspace*{0.3cm}
\includegraphics[height=5.5cm,width=7.5cm]{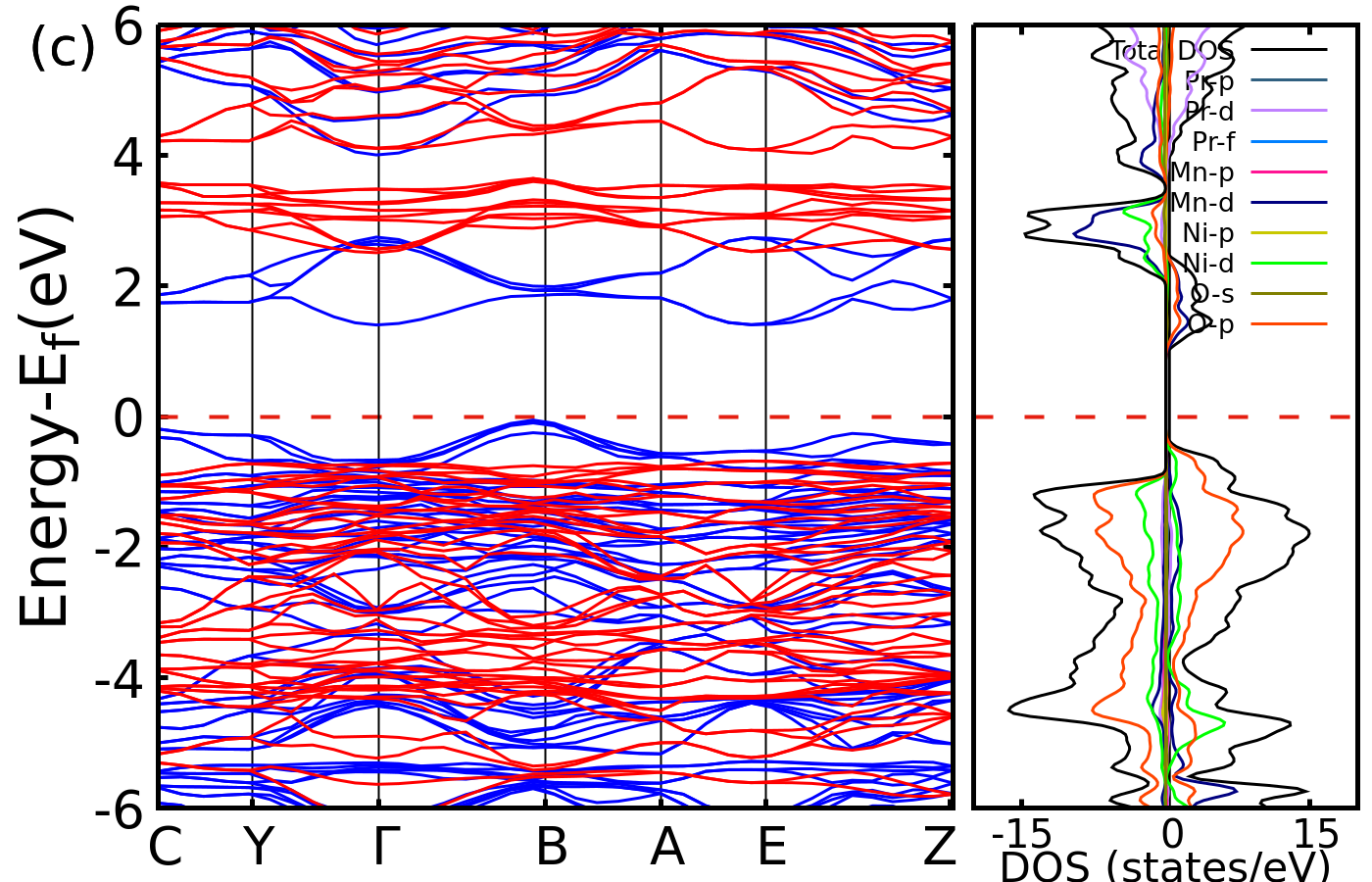}
\includegraphics[height=5.5cm,width=7.5cm]{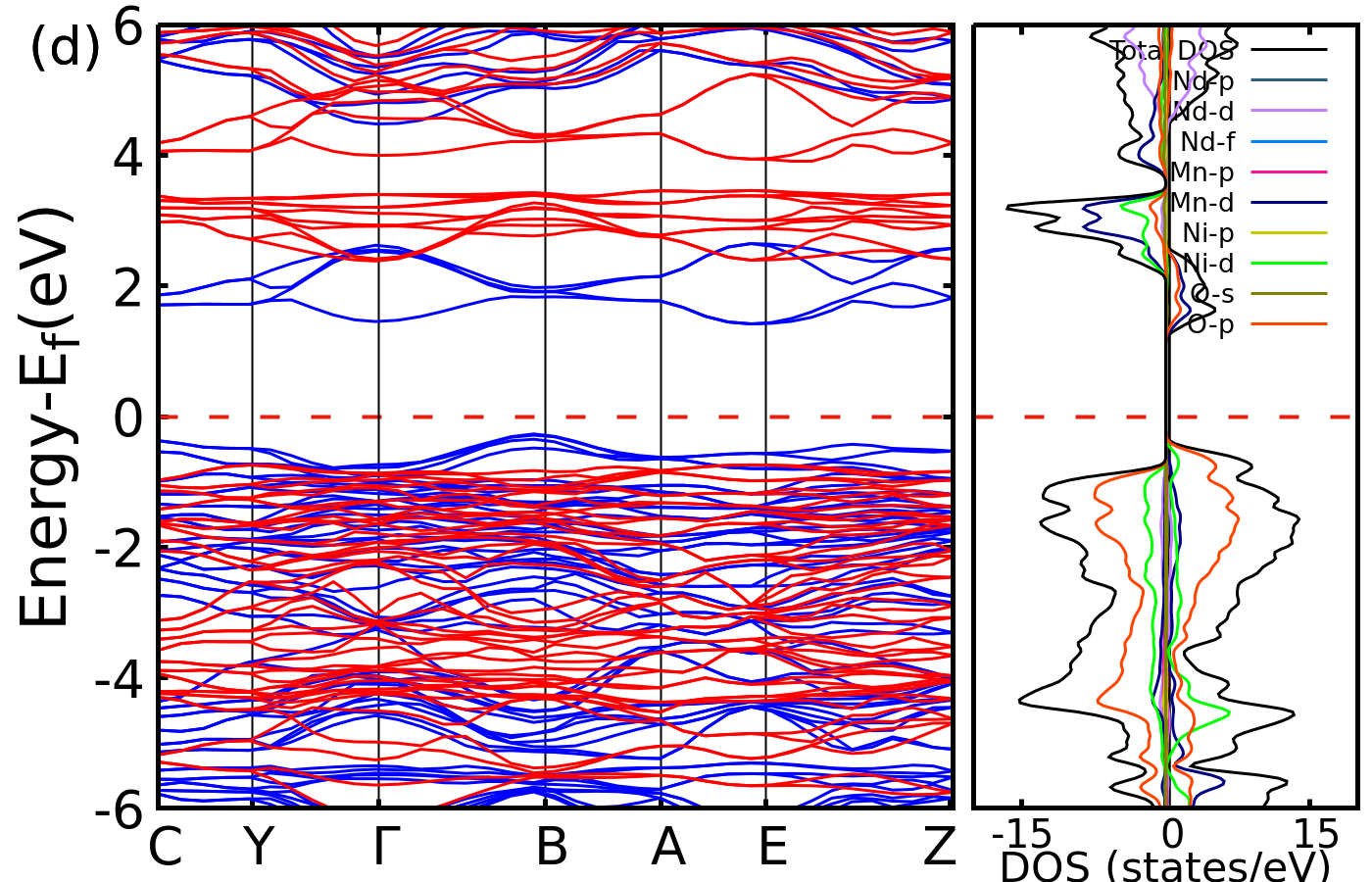}\\ \vspace*{0.3cm}
\end{center}
\end{figure}
\begin{figure}[H]
\begin{center}
\includegraphics[height=5.5cm,width=7.5cm]{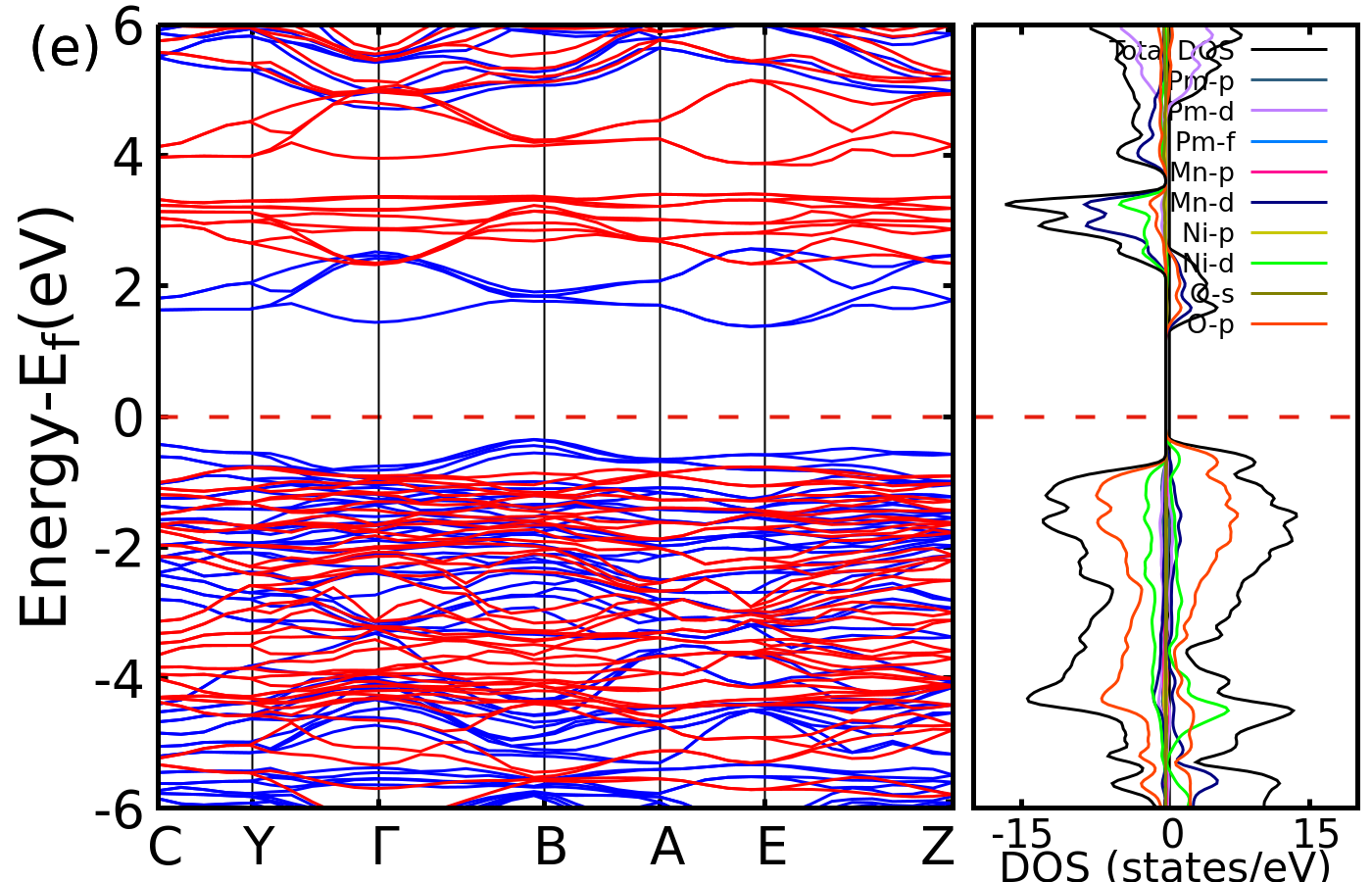}
\includegraphics[height=5.5cm,width=7.5cm]{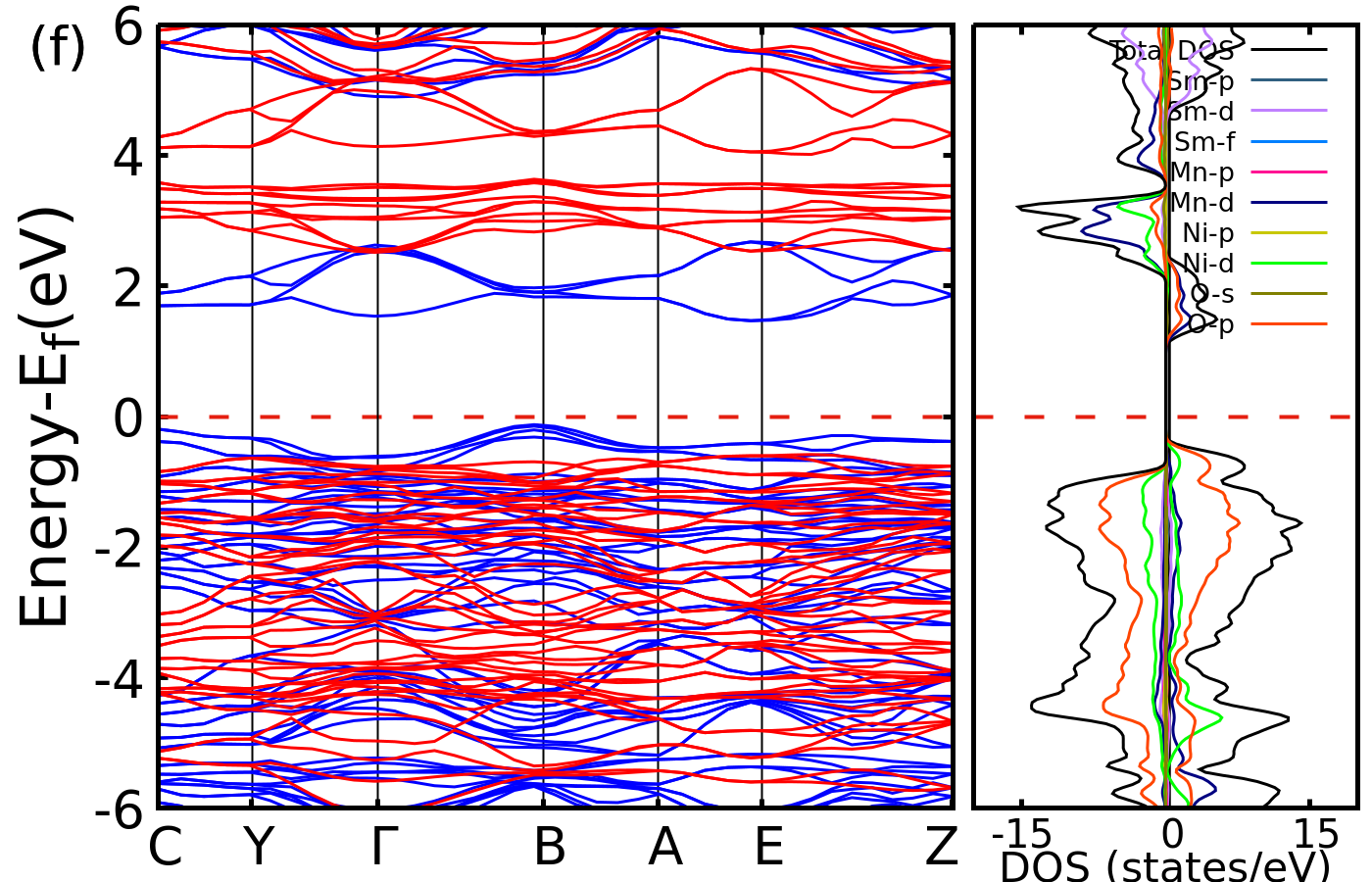}\\ \vspace*{0.3cm}
\includegraphics[height=5.5cm,width=7.5cm]{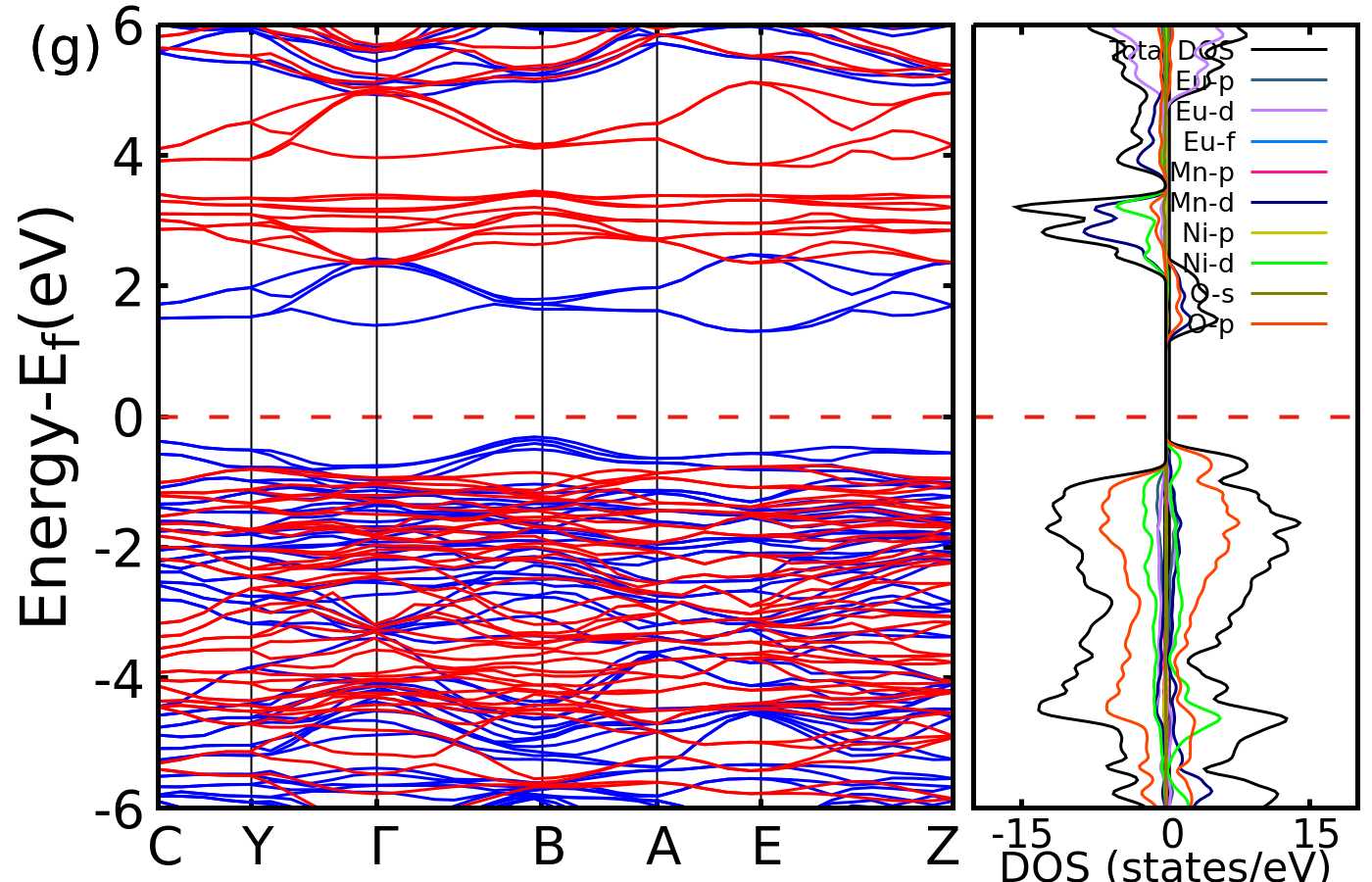}
\includegraphics[height=5.5cm,width=7.5cm]{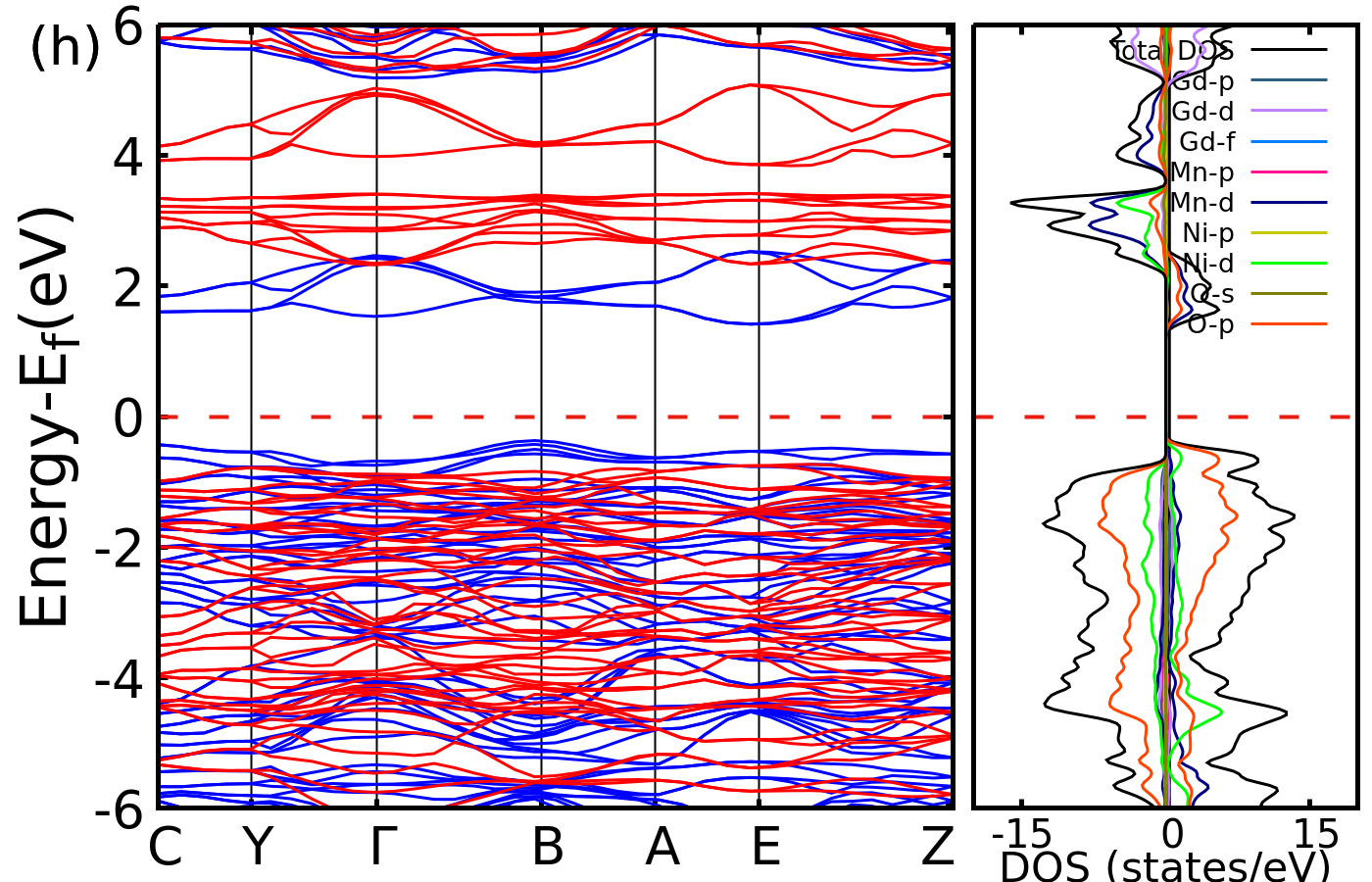}\\ \vspace*{0.3cm}
\includegraphics[height=5.5cm,width=7.5cm]{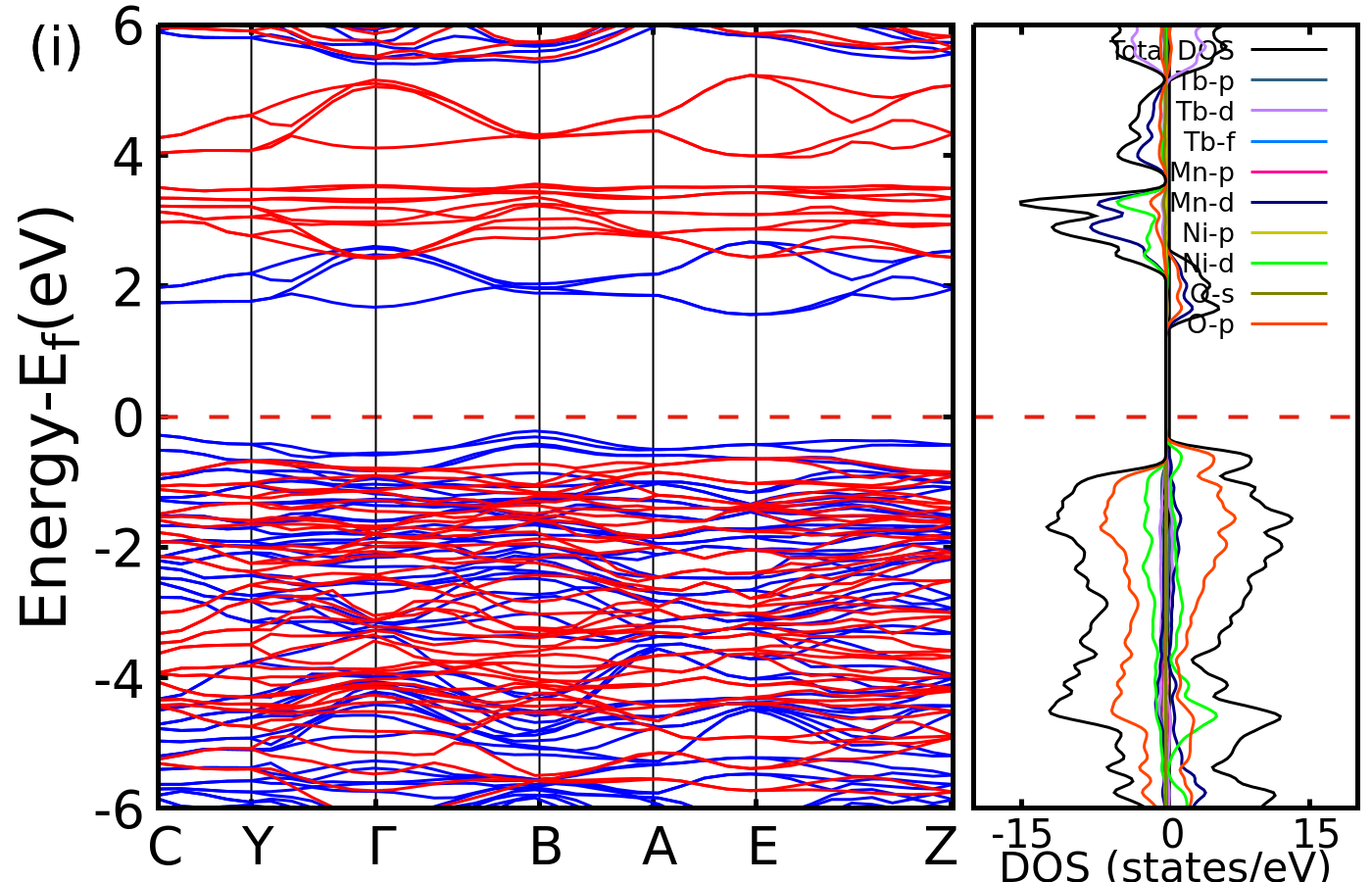}
\includegraphics[height=5.5cm,width=7.5cm]{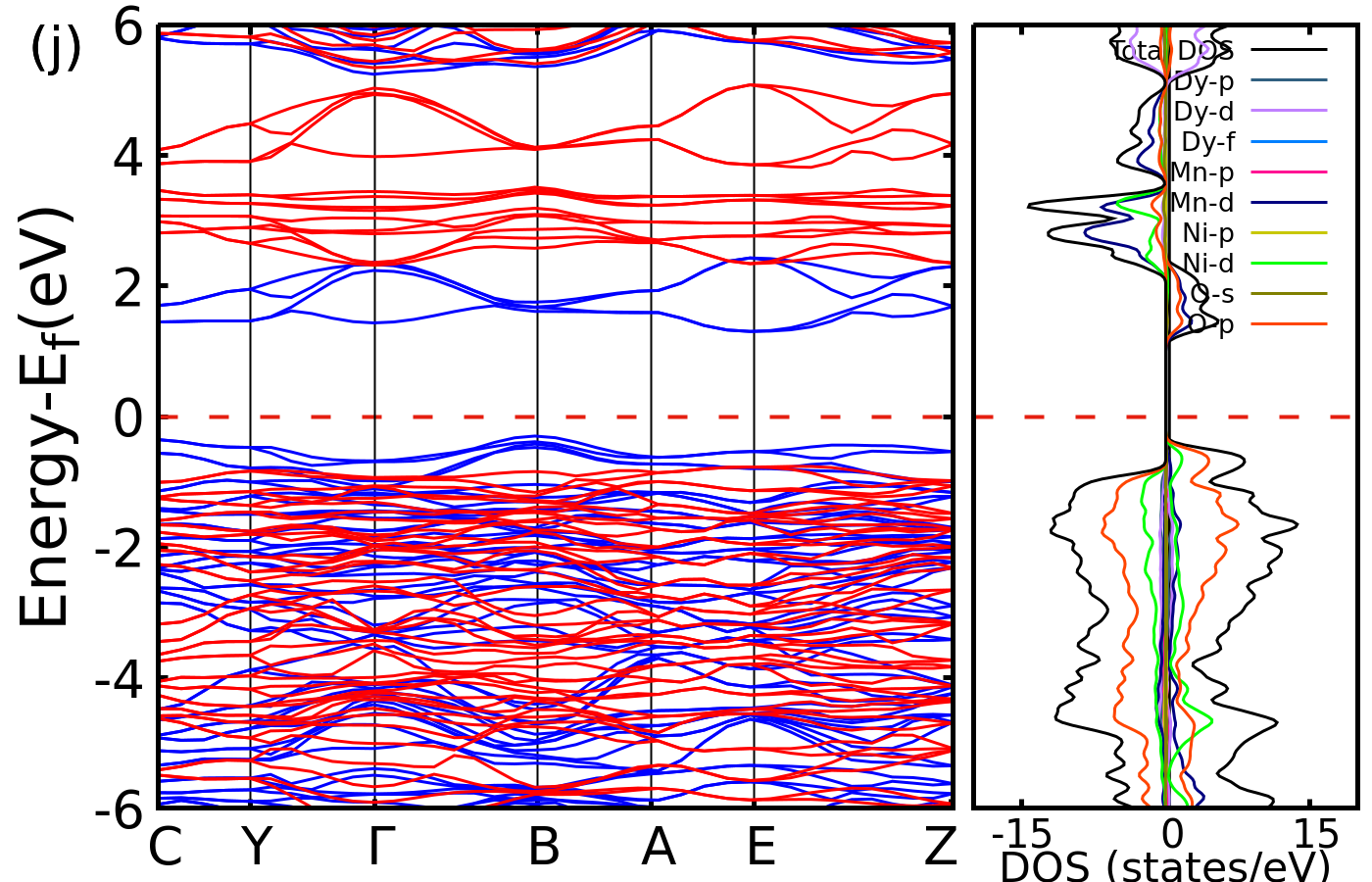}\\ \vspace*{0.3cm}
\includegraphics[height=5.5cm,width=7.5cm]{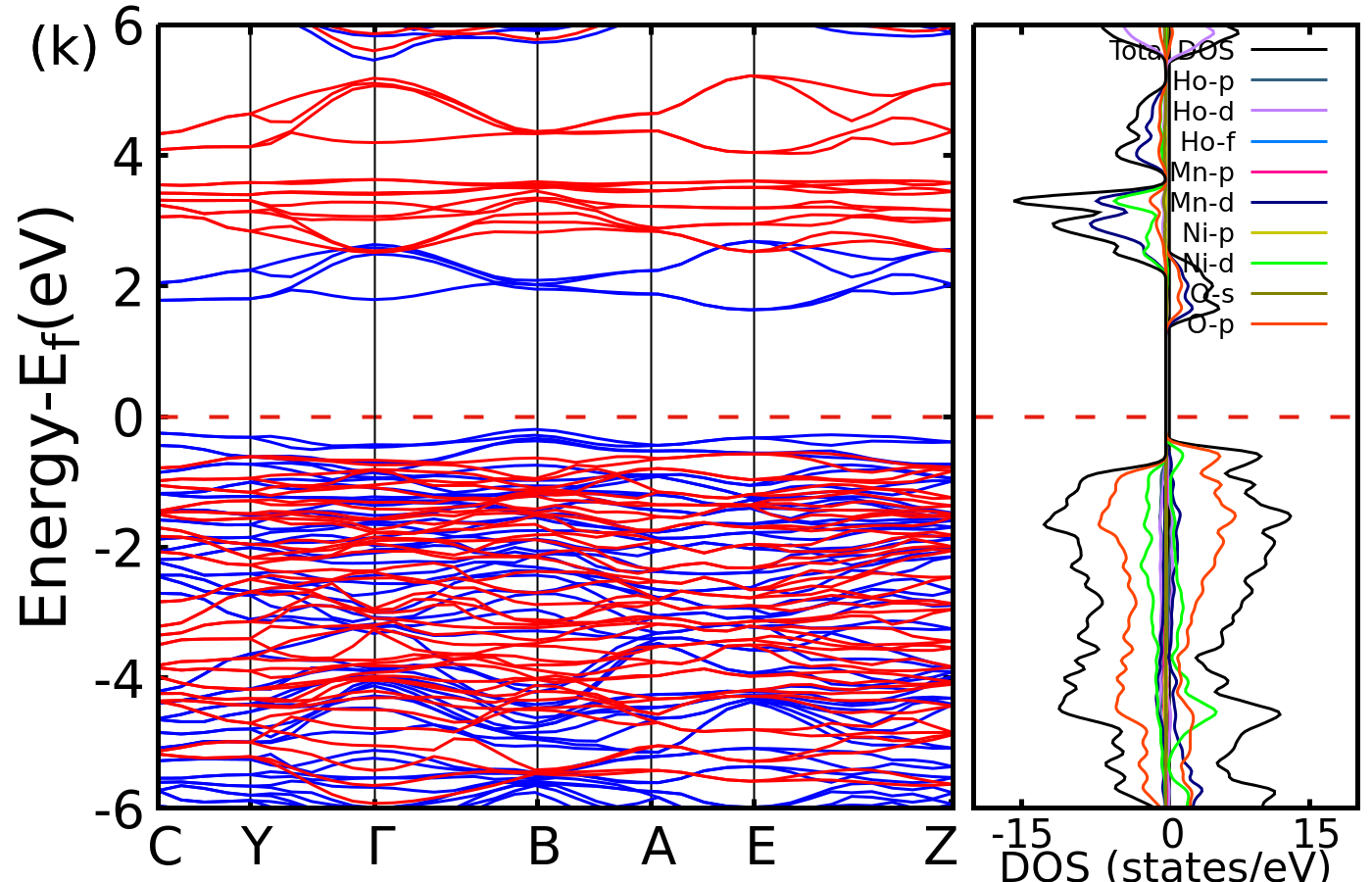}
\includegraphics[height=5.5cm,width=7.5cm]{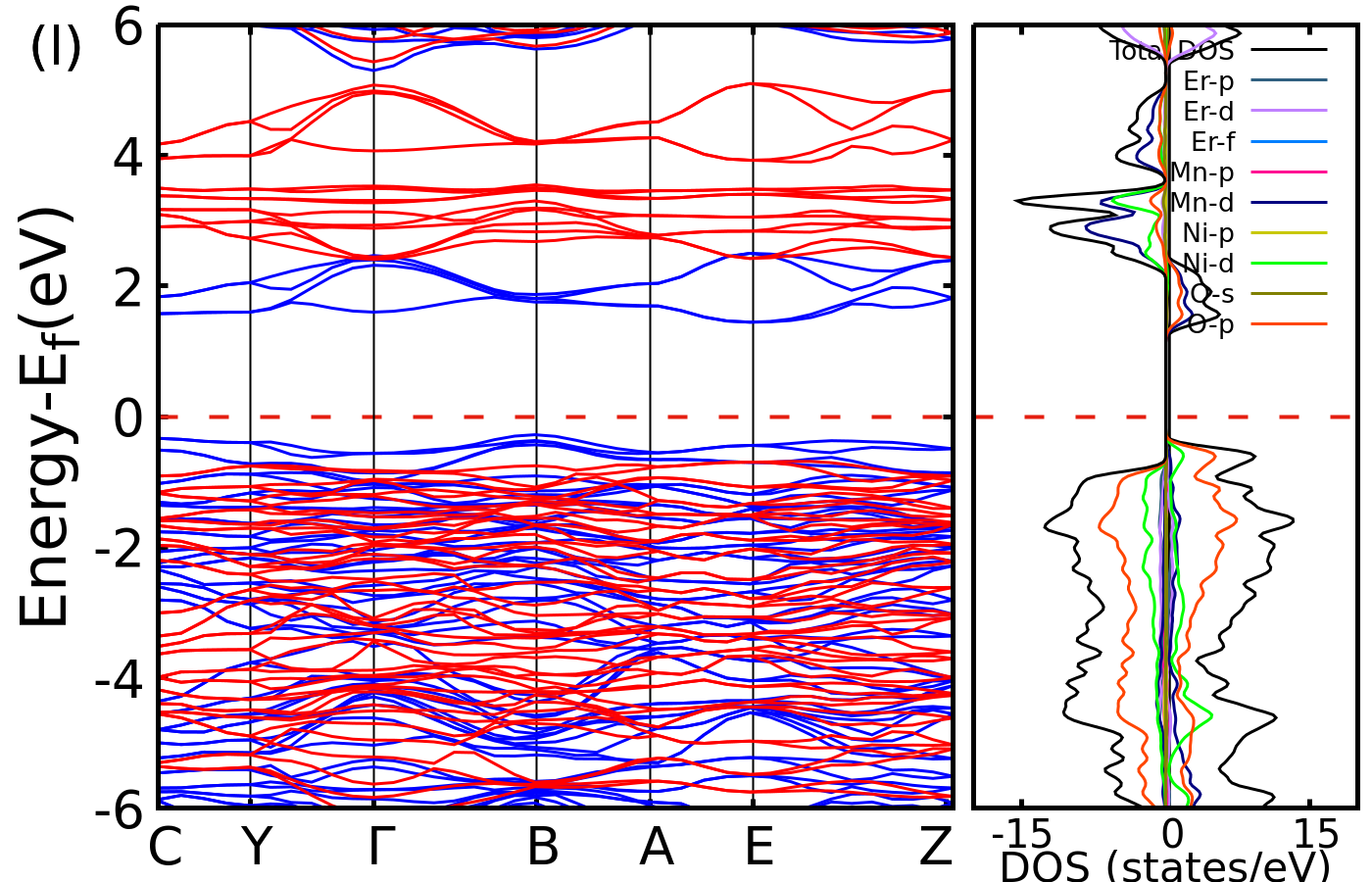}\\ \vspace*{0.3cm}
\end{center}
\end{figure}
\begin{figure}[H]
\begin{center}
\includegraphics[height=5.5cm,width=7.5cm]{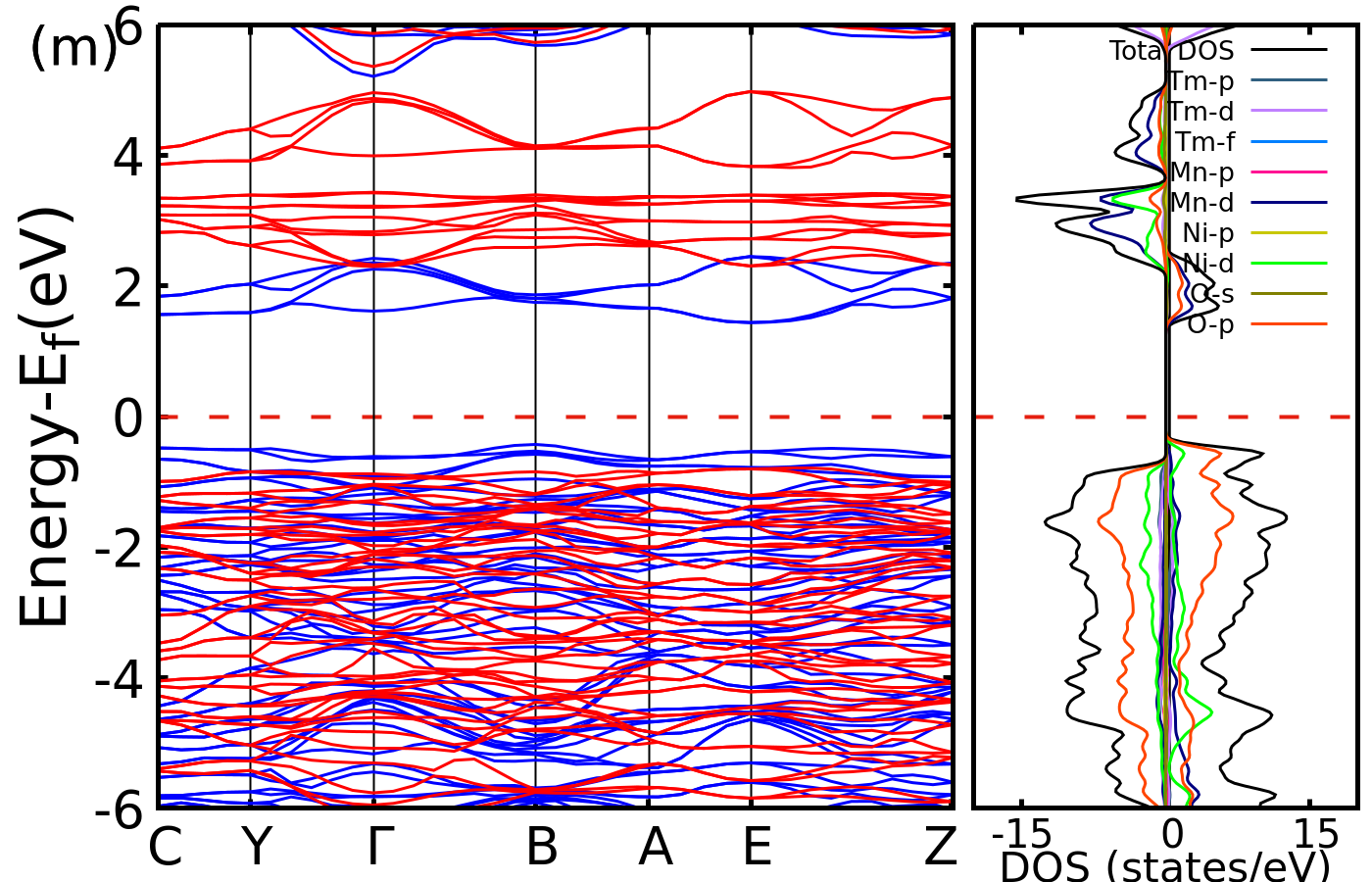}
\includegraphics[height=5.5cm,width=7.5cm]{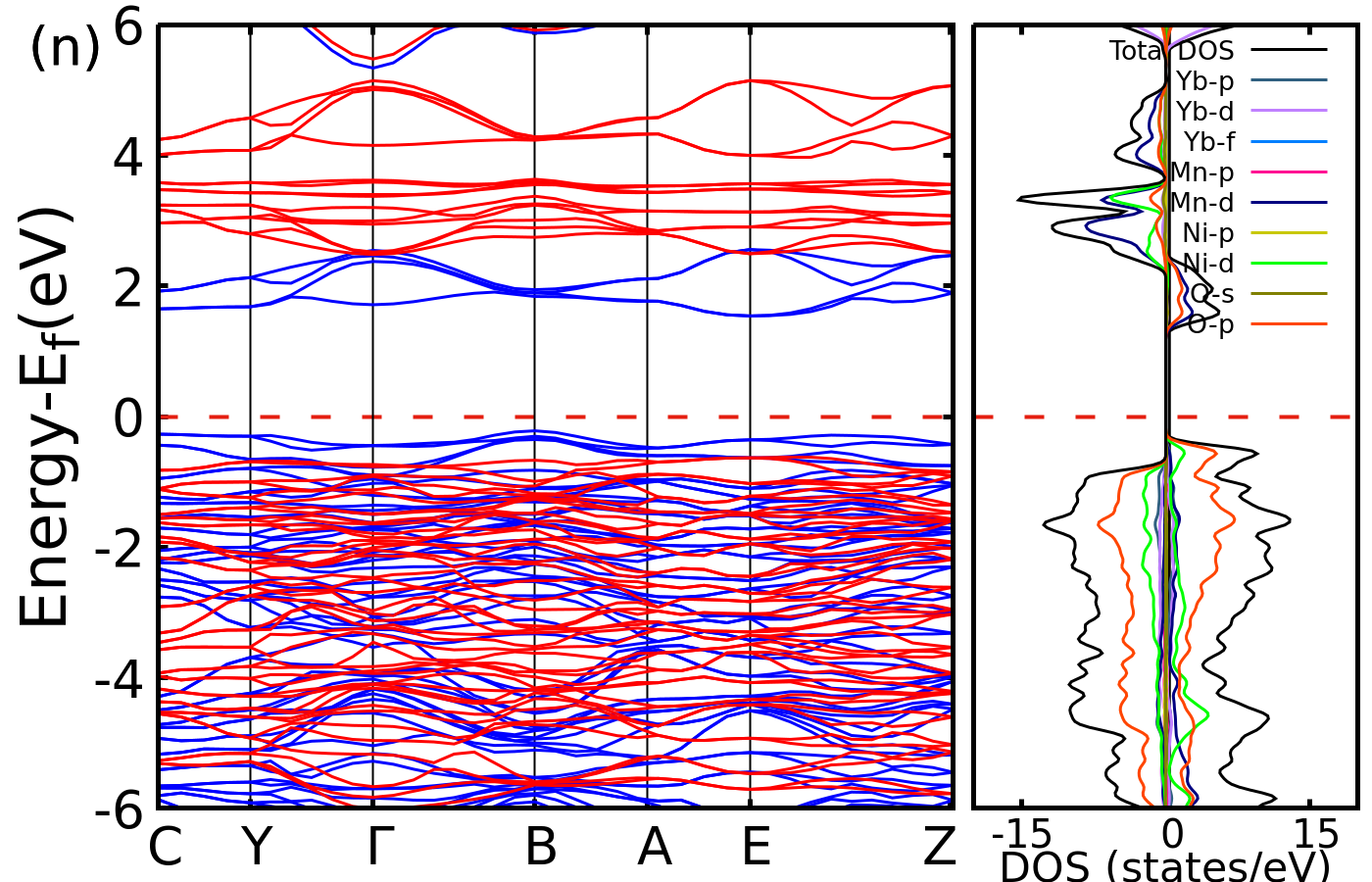}\\ \vspace*{0.3cm}
\includegraphics[height=5.5cm,width=7.5cm]{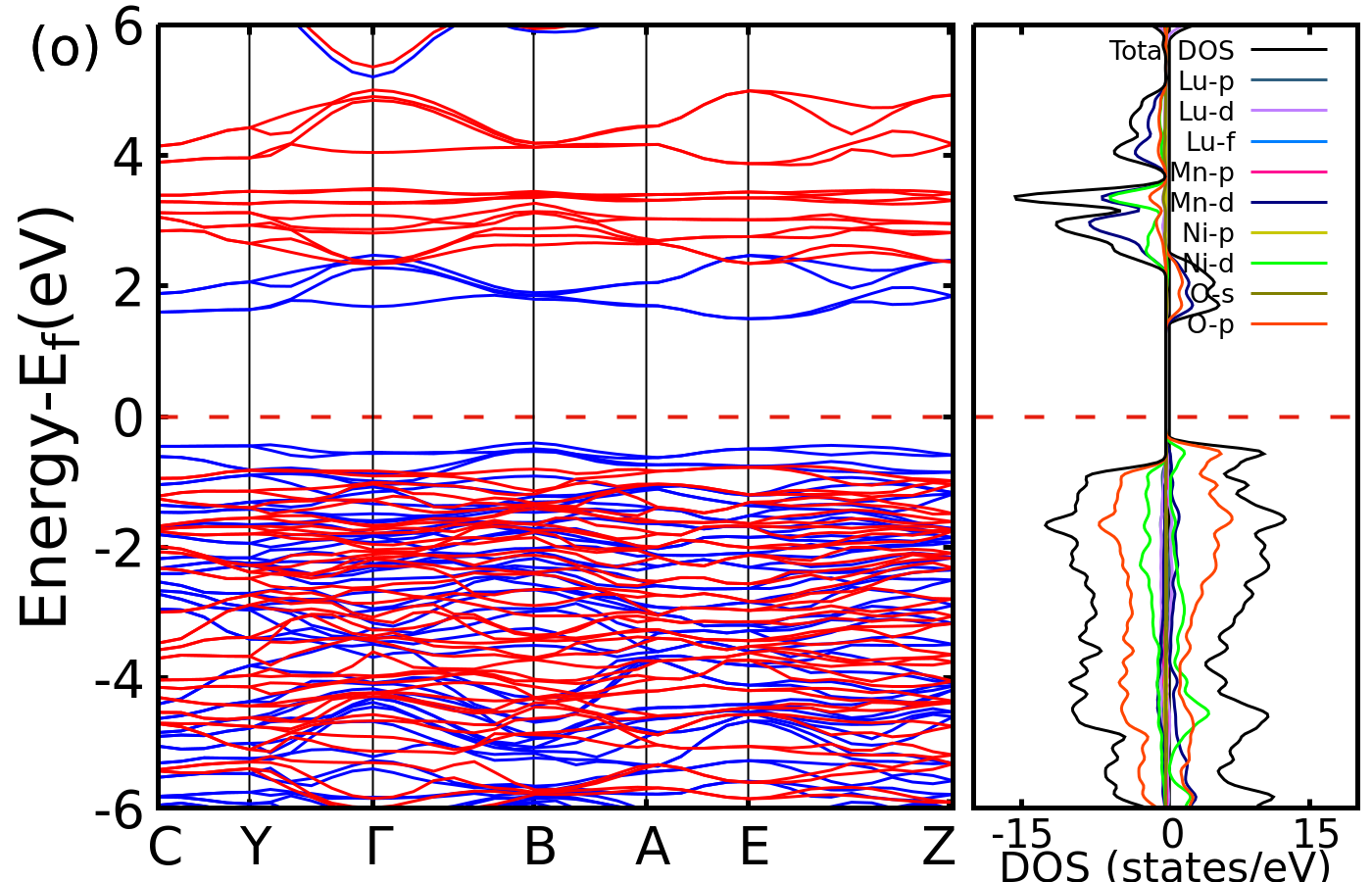}
\includegraphics[height=5.5cm,width=7.5cm]{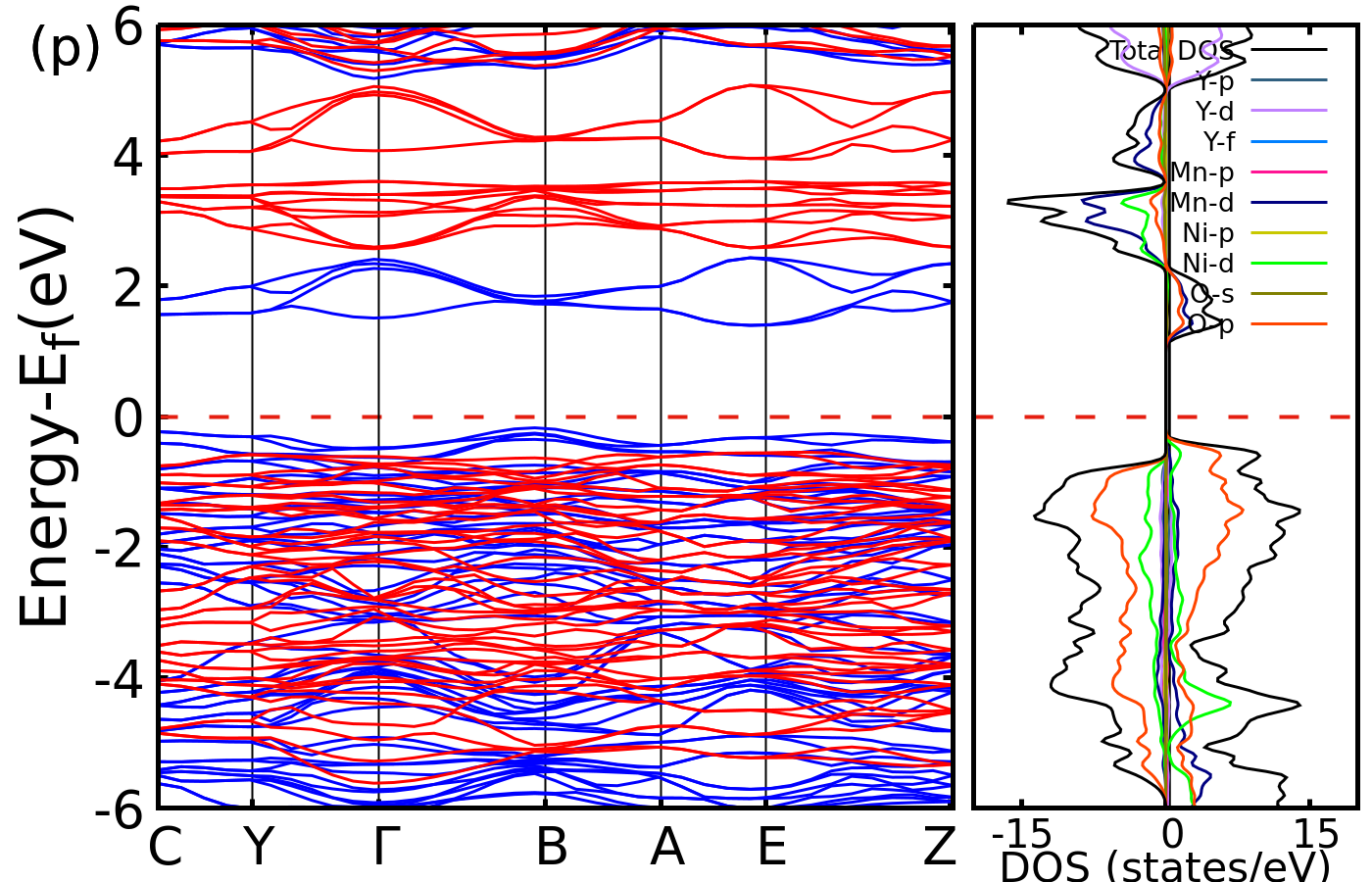}\\ \vspace*{0.3cm}
\caption{Band structures of Re$_2$MnNiO$_{6}$, where Re stands for (a)La, (b)Ce, (c)Pr, (d)Nd, (e)Pm, (f)Sm, (g)Eu, (h)Gd, (i)Tb, (j)Dy, (k)Ho, (l)Er, (m)Tm, (n)Yb, (o)Lu and (p)Y, without considering $f$-electrons in valence band.}
\label{fig:bs6}
\end{center}
\end{figure}
Furthermore, asymmetry of the spin up and spin down near the Fermi level in the density of states plot contribute to the magnetism, making these materials magnetic semiconductors \cite{MFJ16,Kittel86,Adhikari22}. The majority contribution to the density of states is from 2$p$-orbital of oxygen and 3$d$-orbitals of Ni and Mn. Ni-$3$d and O-$2$p shows major contribution to the valence band whereas $d$-orbitals of Mn and Ni dominate in the conduction band. Lanthanides contribution is observed at higher energy, far from Fermi-level, beyond 4 $eV$. Contribution to the density of states from the Re-metal is fairly low at the Fermi level as there was no role of $f$-electrons in valence band for the first set (see Fig. \ref{fig:bs6}).
The total and projected density of states (DOS) analysis reinforces these observations. The valence band displays a distinct peak just below the Fermi level, primarily arising from the hybridization between Ni $t_{2g}$ and O–2$p$ orbitals. In contrast, the conduction band is dominated by Mn $e_g$ states hybridized with O–2$p$ orbitals. Across the rare-earth series, a reduction in the ionic radius of the Re element induces an upward shift of the Mn–3$d$ states, leading to an increase in the band gap. The $p$–$d$ hybridization in the conduction band is predominantly attributed to Mn–3$d$ orbitals, whereas in the valence band it is governed by Ni–3$d$ orbitals, reflecting the fact that $d$–$d$ electronic interactions are intrinsically weaker than $p$–$d$ interactions. The Ni–3$d$ orbitals, possessing a more fully occupied 3$d$ shell at lower energies and in closer proximity to the O–2$p$ states, which enhances orbital overlap and covalency—accounting for their dominant role in the valence band. In contrast, the Mn–3$d$ orbitals, with a less occupied 3$d$ shell, are positioned at higher energies relative to O–2$p$, resulting in a larger energy separation. Consequently, the valence band is primarily characterized by Ni–O $p$–$d$ hybridization, while the conduction band shows a stronger contribution from Ni–Mn interactions in conjunction with rare-earth orbital states \cite{Zhou2017}.

In the second set of calculations, the inclusion of $f$-electrons in the valence band produces substantial modifications in the band dispersion. In the absence of $f$-orbitals, both spin channels exhibit finite band gaps; however, when $f$-orbitals are included, the system displays a spin-polarized electronic structure in which one spin channel remains metallic while the other becomes semiconducting (Table \ref{tab:bandgapsRe6}).
\begin{table}[H]
\begin{center}
\setlength{\tabcolsep}{1.85pt}
\begin{tabular}{|c|c|c|c|c|c|c|c|c|c|c|c|c|c|c|}
\hline 
Re & Ce & Pr & Nd & Sm & Pm & Eu & Gd & Tb & Dy & Ho & Er & Tm & Yb & Lu \\ 
\hline 
Spin-up & 0 & 0 & 0 & 0 & 0 & 0 & 1.77 & 0 & 1.64 & 2.14 & 3.06 & 2.07 & 1.76 & 1.87 \\ 
\hline 
Spin-down & 2.08 & 3.26 & 3.18 & 2.53 & 0 & 3.07 & 2.98 & 0 & 0 & 0 & 0 & 0 & 0 & 3.05 \\ 
\hline 
\end{tabular} 
\caption{Bandgaps (in eV) for RMNO with $f$-electrons.}
\label{tab:bandgapsRe6}
\end{center}
\end{table}
The presence of nearly flat bands, characterized by low group velocity and high effective mass, reflects the localized nature of $d$- and $f$-electron states. Spin-polarized exchange interactions lift the spin degeneracy, with the resulting energy splitting being directly proportional to the initial magnetic moment.
\begin{figure}[H]
\begin{center}
\includegraphics[height=5.5cm,width=7.5cm]{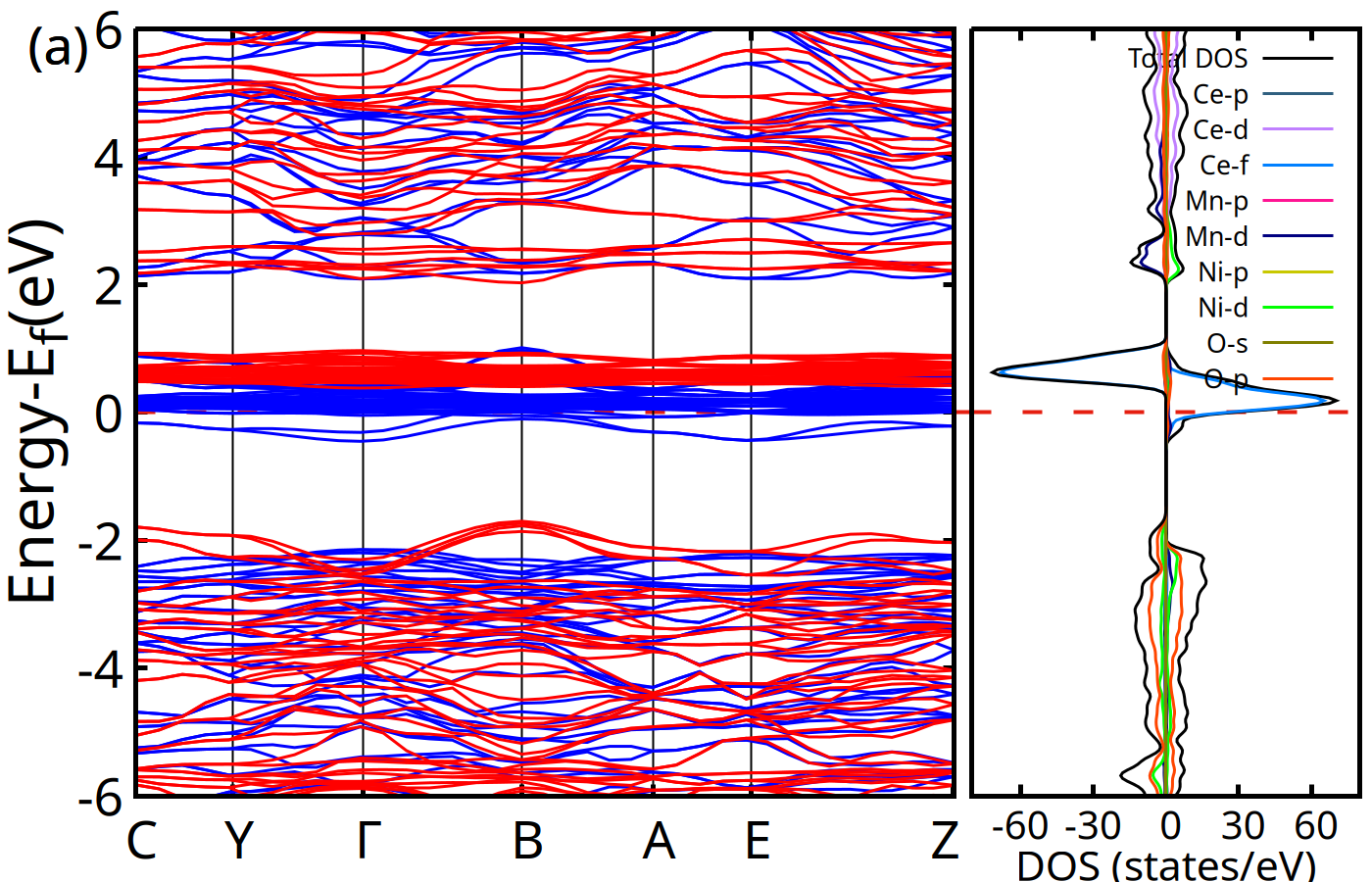}
\includegraphics[height=5.5cm,width=7.5cm]{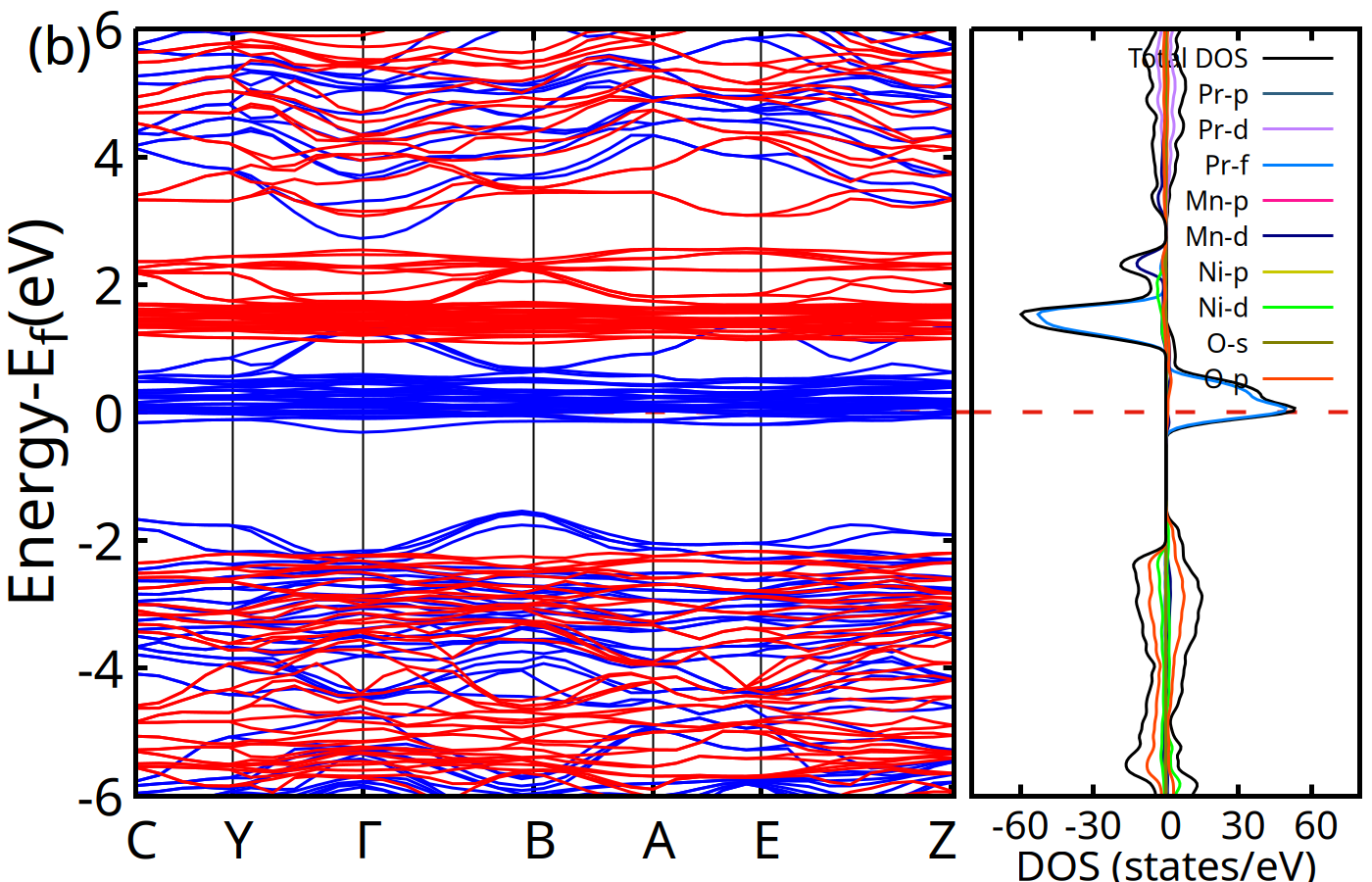}\\ \vspace*{0.3cm}
\includegraphics[height=5.5cm,width=7.5cm]{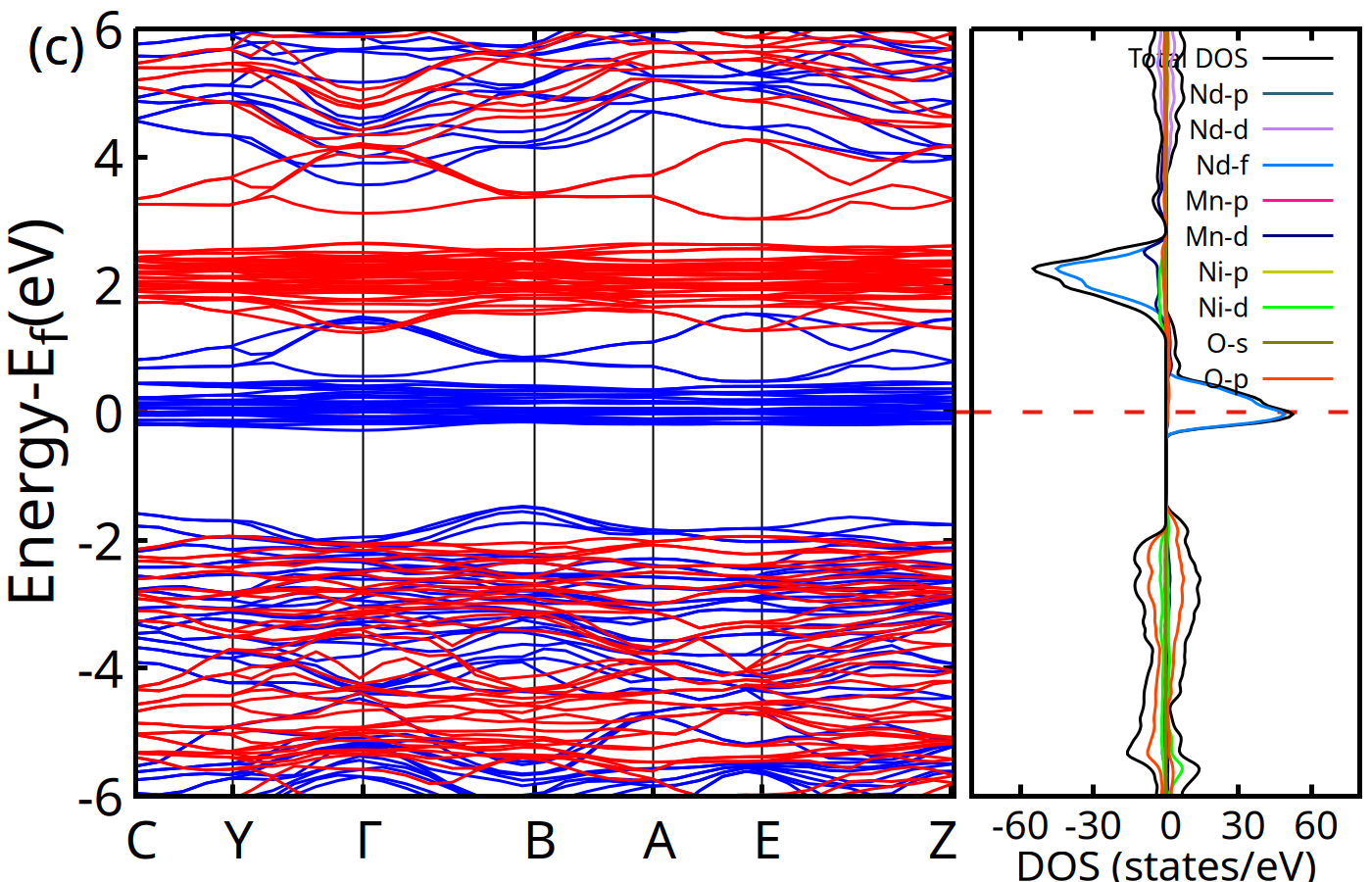}
\includegraphics[height=5.5cm,width=7.5cm]{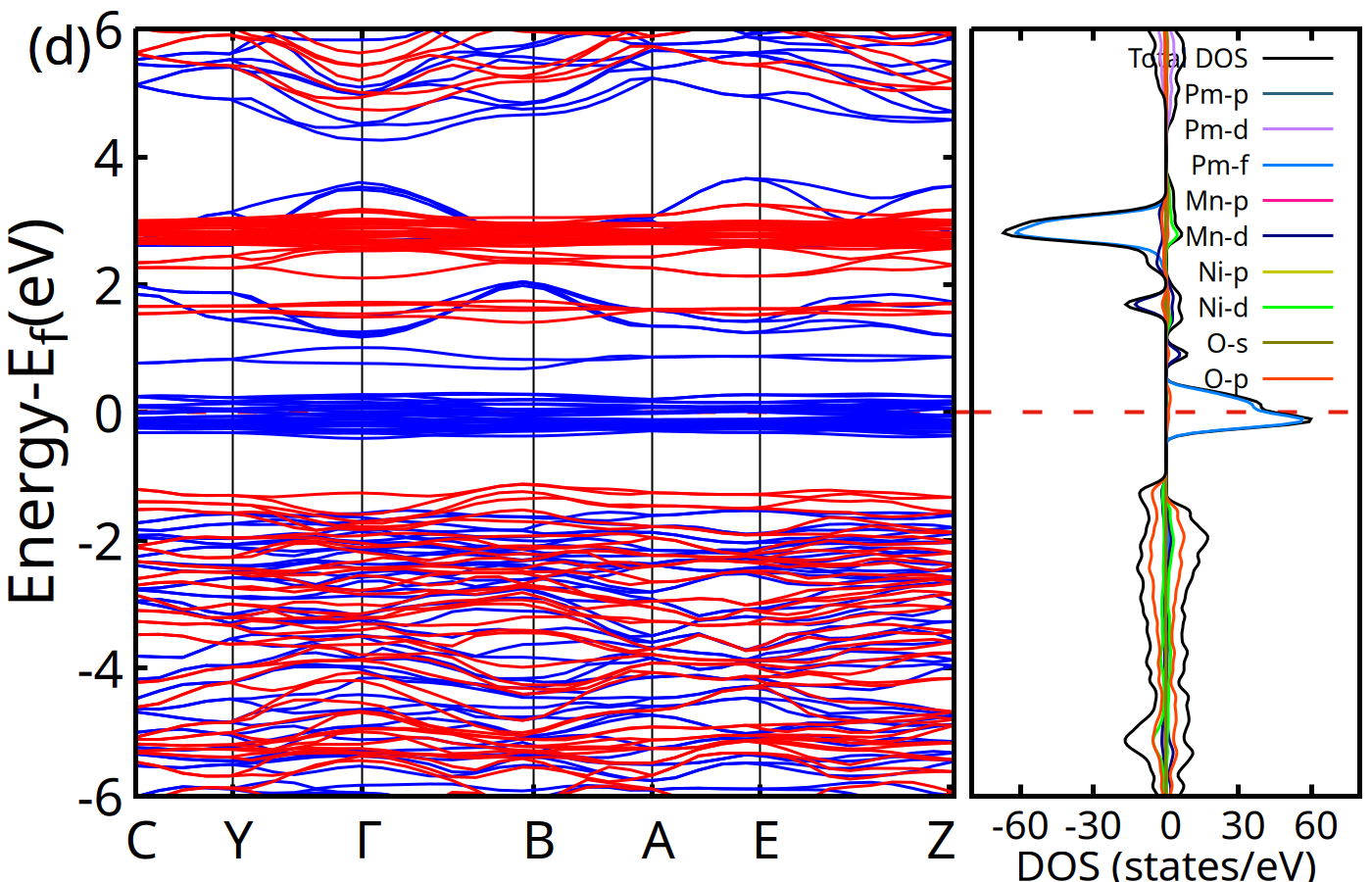}\\ \vspace*{0.3cm}
\includegraphics[height=5.5cm,width=7.5cm]{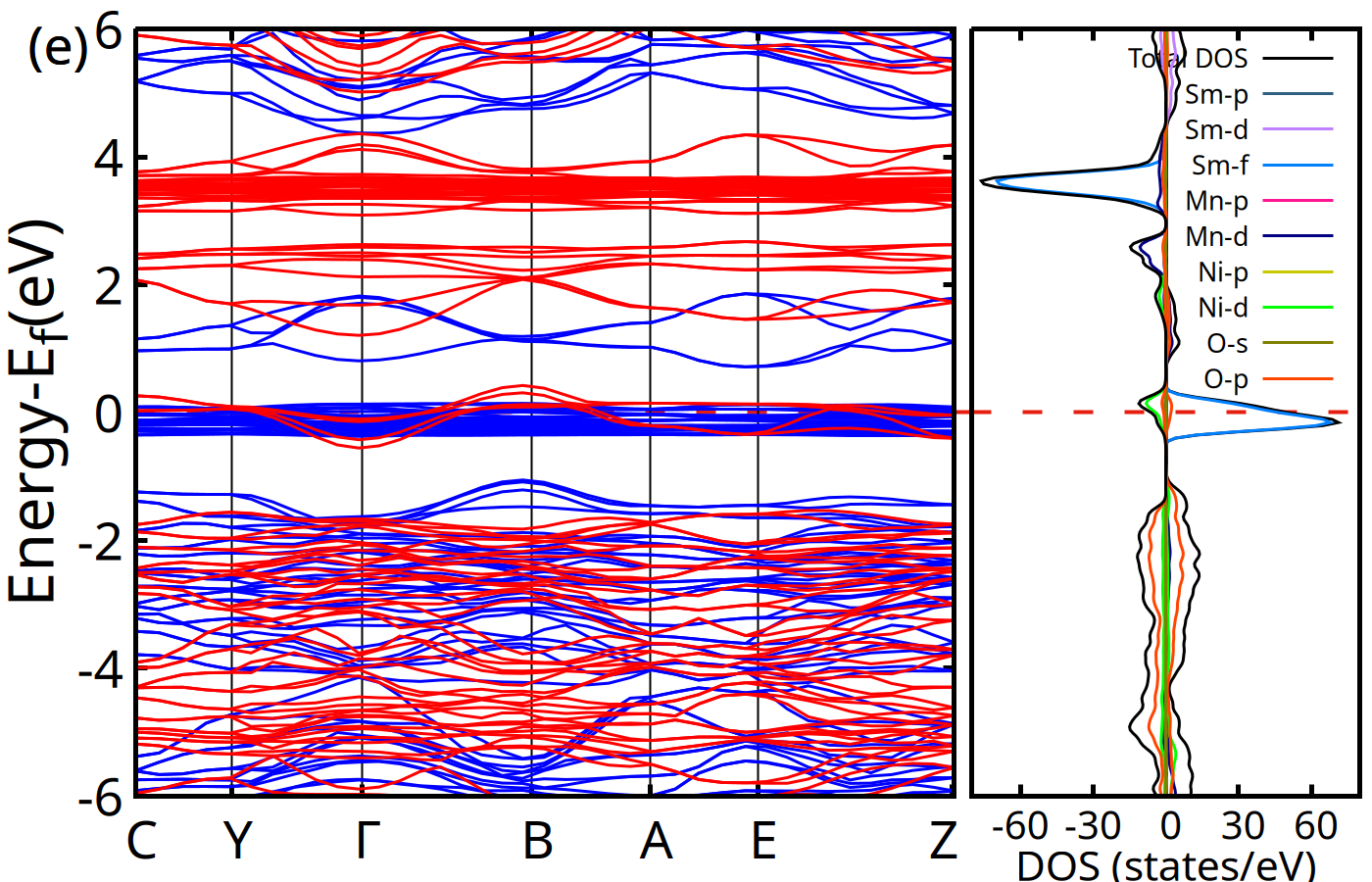}
\includegraphics[height=5.5cm,width=7.5cm]{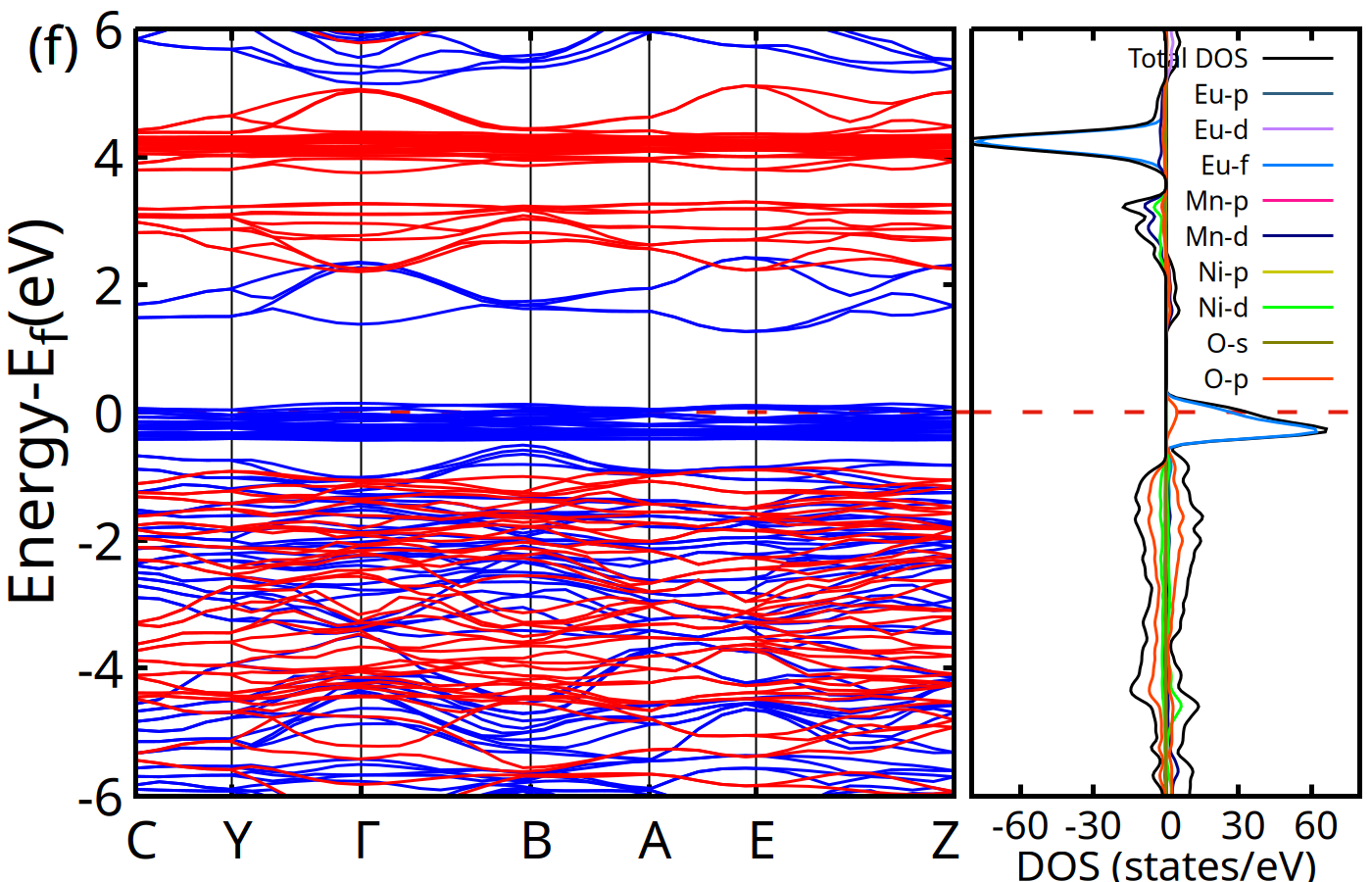}\\ \vspace*{0.3cm}
\end{center}
\end{figure}
\begin{figure}[H]
\begin{center}
\includegraphics[height=5.5cm,width=7.5cm]{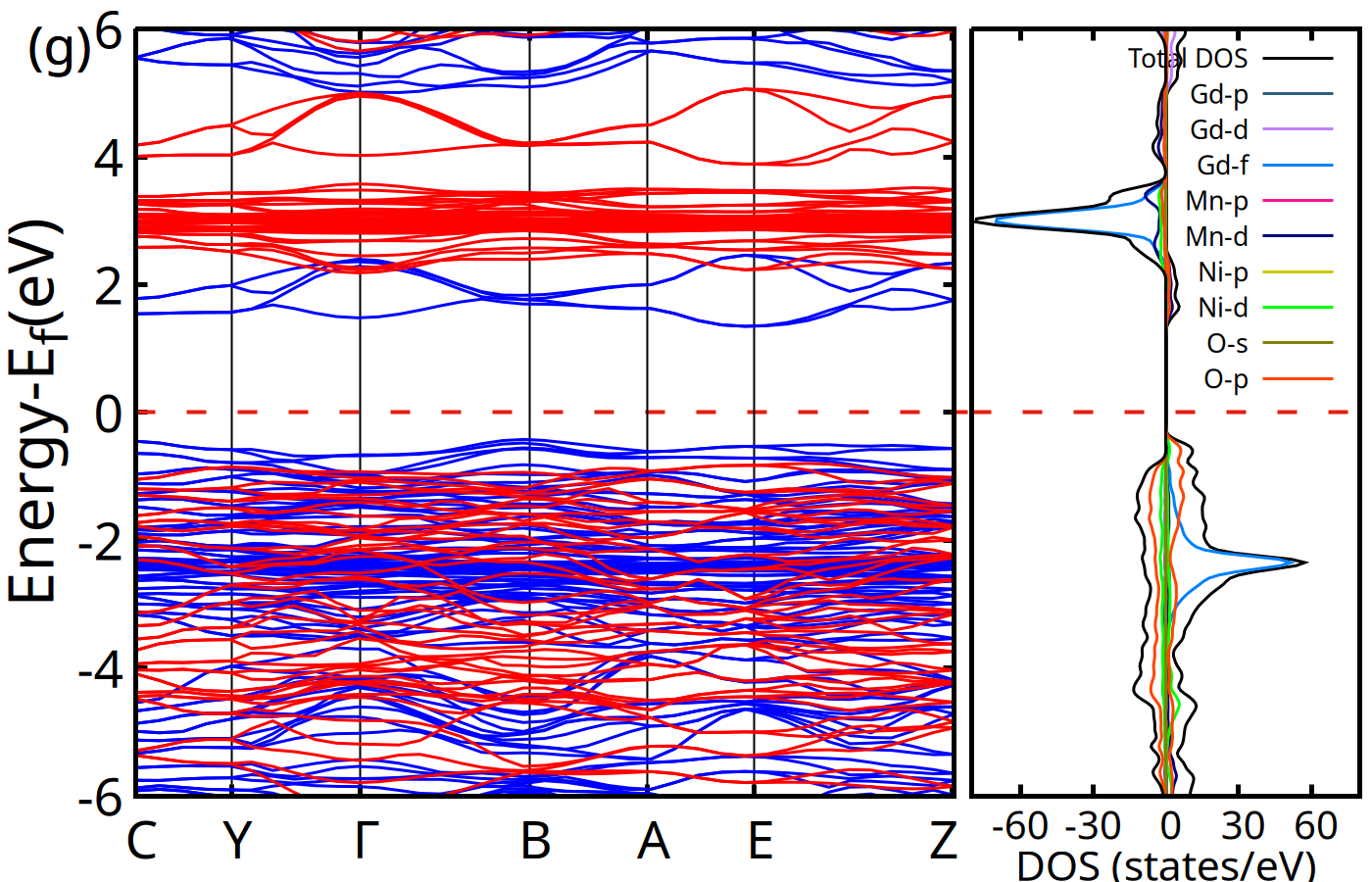}
\includegraphics[height=5.5cm,width=7.5cm]{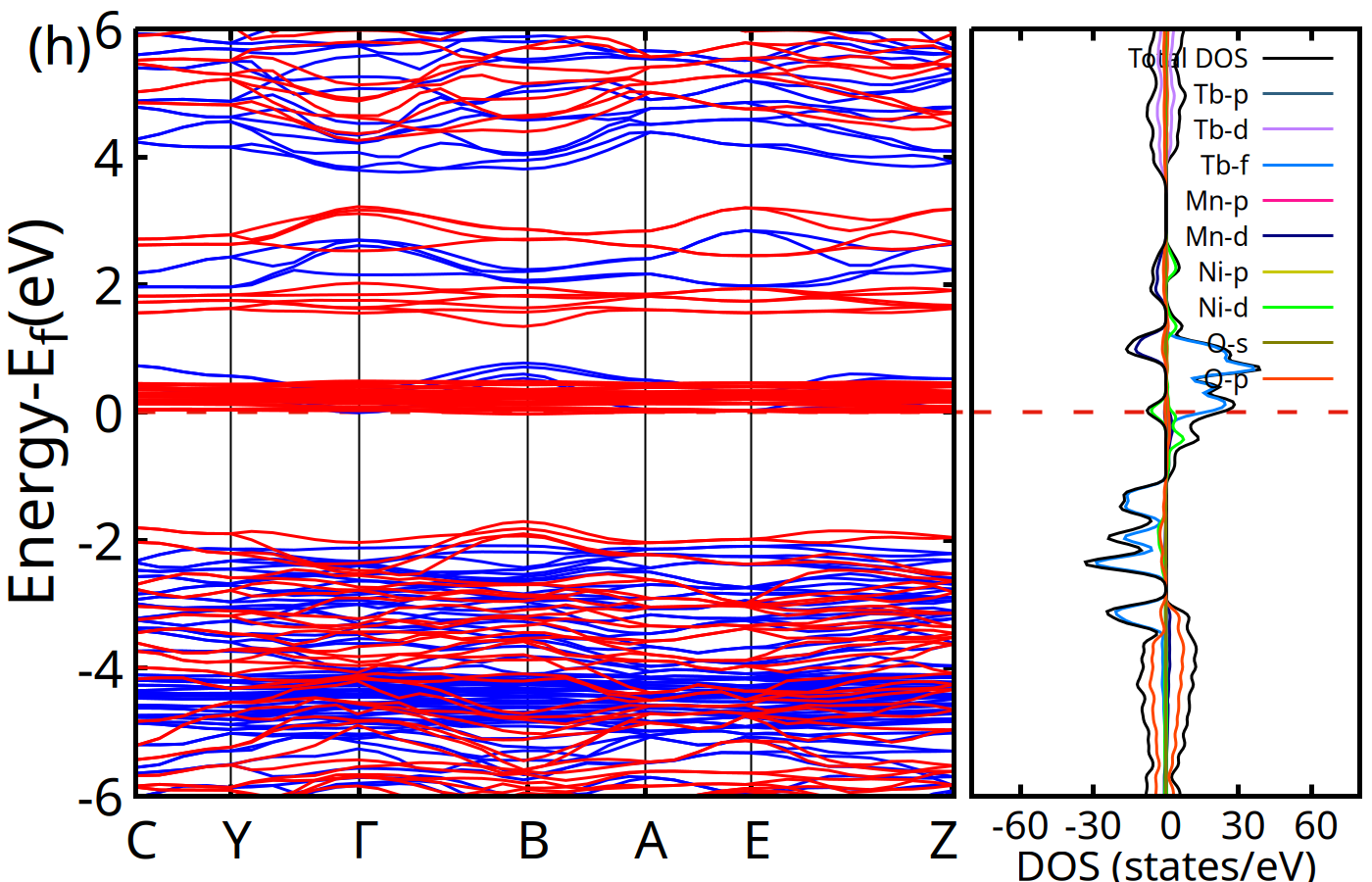}\\ \vspace*{0.3cm}
\includegraphics[height=5.5cm,width=7.5cm]{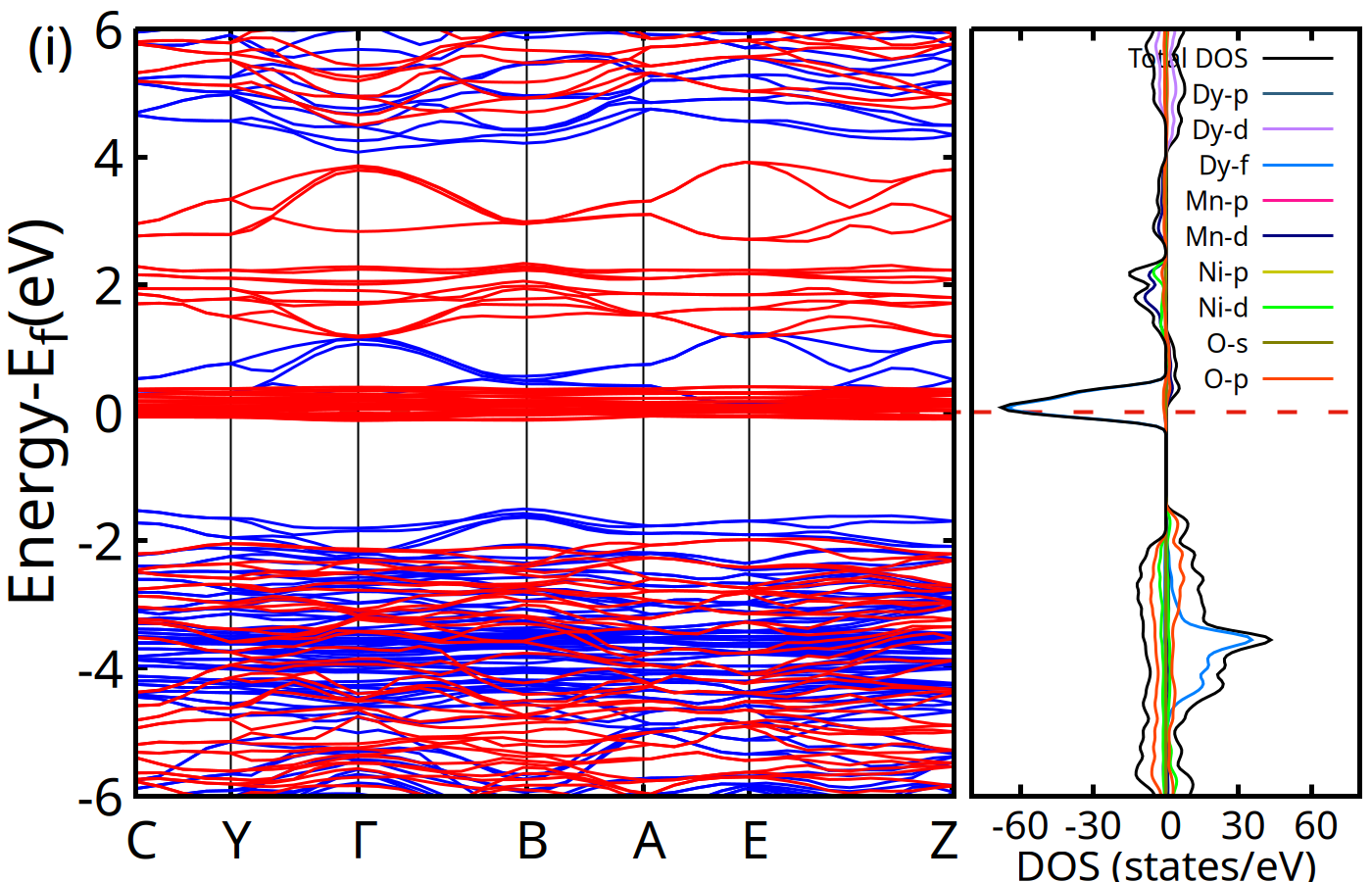} 
\includegraphics[height=5.5cm,width=7.5cm]{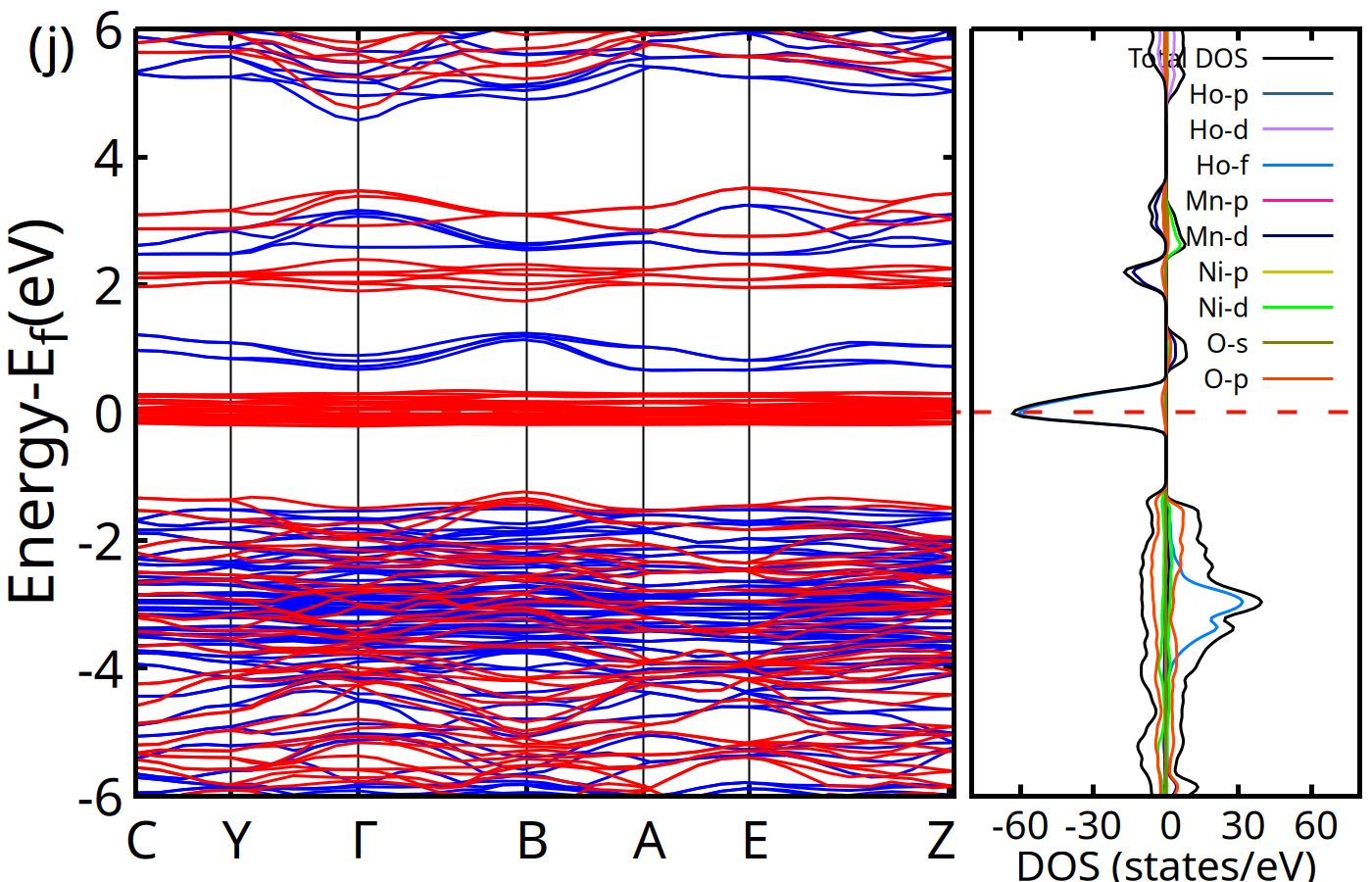}\\ \vspace*{0.3cm}
\includegraphics[height=5.5cm,width=7.5cm]{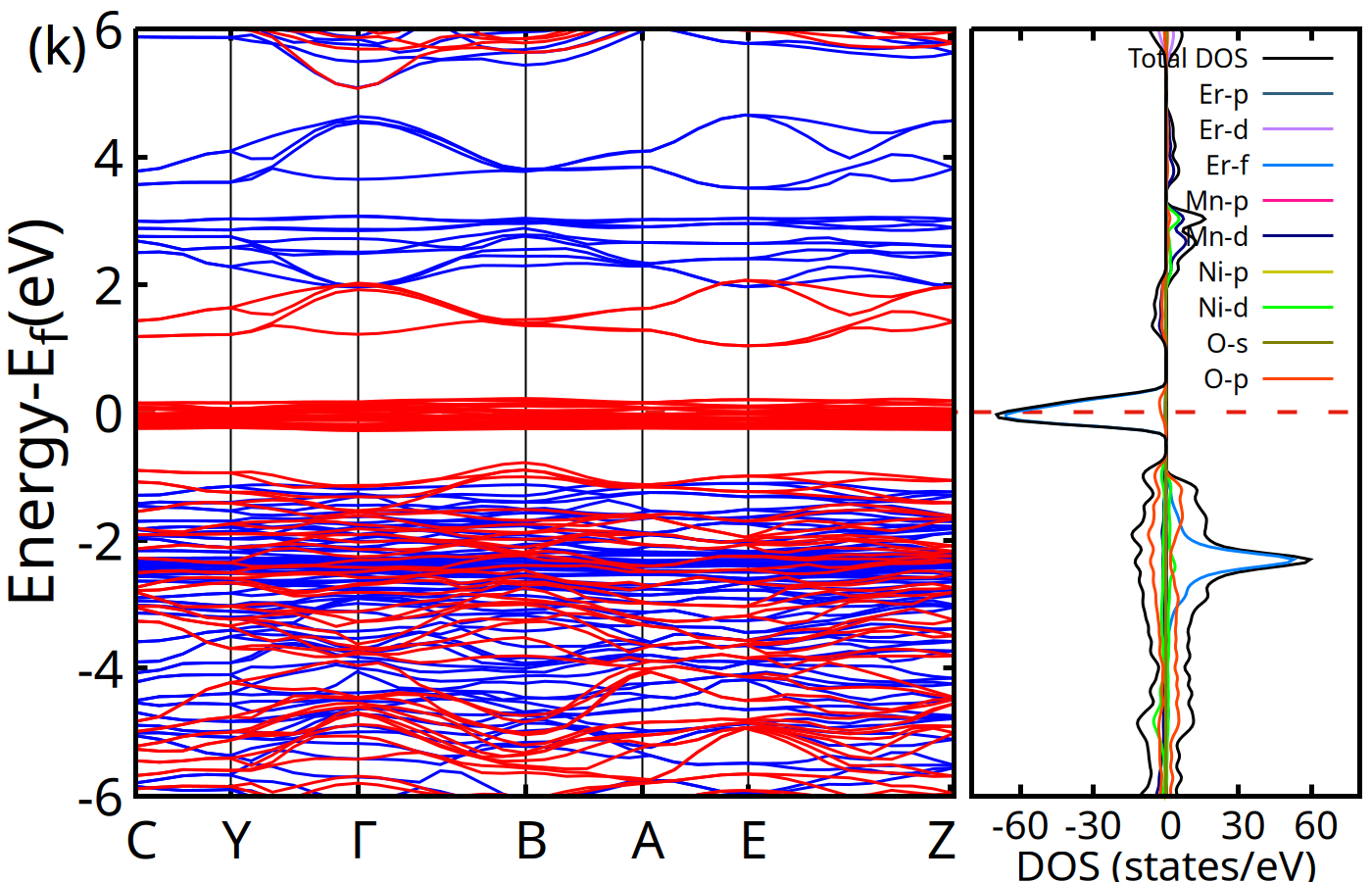}
\includegraphics[height=5.5cm,width=7.5cm]{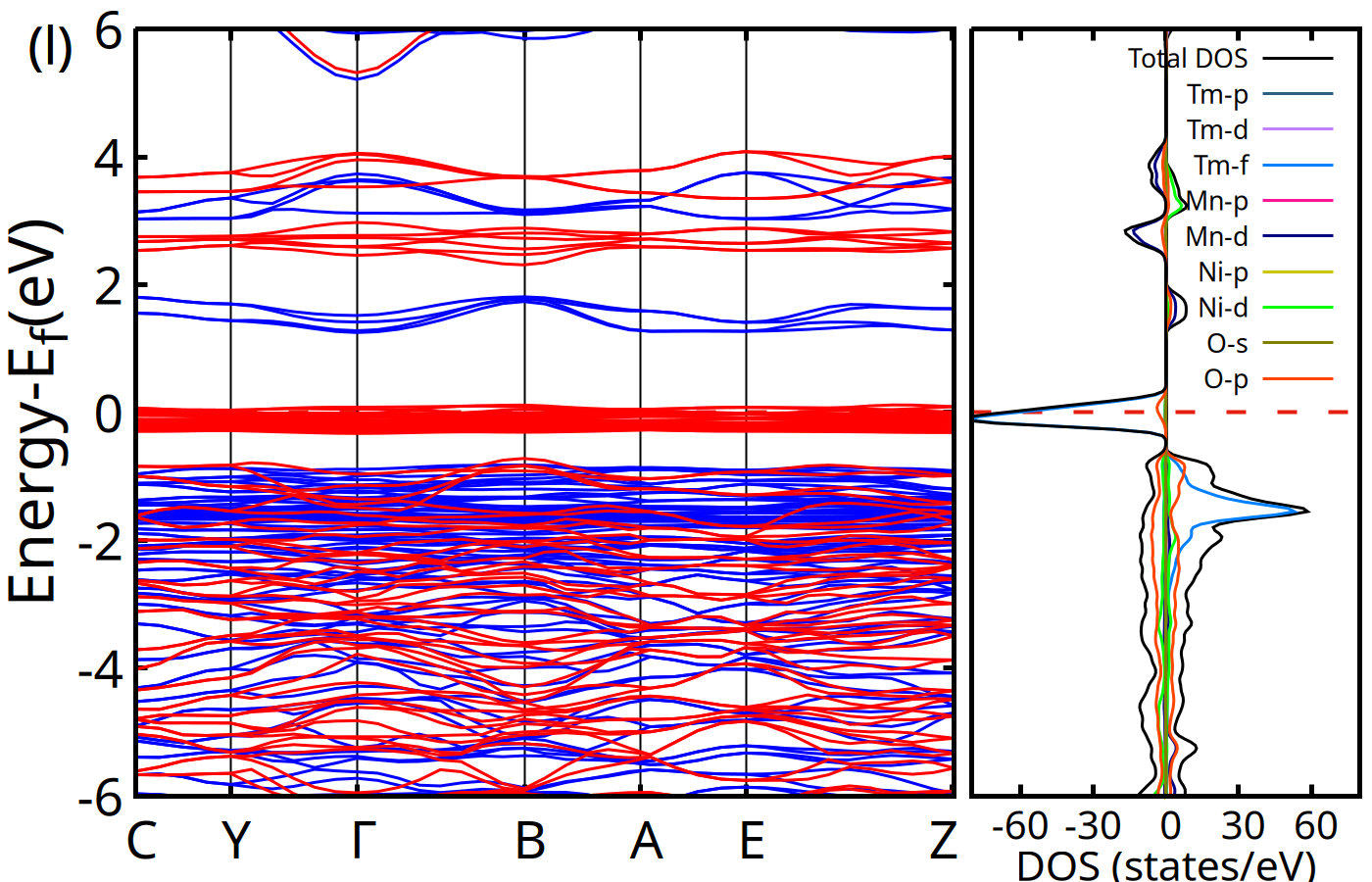}\\ \vspace*{0.3cm}
\includegraphics[height=5.5cm,width=7.5cm]{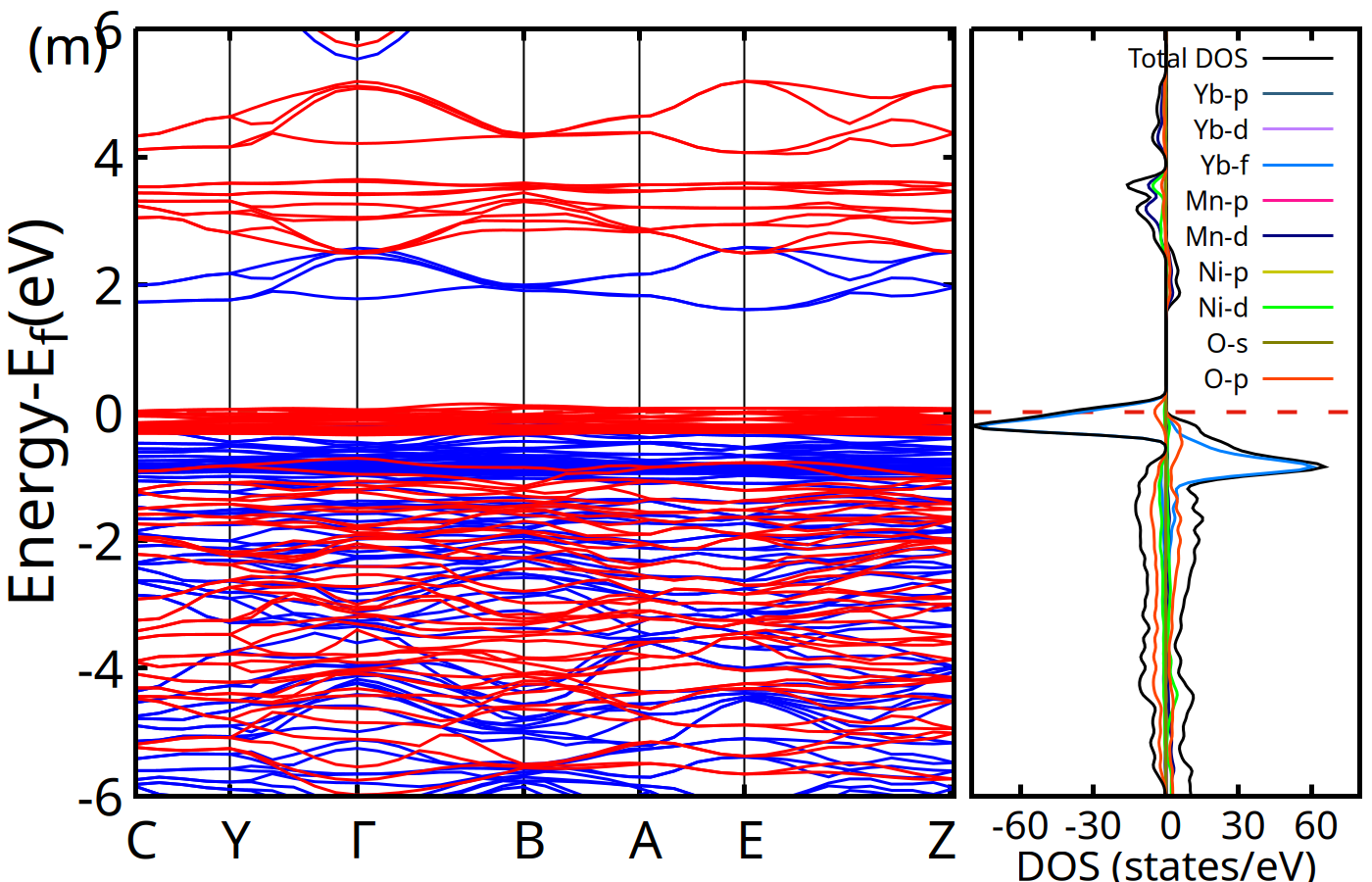}
\includegraphics[height=5.5cm,width=7.5cm]{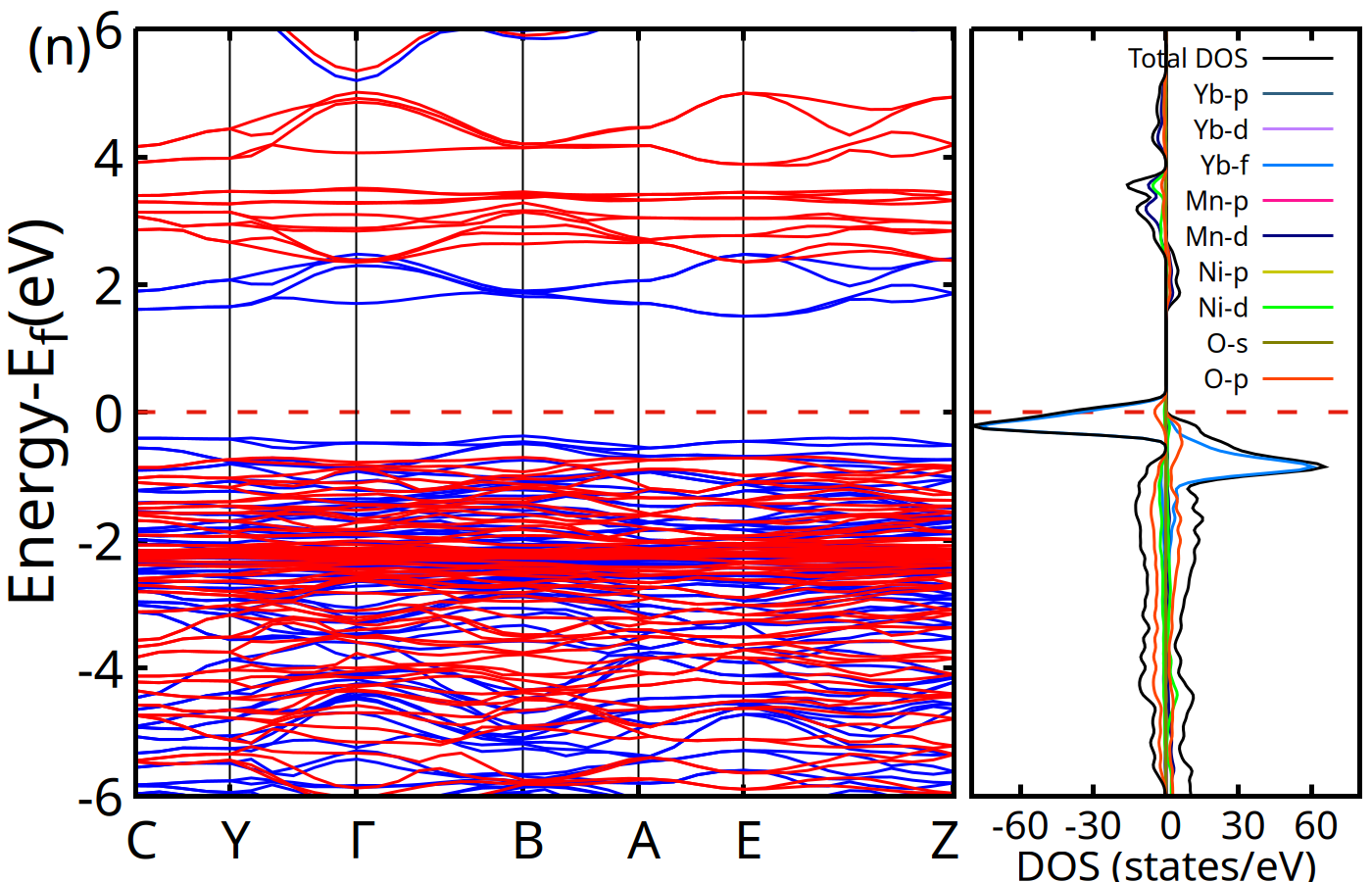}\\ \vspace*{0.3cm}
\caption{Band structures of Re$_2$MnNiO$_{6}$, where Re stands for (a)Ce, (b)Pr, (c)Nd, (d)Pm, (e)Sm, (f)Eu, (g)Gd, (h)Tb, (i)Dy, (j)Ho, (k)Er, (k)Tm, (m)Yb and (n)Lu with $f$-electrons in valence band.}
\label{fig:bs6_Re}
\end{center}
\end{figure}
A pronounced density of bands in the vicinity of the Fermi level originates from the hybridization of $f$-orbitals with the orbitals of other constituent atoms. The number of states in this dense region exceeds the nominal count of valence $f$-electrons, indicating strong $f$-hybridization that plays a decisive role in determining the band gap (Fig. \ref{fig:bs6_Re}). As $f$-hybridization arises from weak interactions between low-energy orbitals, its effects are most prominent near the Fermi level \cite{Takahashi25}. This type of weak hybridization is known as Kondo hybridization, which can open a narrow gap at the Fermi energy. Consequently, the system develops narrow quasiparticle bands near the Fermi level, corresponding to the dense spin-resolved states observed around 0 $eV$\cite{Frantzeskakis13}.
Overall, the electronic structure analysis, including band gap evaluation, orbital-resolved density of states, and spin asymmetry, revealed critical insights into the material's semiconducting metallic behavior, and potential magnetic ordering, and bonding characteristics. It also indicates the strong metal-oxygen hybridization owing to the electronic configuration of oxygen and its electronegative nature. 

\subsection{Phonon Dispersion}
\begin{figure}[H]
\begin{center}
\includegraphics[height=5.5cm,width=7.5cm]{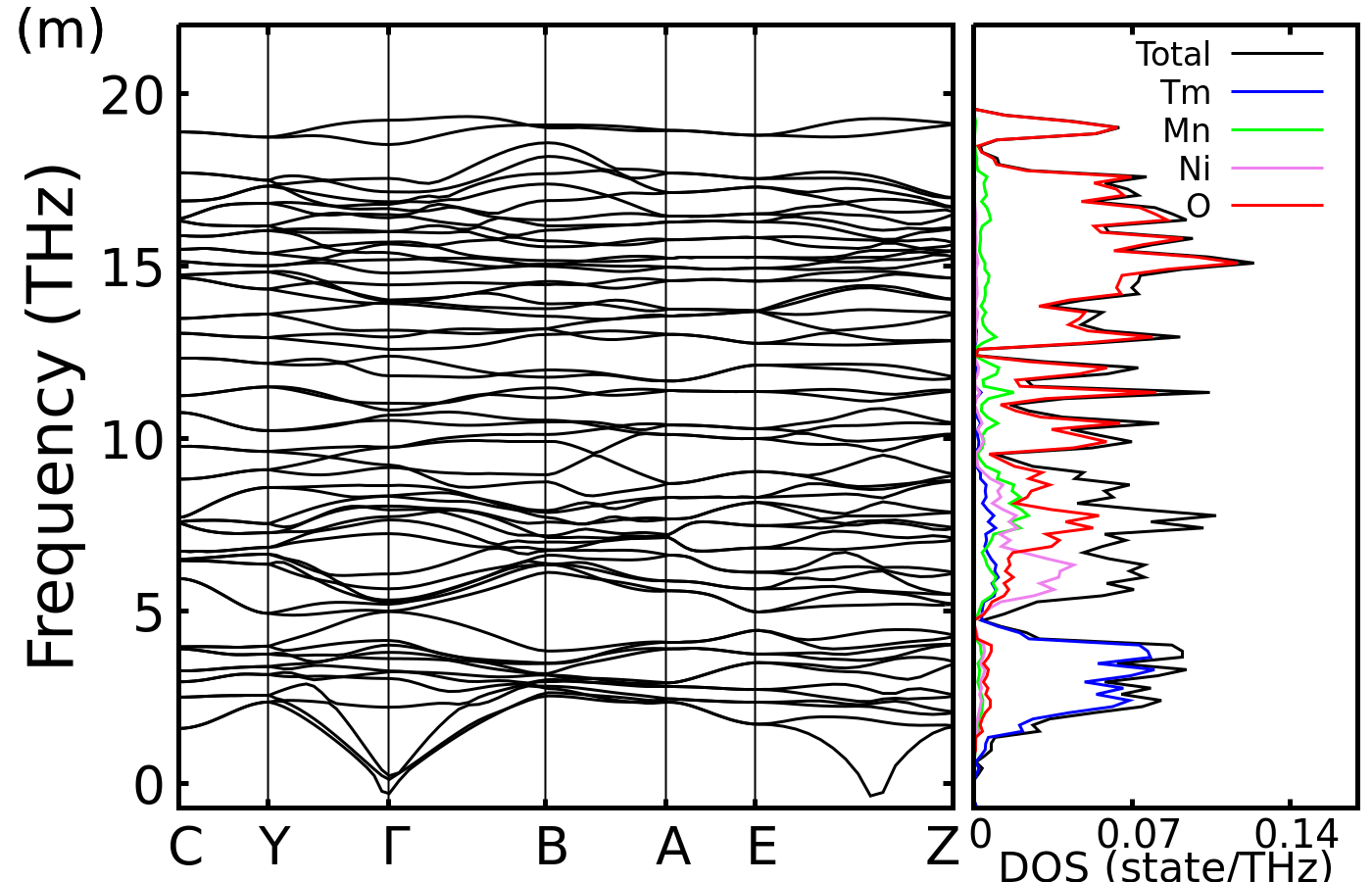}
\includegraphics[height=5.5cm,width=7.5cm]{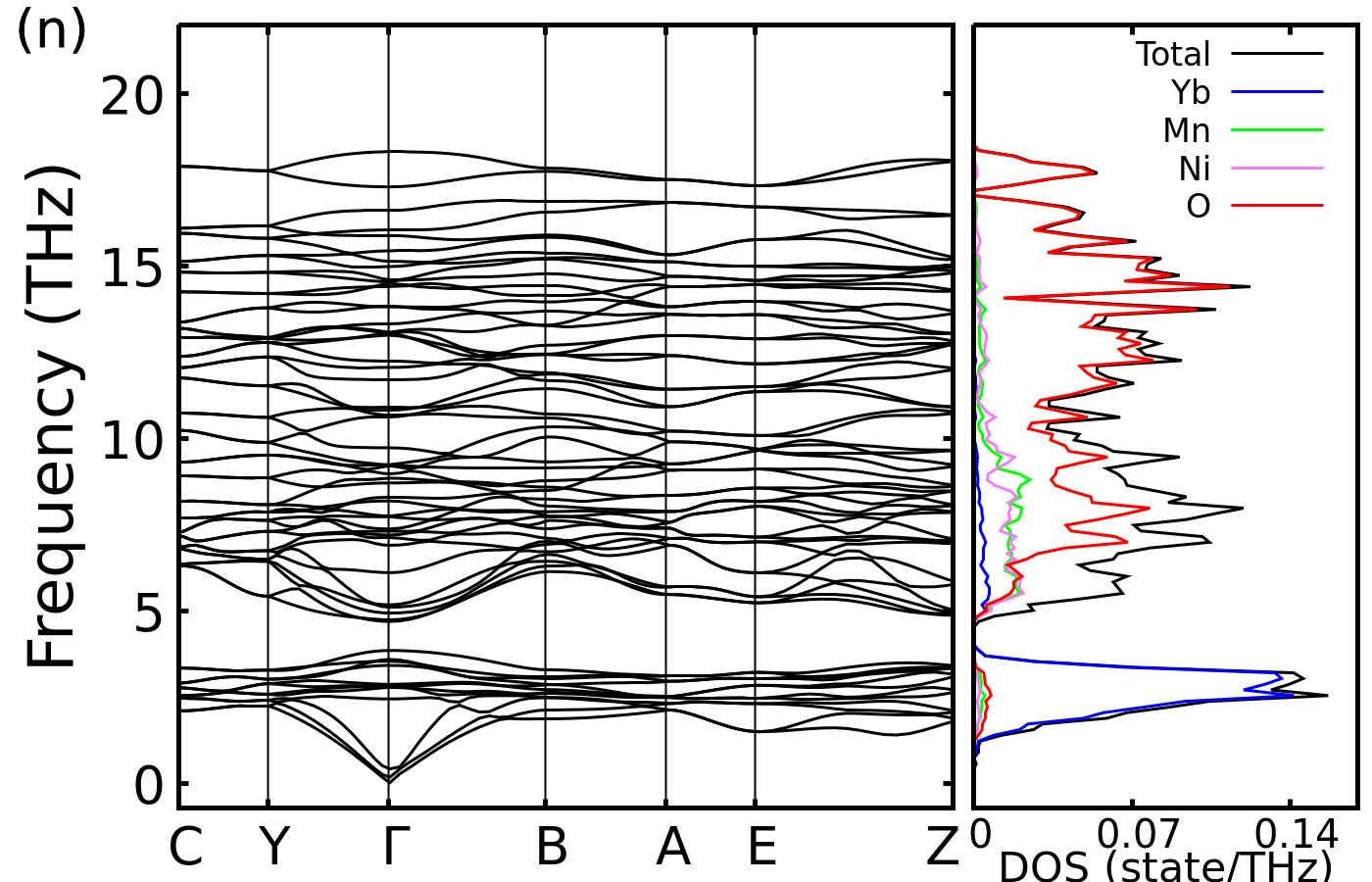}\\ \vspace*{0.3cm}
\includegraphics[height=5.5cm,width=7.5cm]{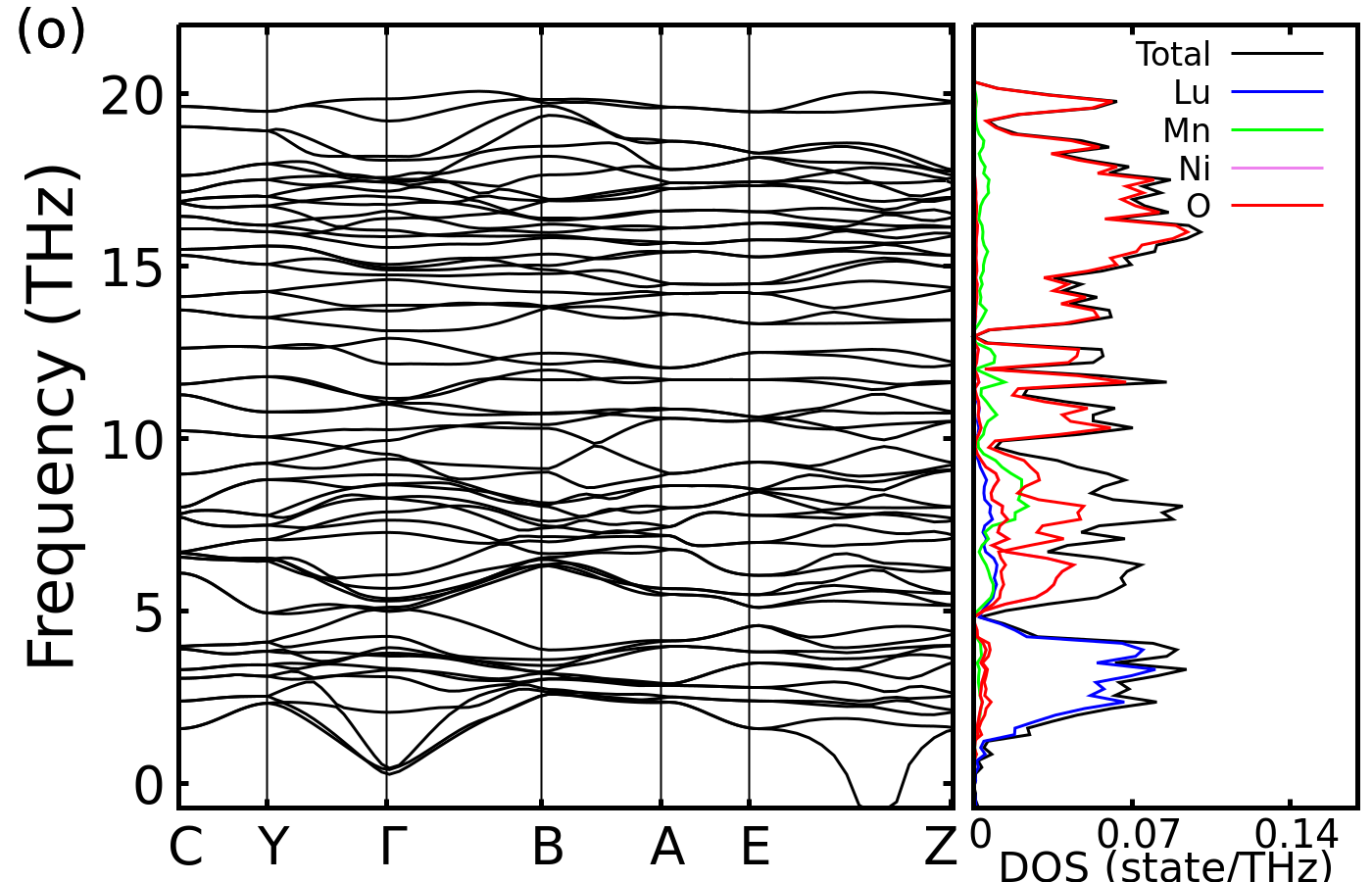}
\includegraphics[height=5.5cm,width=7.5cm]{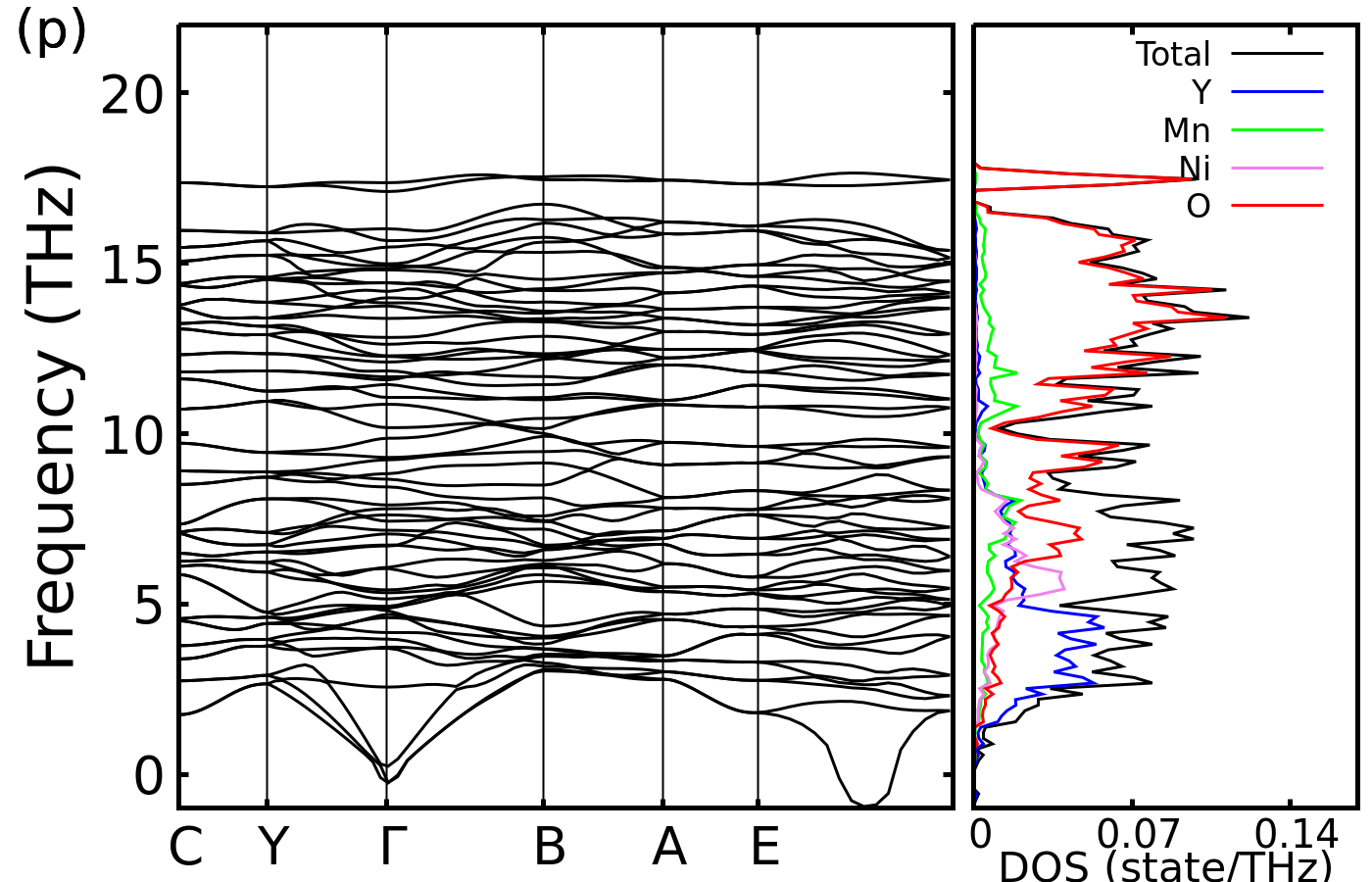}\\
\caption{Phonon band structures of Re$_2$MnNiO$_{6}$, where Re stands for (a)La, (b)Ce, (c)Pr, (d)Nd, (e)Pm, (f)Sm, (g)Eu, (h)Gd, (i)Tb, (j)Dy, (k)Ho, (l)Er, (m)Tm, (n)Yb, (o)Lu and (p)Y.}
\label{fig:phonon6}
\end{center}
\end{figure}

The phonon dispersion curve along with the phonon density of states is presented in \ref{fig:phonon6}. As is evident from the plots, there are no imaginary modes for Pr$_2$MnNiO$_6$ and Yb$_2$MnNiO$_6$, and very few imaginary modes for other elements. Further absence of negative modes at gamma points of the phonon-dispersion plot implies these structures are highly stable. The few imaginary modes present are due to computational limitations such as insufficient k-point sampling, finite size effects of supercells done to optimize computational demands and numerical noise in force constants and do not influence dynamical stability of the structure. The structures, therefore, will not undergo phase transitions. It was observed that the rare earth metals have most bands at low frequencies. The phonon bands of the transition metals have slightly higher frequencies. Ni has bands of frequencies higher than the Re metal and lower than that of Mn. Oxygen has the bands with highest frequencies. There are 3 acoustic modes and 57 optical modes. Thus the rare earth metal is 

\begin{figure}[H]
\begin{center}
\includegraphics[height=5.5cm,width=7.5cm]{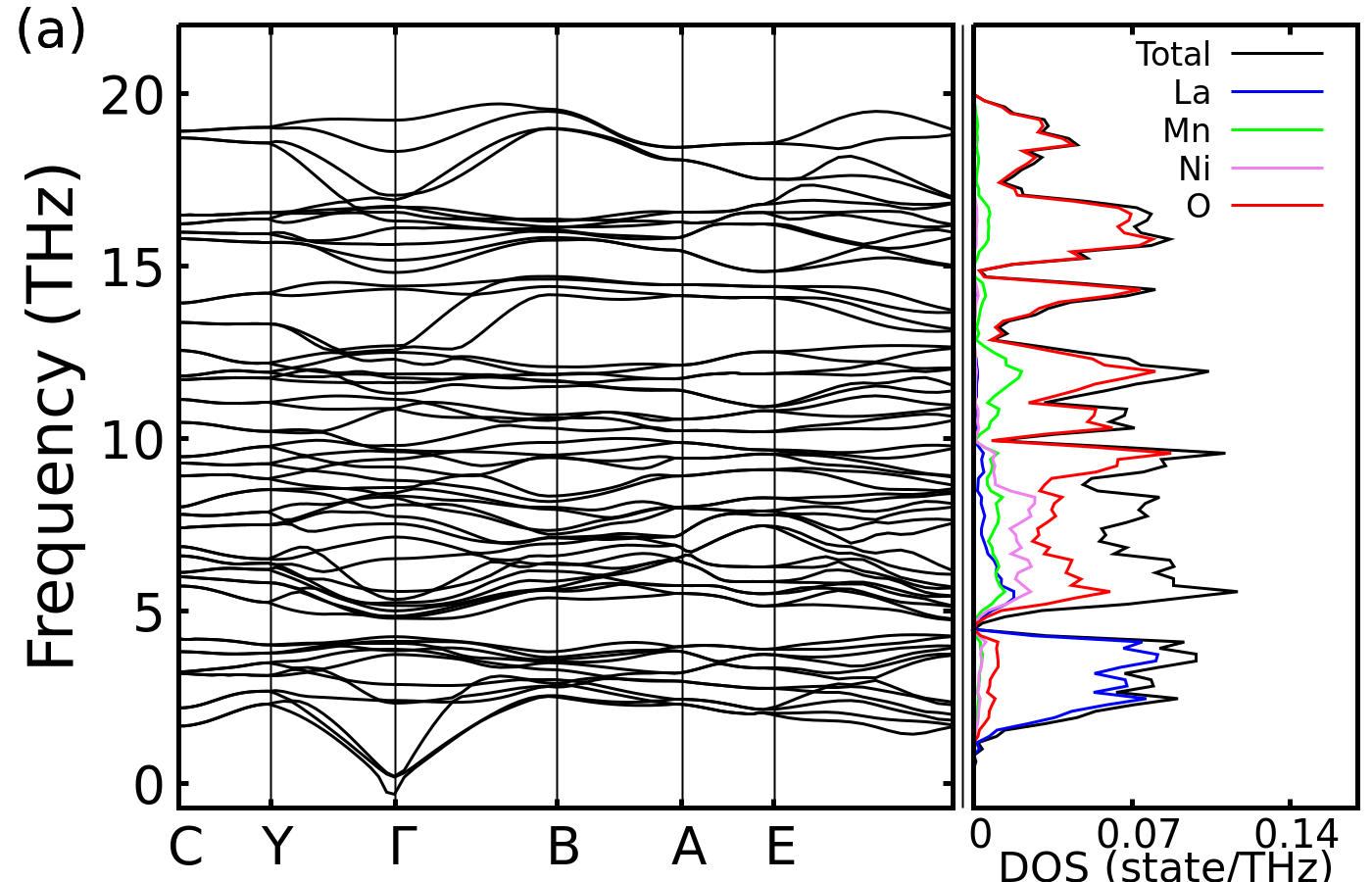}
\includegraphics[height=5.5cm,width=7.5cm]{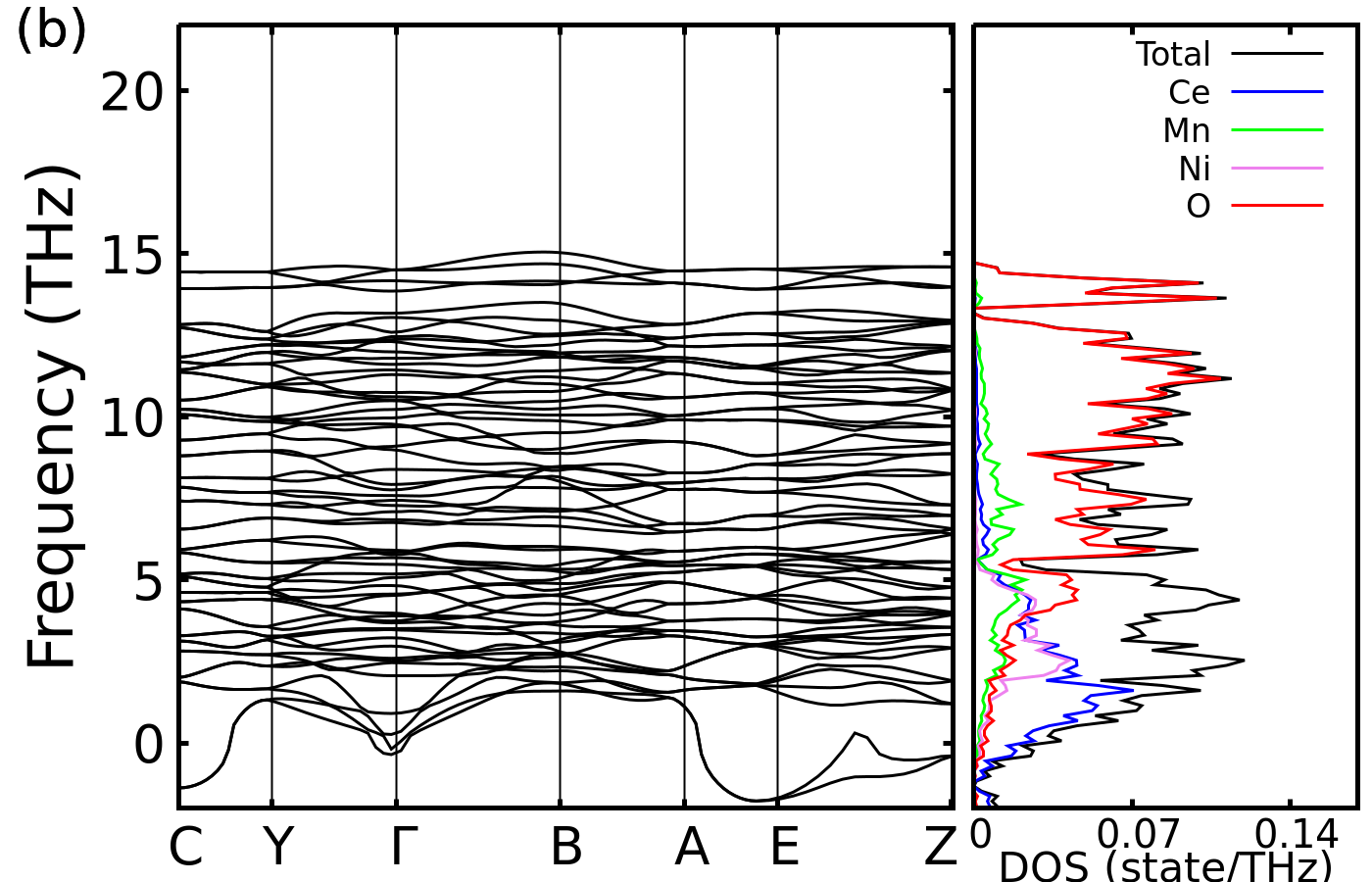}\\ \vspace*{0.3cm}
\includegraphics[height=5.5cm,width=7.5cm]{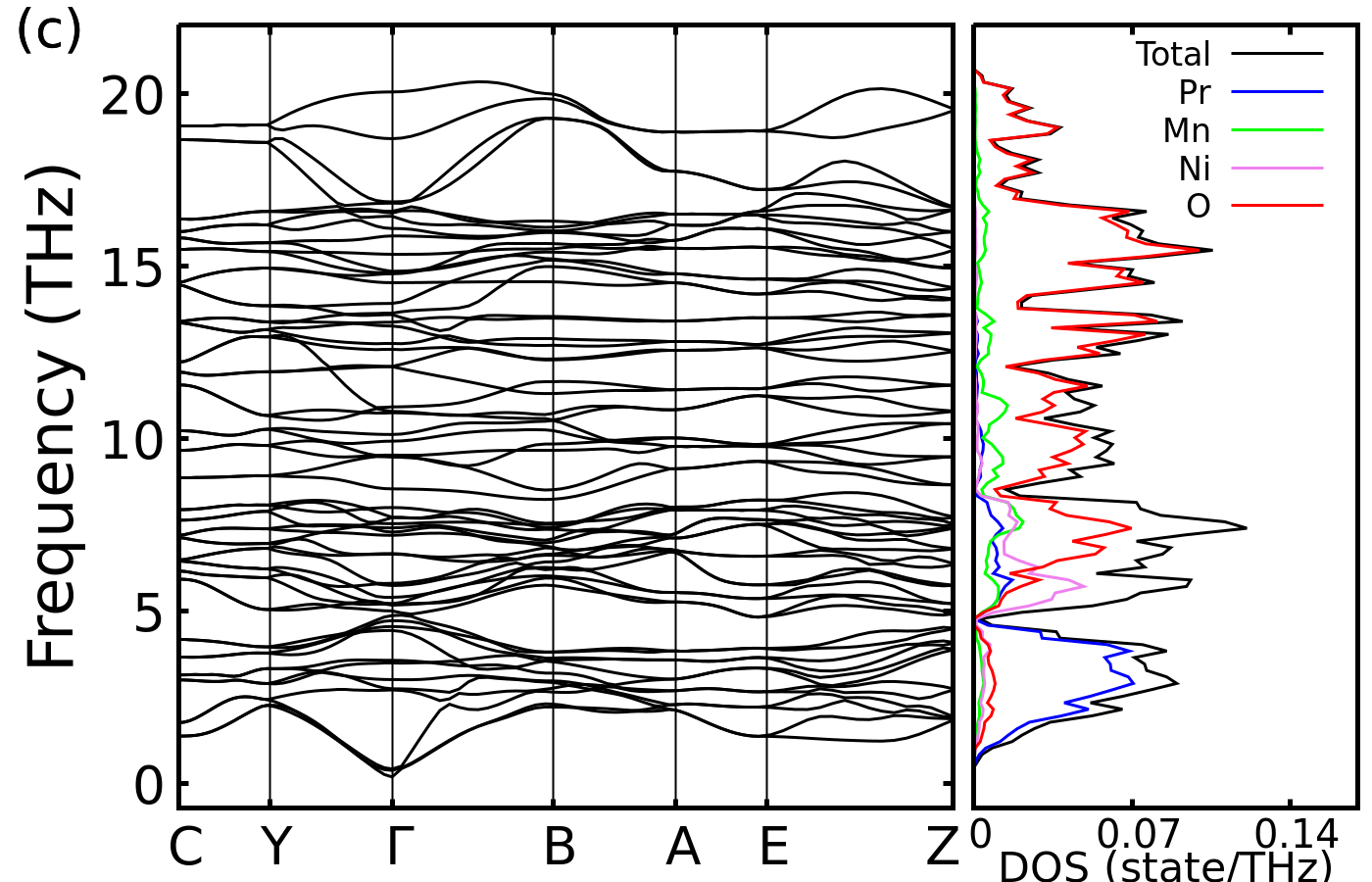}
\includegraphics[height=5.5cm,width=7.5cm]{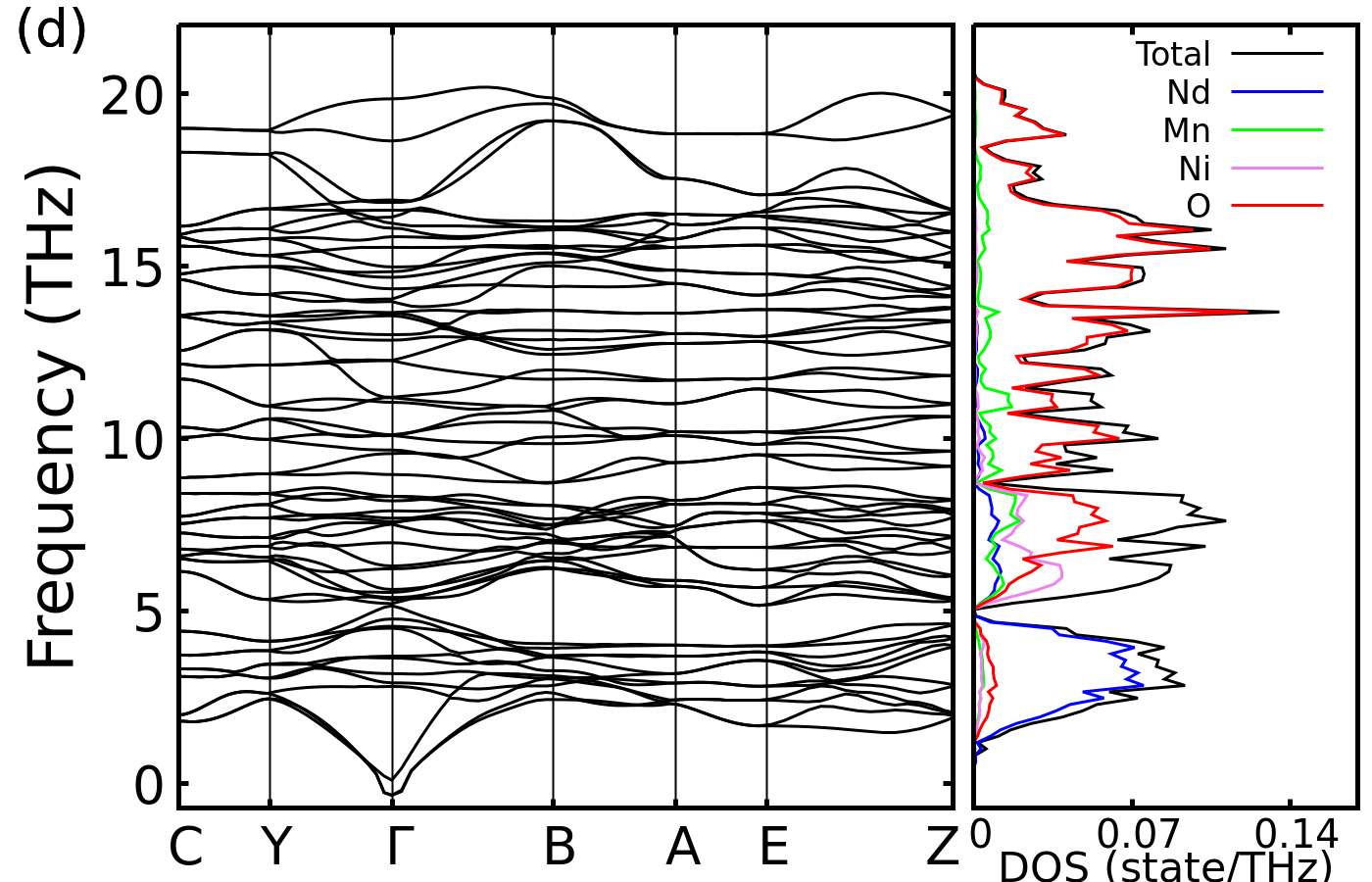}\\ \vspace*{0.3cm}
\end{center}
\end{figure}
\begin{figure}[H]
\begin{center}
\includegraphics[height=5.5cm,width=7.5cm]{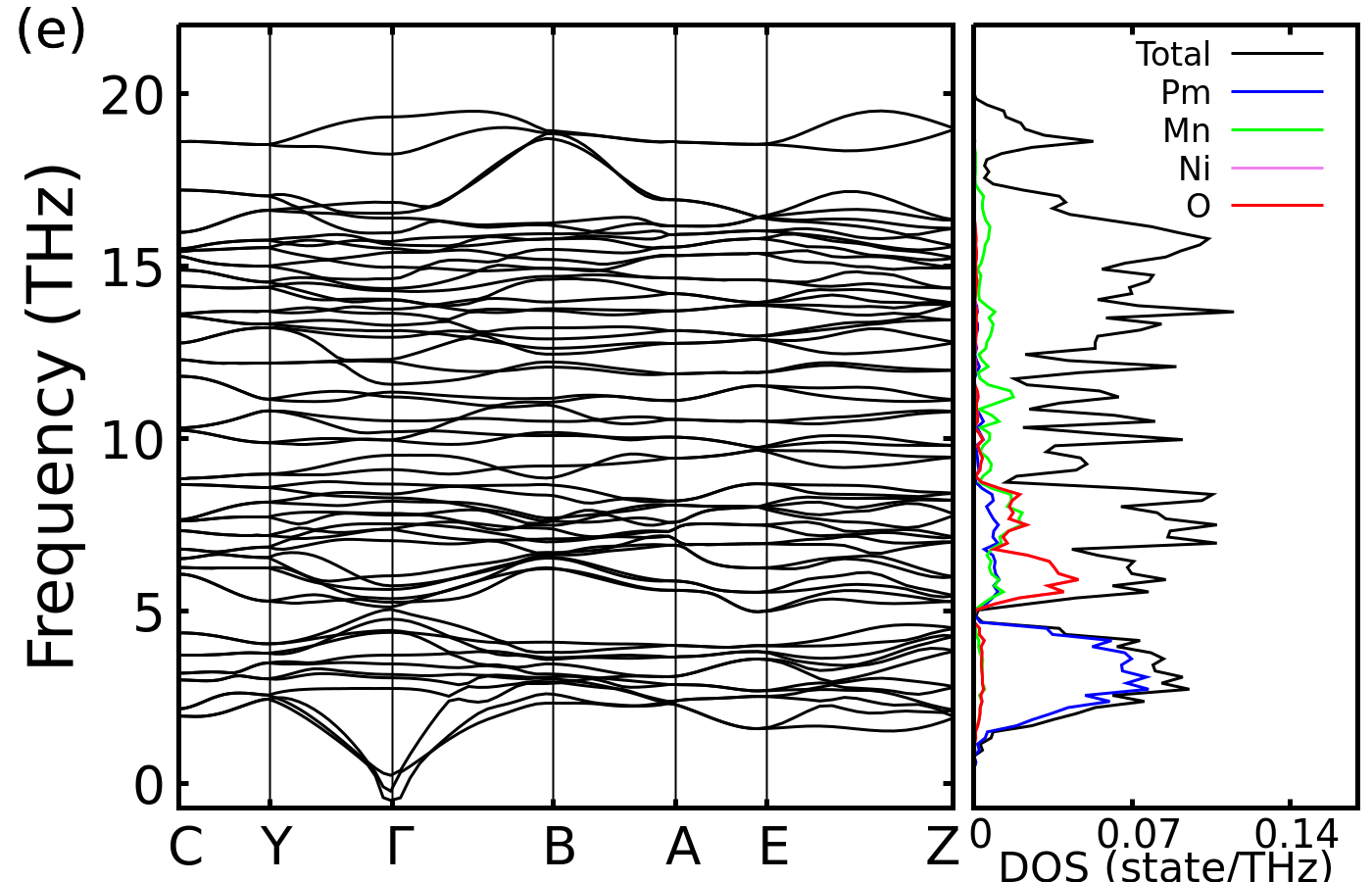}
\includegraphics[height=5.5cm,width=7.5cm]{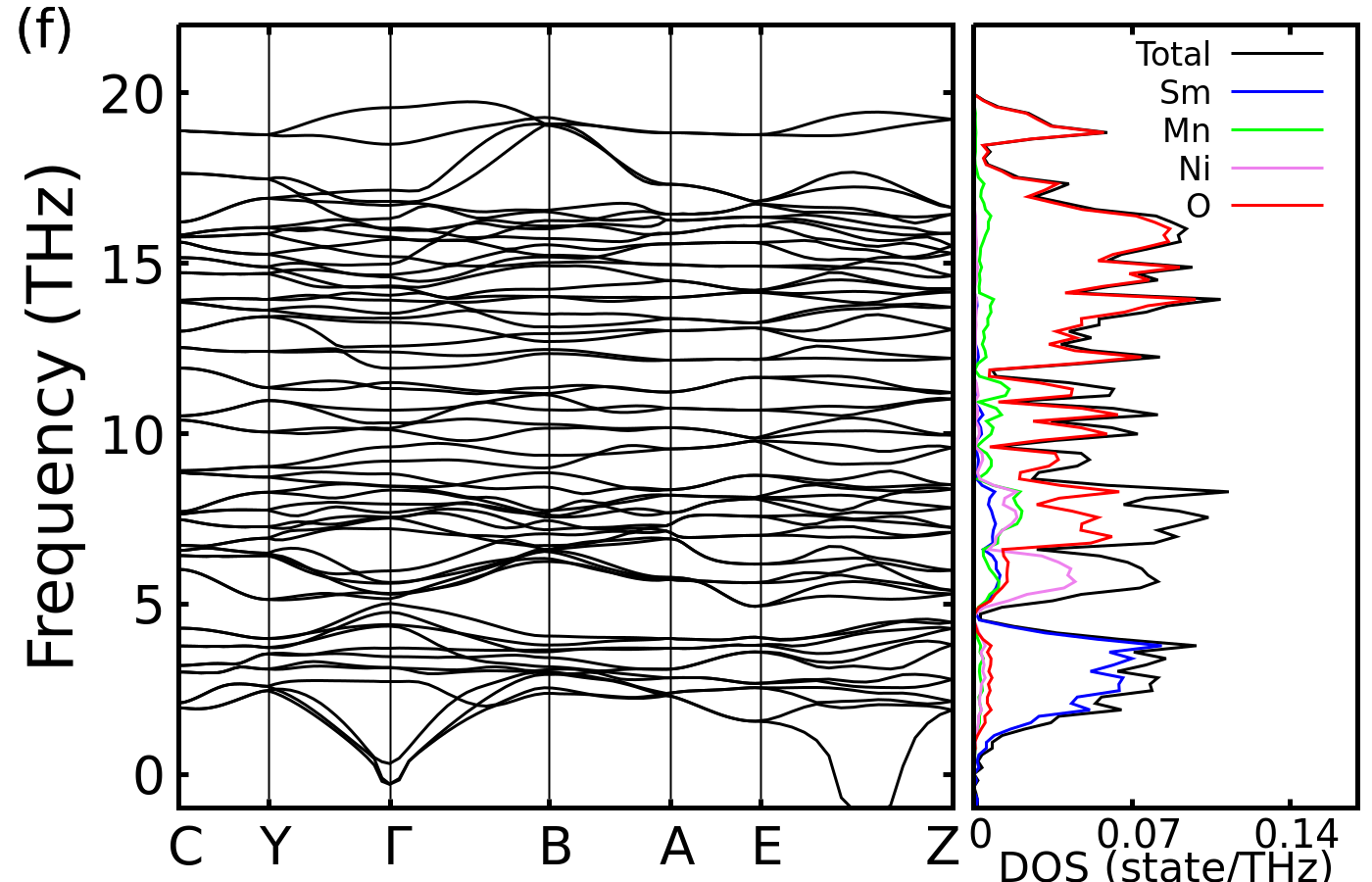}\\ \vspace*{0.3cm}
\includegraphics[height=5.5cm,width=7.5cm]{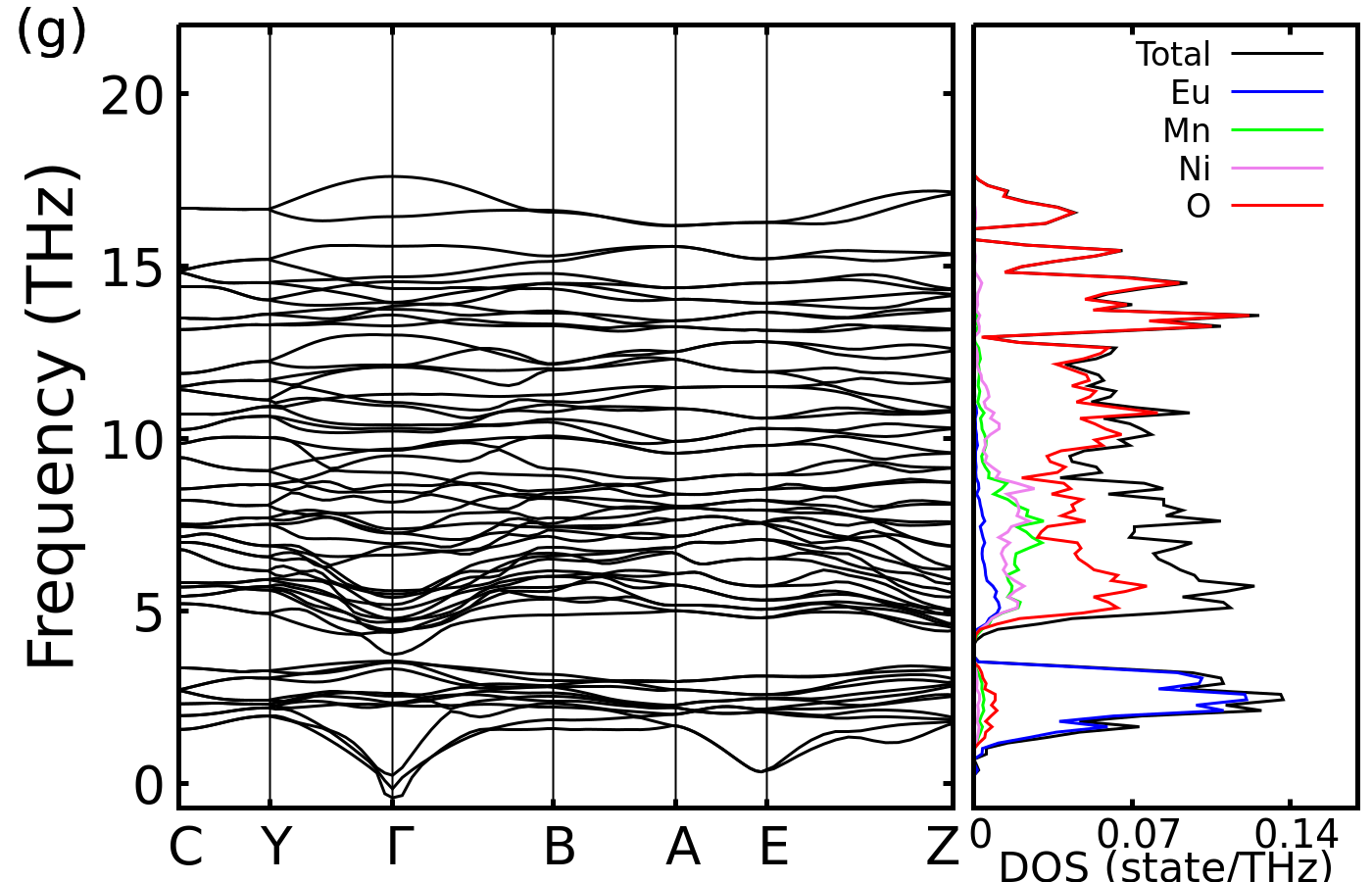}
\includegraphics[height=5.5cm,width=7.5cm]{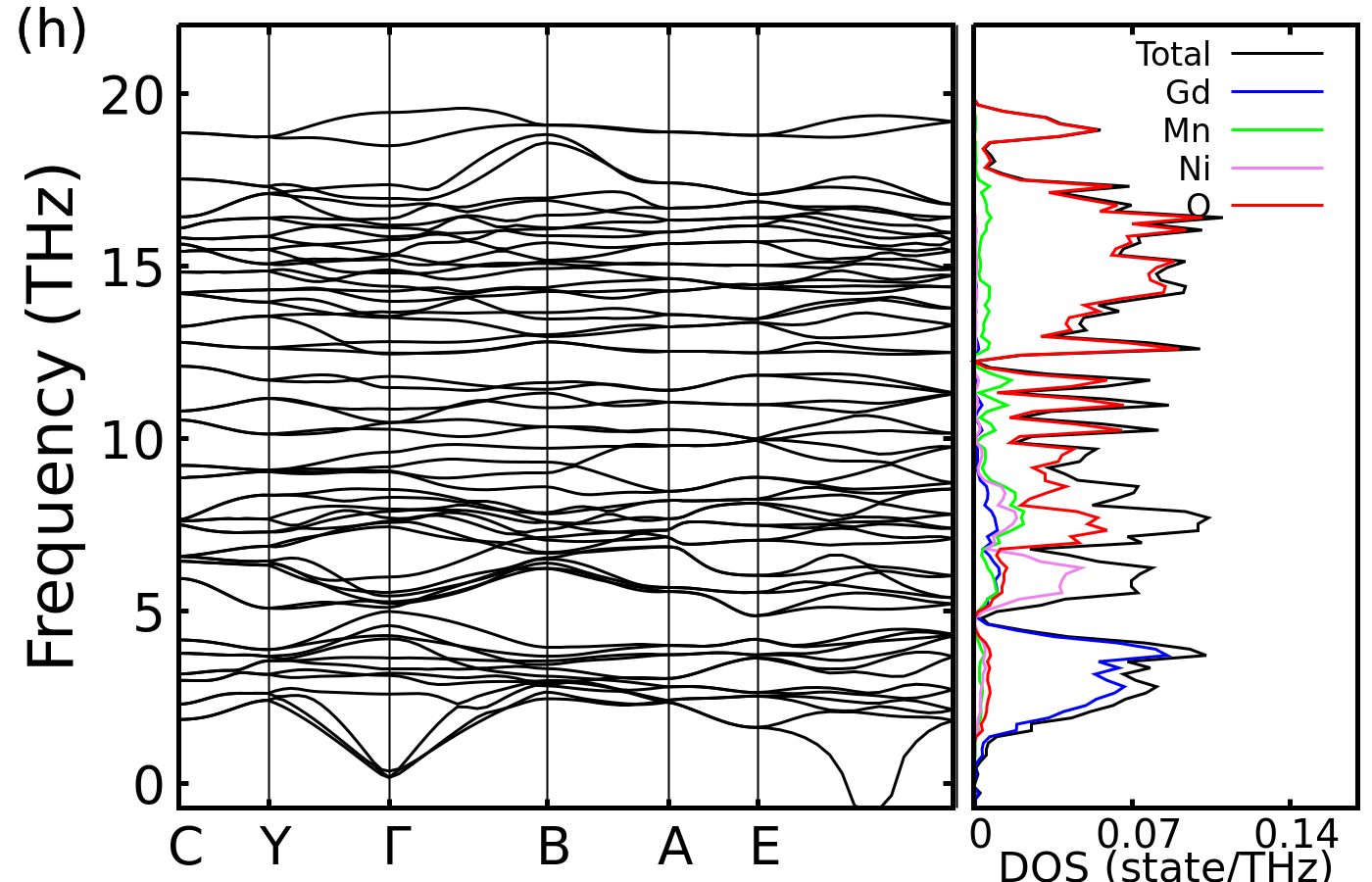}\\ \vspace*{0.3cm}
\includegraphics[height=5.5cm,width=7.5cm]{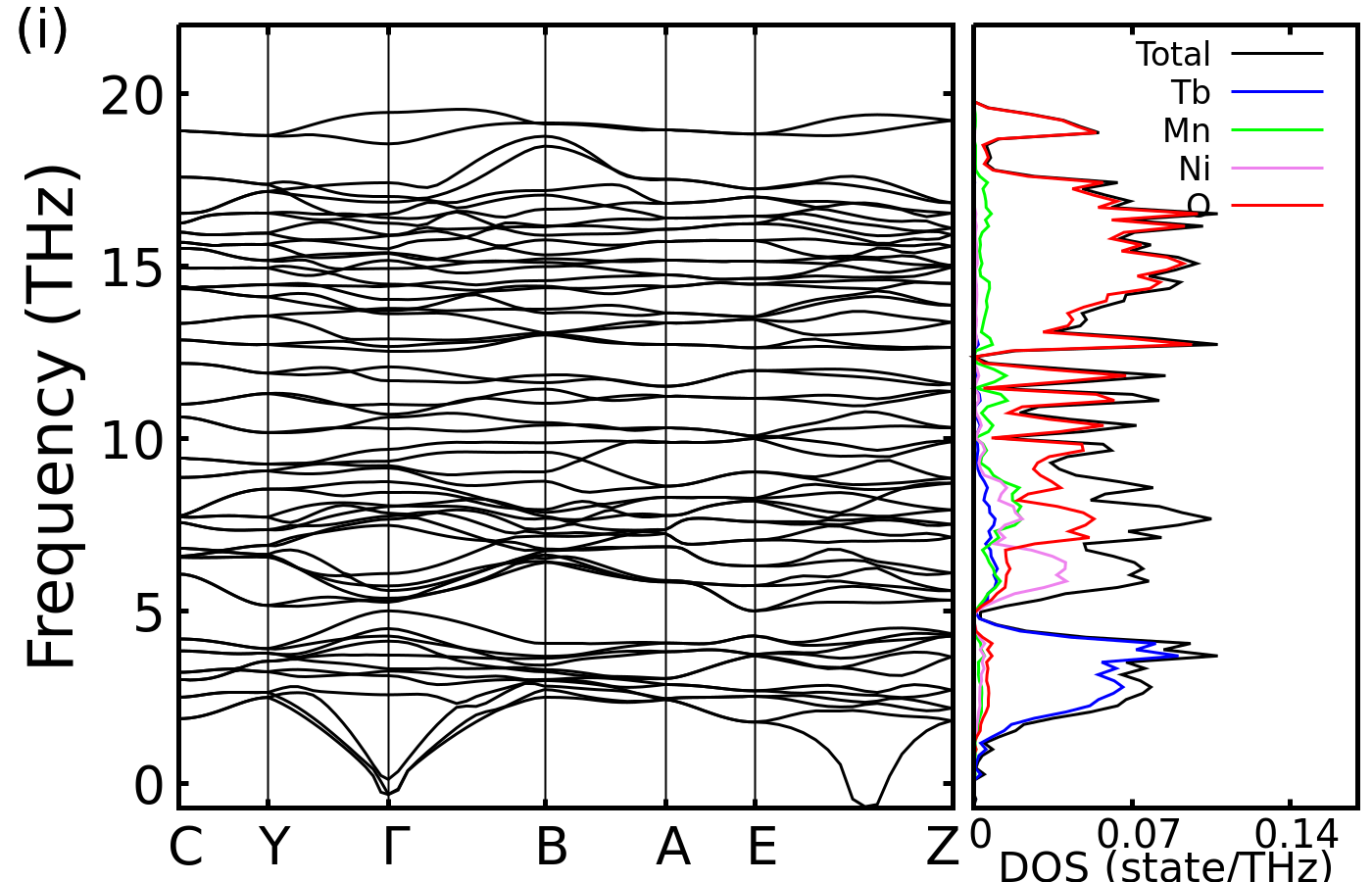}
\includegraphics[height=5.5cm,width=7.5cm]{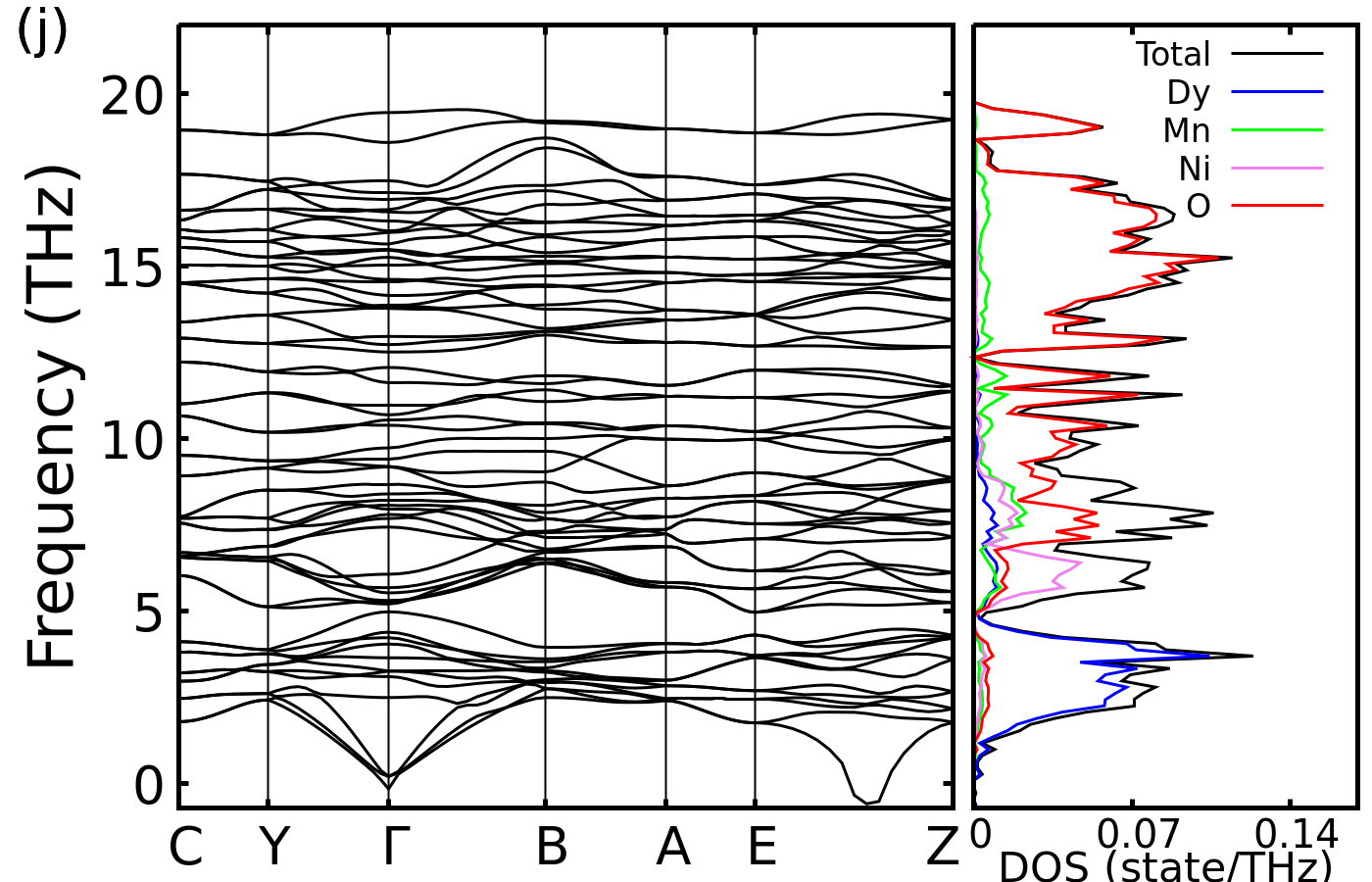}\\ \vspace*{0.3cm}
\includegraphics[height=5.5cm,width=7.5cm]{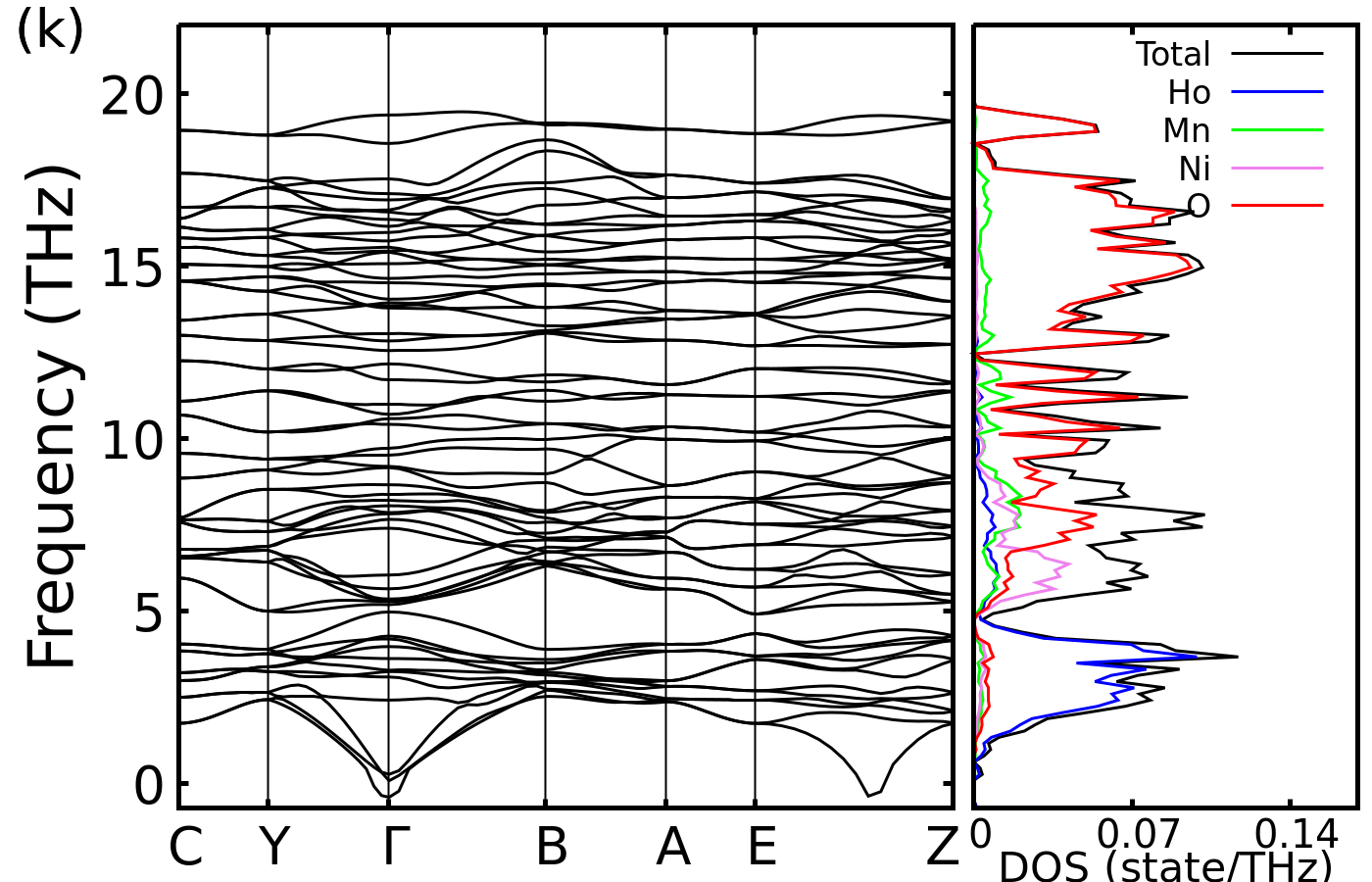}
\includegraphics[height=5.5cm,width=7.5cm]{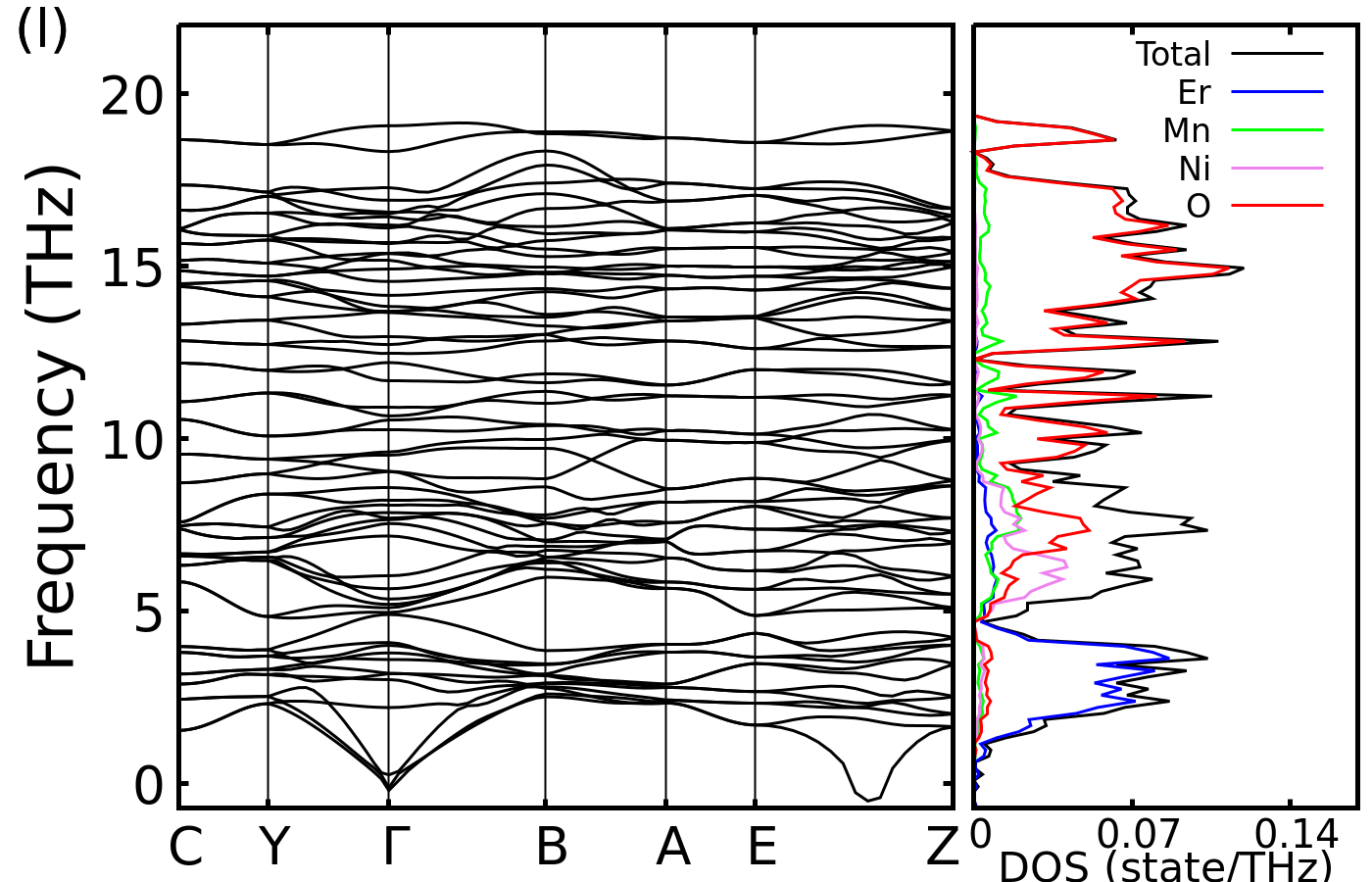}\\ \vspace*{0.3cm}
\end{center}
\end{figure}

responsible for low frequency acoustic modes. Ni, Mn, and O contribute to optical modes. Phonon-dispersion relations, calculated via density functional perturbation theory, show no imaginary frequencies across the Brillouin zone, confirming dynamical stability in both structural variants. Acoustic branches at low frequencies involve coupled motions of heavy Re and transition-metal cations, while high-frequency optical modes originate from internal vibrations of the NiO$_6$ and MnO$_6$ octahedra. From the phonon density of states, we compute vibrational thermodynamic functions: the Helmholtz free energy, vibrational entropy, and the constant-volume heat capacity $C_V$. The heat capacity approaches the Dulong-Petit limit by approximately 300 K, following expected Debye-Einstein behavior.

\subsection{Optical Properties} 
The optical parameters of perovskite materials are crucial to investigate due to their promising potential in photovoltaic and optoelectronic applications. When these materials interact with electromagnetic radiation, the interaction is governed by Maxwell’s equations. One of the key quantities used to describe the optical behavior of solids is the complex dielectric function, defined as:
$$
\varepsilon(\omega) = \varepsilon_1(\omega) + i \varepsilon_2(\omega)
$$

\begin{figure}[H]
\begin{center}
\includegraphics[height=5.5cm,width=7.5cm]{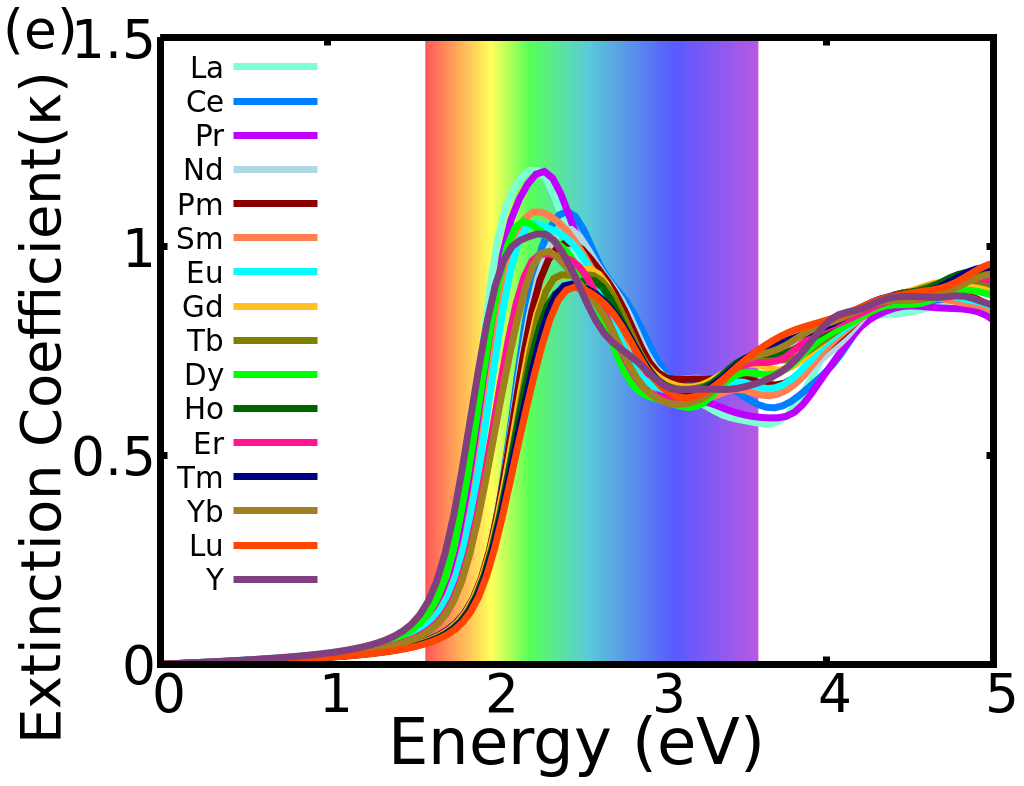}
\includegraphics[height=5.5cm,width=7.5cm]{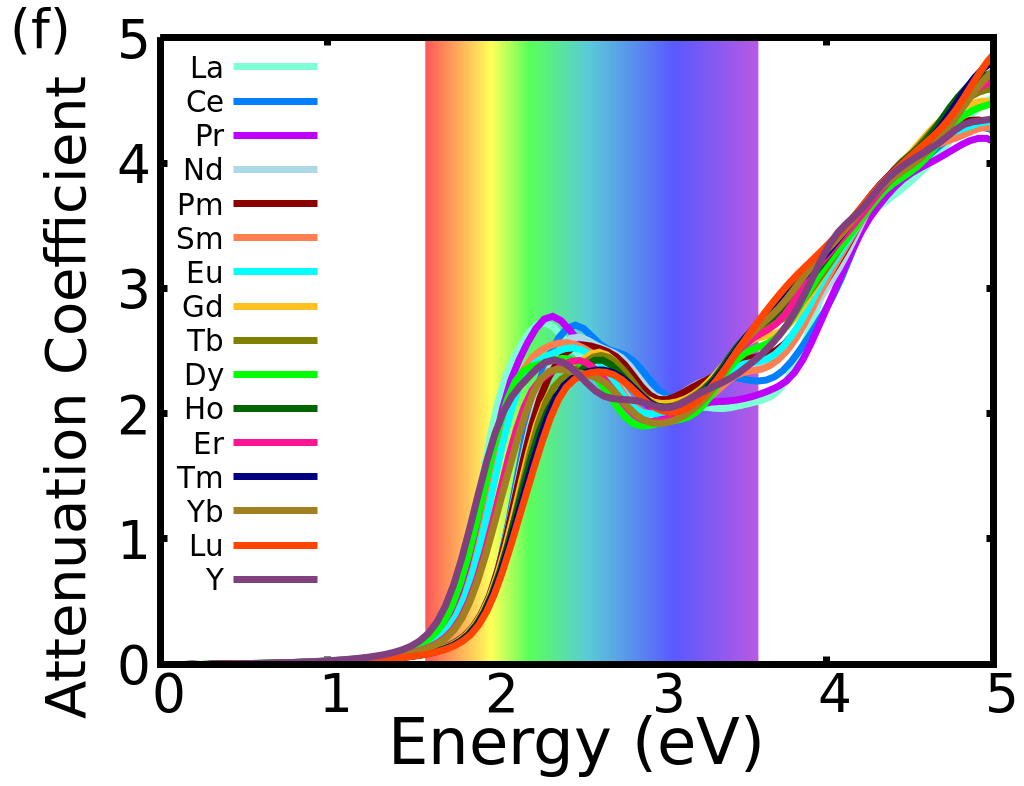}\\
\includegraphics[height=5.5cm,width=7.5cm]{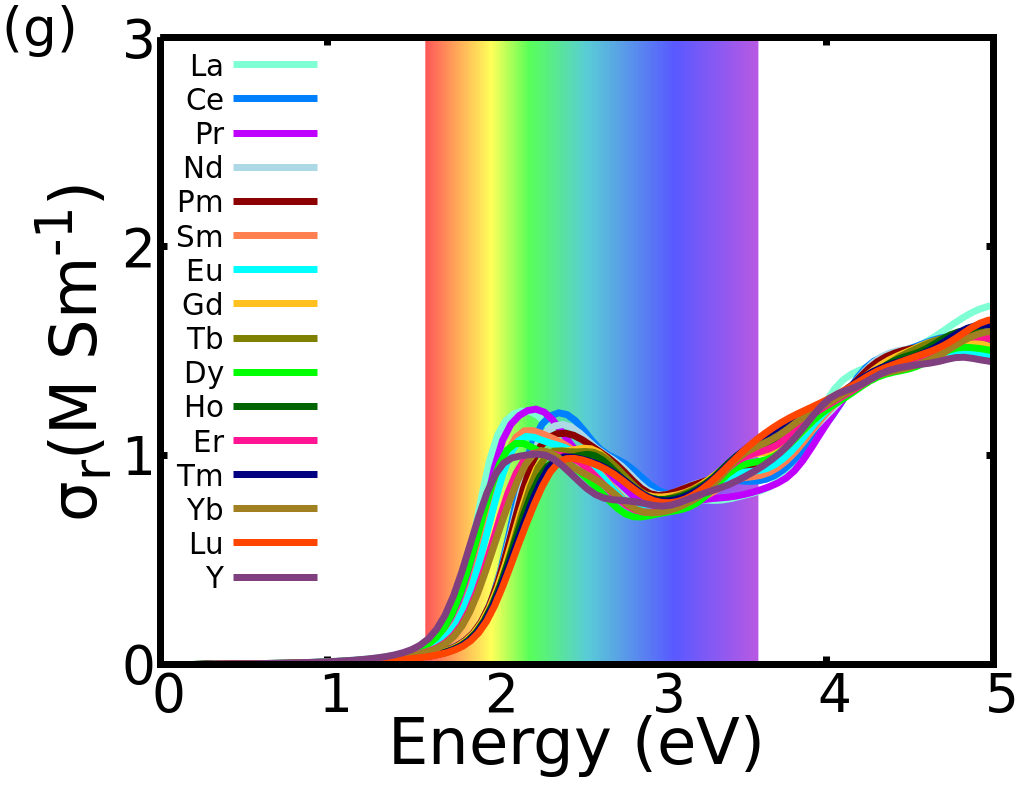}\includegraphics[height=5.5cm,width=7.5cm]{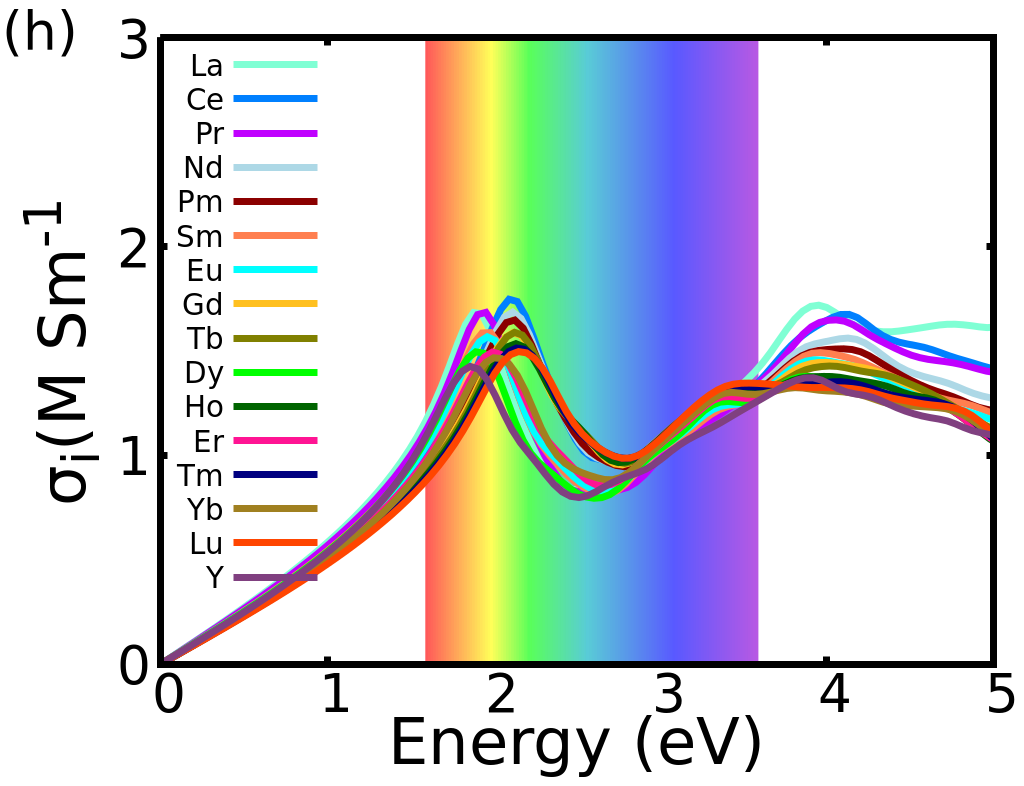}\\
\includegraphics[height=5.5cm,width=7.5cm]{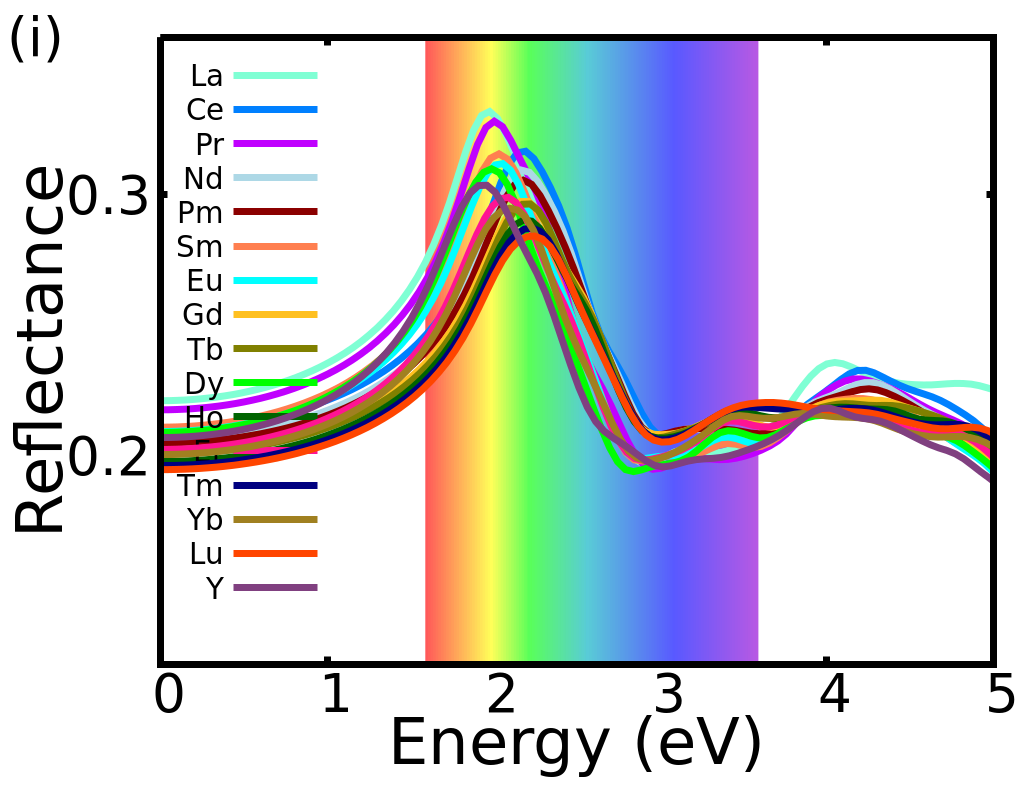}
\includegraphics[height=5.5cm,width=7.5cm]{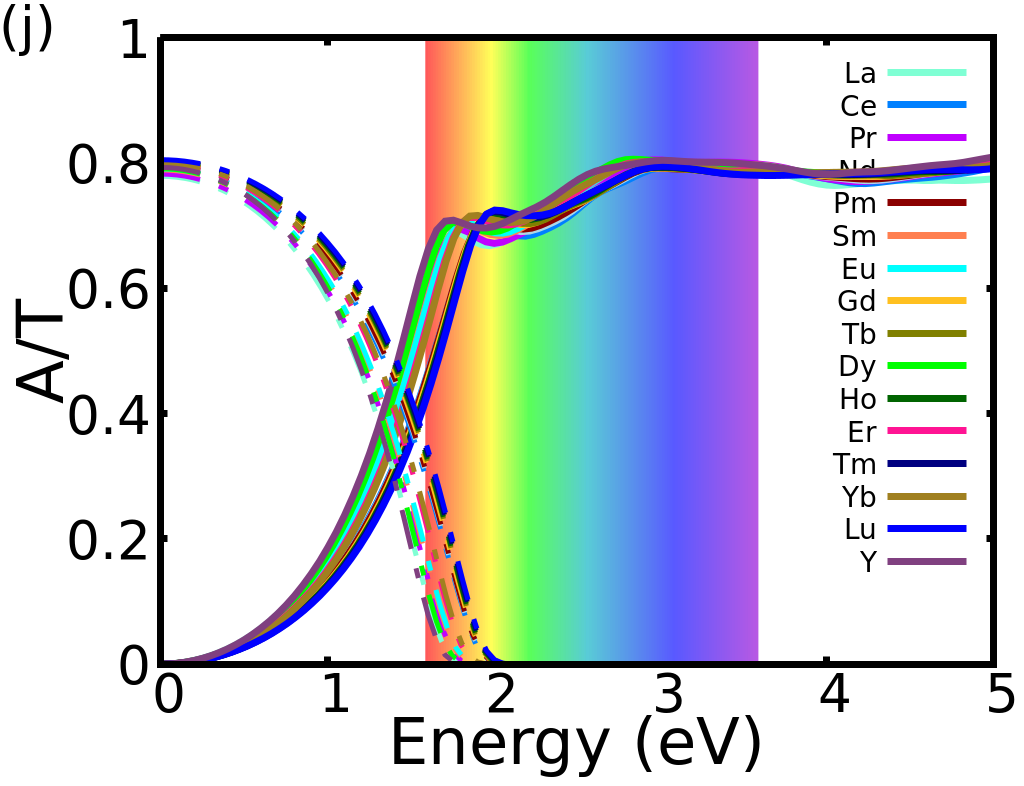}\\
\caption{Optical properties of RE$_2$NiMnO$_6$: (a,b)Real and Imaginary parts of dielectric function, (c) Refractive index, (d) Absorption coefficient, (e,f)Extinction and Attenuation coefficient, (g,h)Real and Imaginary parts of conductivity, (i) Reflectance, (j)Absorbance (solid line) and Transmittance (broken line).}
\label{fig:optical6}
\end{center}
\end{figure}

This function depends on the frequency of incident radiation and not a fixed value. Using the dielectric function, other important optical constants can be derived, including the refractive index $\eta(\omega)$, reflectivity $R(\omega)$, absorption coefficient $\alpha(\omega)$, and the extinction coefficient, all of which are typically calculated over the photon energy range of 0-5 $eV$. They exhibit almost similar optical behaviour. 
	The real part of the dielectric function $\varepsilon_1(\omega)$, computed using the Kramers-Kronig relation, is presented in Fig. \ref{fig:optical6}(a). Table. \ref{tab:optical6} indicates a notable variation in the static dielectric constant $\varepsilon_r(0)$ for materials. This trend can be attributed to changes in the electronic structures of the materials indicating their semiconducting characteristics. This lists the calculated absorption coefficient ($\alpha$), refractive index ($\eta$) and static dielectric constants $\varepsilon_r(0)$ for all compounds. These values are consistent with Penn’s model, which states that the static dielectric function is inversely proportional to the band gap. In other words, a larger band gap corresponds to a lower $\varepsilon_1(0)$, and vice versa \cite{Diego15}. The real part $\varepsilon_r(\omega)$ reflects the material’s ability to polarize in response to an external electromagnetic field; as the incident photon energy increases, $\varepsilon_r(\omega)$ approaches its peak, indicating maximum polarization. 

The imaginary part of the dielectric function $\varepsilon_i(\omega)$, which is associated with the electronic transitions between occupied and unoccupied states, is plotted in Fig. \ref{fig:optical6}(b). It provides insights into the interband optical transitions and absorption mechanisms in these materials.The absorption coefficient $\alpha(\omega)$, shown in Fig. \ref{fig:optical6}(d), indicates how much light is absorbed by the material at different photon energies. All compounds show absorption onsets before the visible range, suggesting significant transparency in the visible spectrum. Among them, La$_2$MnNiO$_6$ exhibits stronger absorption, supporting its superior semiconducting nature. The optical conductivity $\sigma(\omega)$, a measure of how easily a material that responds to optical excitation, is presented in Fig. \ref{fig:optical6}(g). At low photon energies, the real part of the optical conductivity is nearly zero for all compounds. Around 1.13 eV, $\sigma(\omega)$ begins to increase, reaching a maximum for La$_2$MnNiO$_6$ at lower photon energies followed by other compounds following a similar trend as the calculated band gap values. Fig. \ref{fig:optical6}(i) displays the reflectance spectra R$(\omega)$, which are influenced by both the refractive index and the extinction coefficient. The refractive index indicates how much light bends when entering the material, while the extinction coefficient describes how much light is absorbed within it. The optical analysis of this series of compounds reveals that the refractive index exhibits a pronounced maximum at approximately 3.6$eV$photon energy for La$_2$MnNiO$_6$, whereas the extinction coefficient reaches its peak near 1.23eV. At 0 eV, La$_2$MnNiO$_6$ compound shows the highest reflectance (about 22\%) and Lu$_2$MnNiO$_6$ shows the lowest reflectance (about 19.4\%). This makes the RMNO compounds a suitable candidate for absorbing layers or dielectric components in photovoltaic devices, where higher reflectance can enhance light trapping.

\begin{figure}[H]
\begin{center}
\includegraphics[height=5.5cm,width=7.5cm]{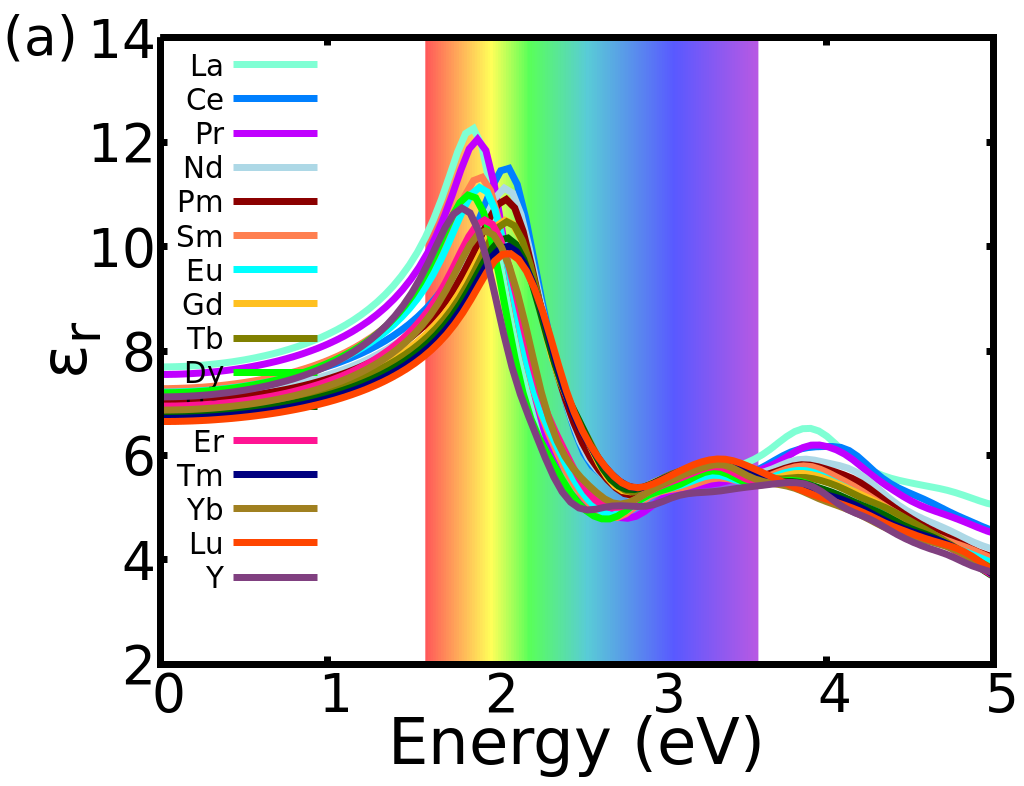}
\includegraphics[height=5.5cm,width=7.5cm]{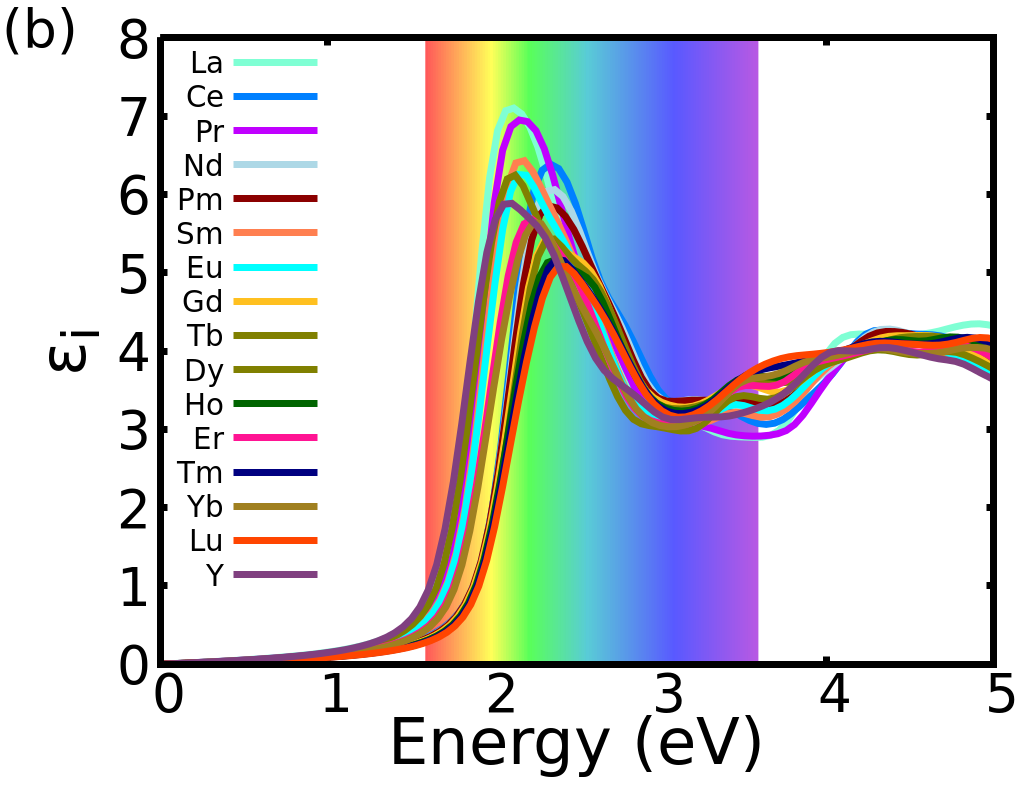}\\
\includegraphics[height=5.5cm,width=7.5cm]{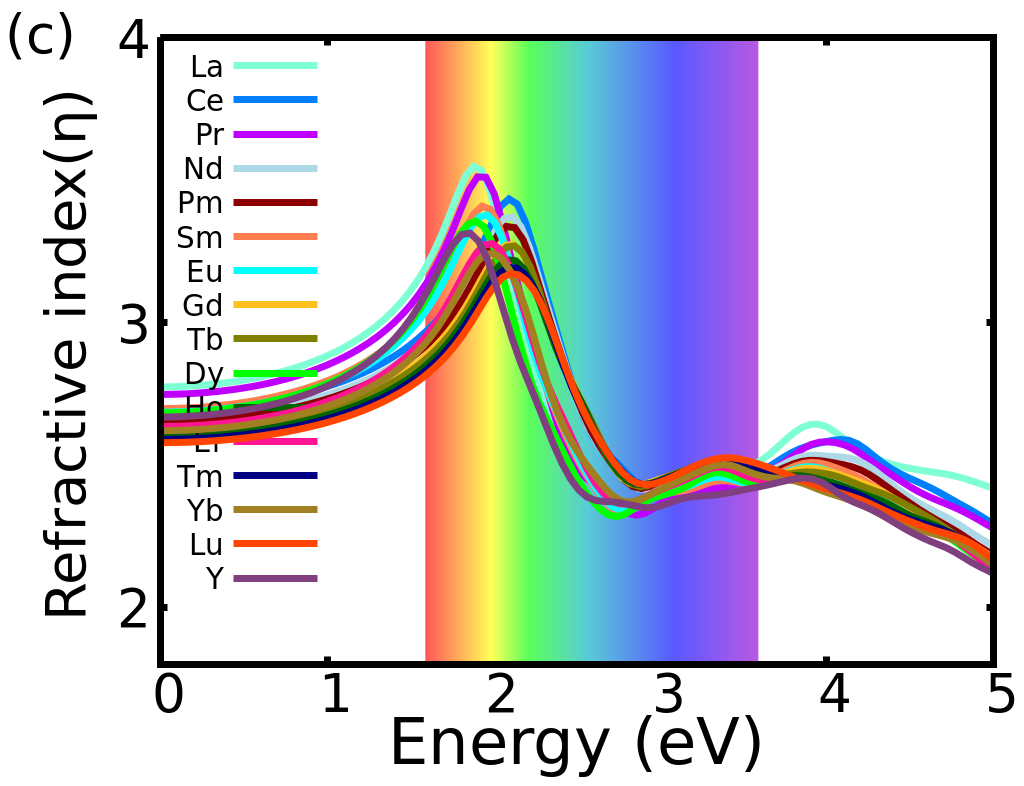}
\includegraphics[height=5.5cm,width=7.5cm]{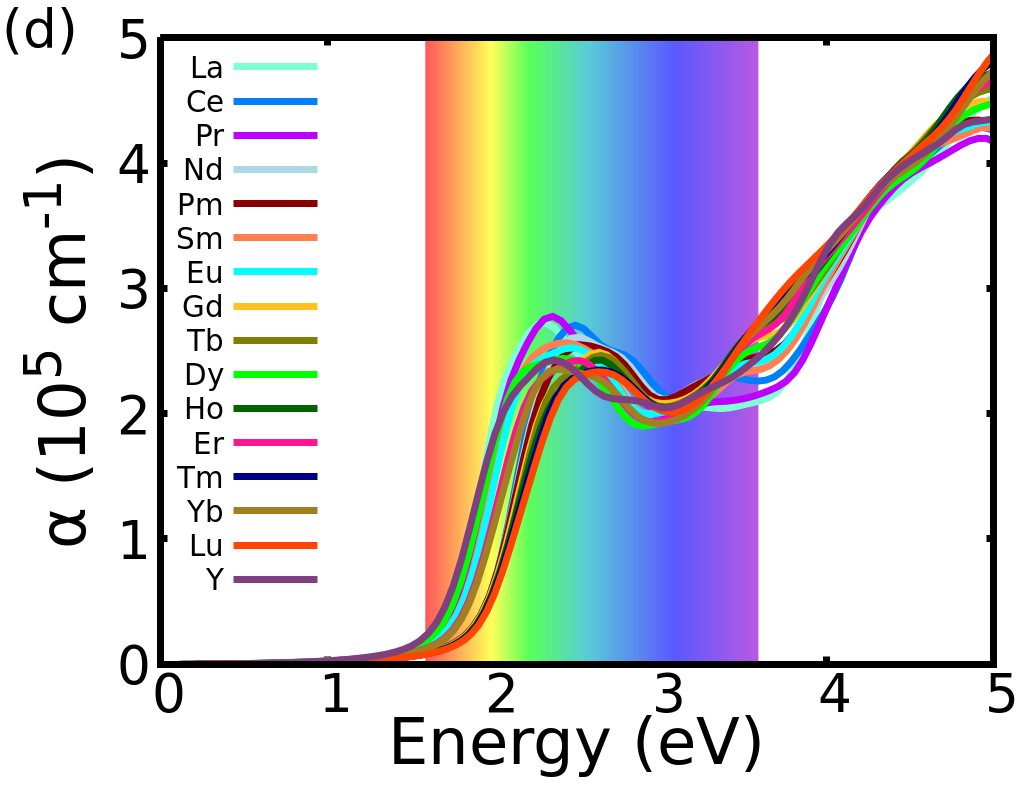}\\
\end{center}
\end{figure}

Fig. \ref{fig:optical6}(j) illustrates the transmittance and absorbance spectra, respectively for the series of compounds. As the photon energy increases from 0 to 5 eV, absorbance increases significantly, while transmittance decreases. This behavior is characteristic of optical materials that transition from transparent to absorbing regimes as the photon energy surpasses their band gaps.
\begin{table}
\begin{center}
\setlength{\tabcolsep}{15pt}
\caption{Static dielectric constant, Max absorption in visible region, absorption edge and refractive index for RMNO.}
\begin{tabular}{|c|c|c|c|c|}
\hline 
\textbf{RE$_2$NiMnO$_6$} & $\varepsilon_{r}$ & $\alpha_{max}$ & $\alpha_{edge}$ & $\eta$ \\ 
\hline 
La & 7.262 & 7.102 & 1.55 & 2.774 \\ 
\hline 
Ce & 7.552 & 6.39 & 1.798 & 2.694 \\ 
\hline 
Pr & 7.552 & 6.95 & 1.609 & 2.748 \\ 
\hline 
Nd & 7.112 & 6.063 & 1.796 & 2.667 \\ 
\hline 
Pm & 7.040 & 5.85 & 1.796 & 2.653 \\ 
\hline 
Sm & 7.280 & 6.43 & 1.632 & 2.698 \\ 
\hline 
Eu & 7.214 & 6.255 & 1.639 & 2.685 \\ 
\hline 
Gd & 6.912 & 5.42 & 1.805 & 2.629 \\ 
\hline 
Tb & 6.874 & 5.43 & 1.823 & 2.621 \\ 
\hline 
Dy & 7.20 & 6.246 & 1.579 & 2.683 \\ 
\hline 
Ho & 6.768 & 5.16 & 1.816 & 2.601 \\ 
\hline 
Er & 6.946 & 5.68 & 1.681 & 2.634 \\ 
\hline 
Tm & 6.708 & 5.17 & 1.826 & 2.588 \\ 
\hline 
Yb & 6.874 & 5.64 & 1.693 & 2.619 \\ 
\hline 
Lu & 6.649 & 5.07 & 1.836 & 2.578 \\ 
\hline 
Y & 7.122 & 5.88 & 1.536 & 2.668 \\
\hline 
\end{tabular} 
\label{tab:optical6}
\end{center}
\end{table}
%
%
%
Optical properties are derived from the frequency-dependent complex dielectric function and computed within the random-phase approximation. All RE$_2$NiMnO$_6$ materials exhibit strong absorption onsets at their respective band edges, with prominent interband transition peaks between 3 and 5 eV. The static refractive index $\eta_0$ decreases from approximately 2.7 for La$_2$NiMnO$_6$ to around 2.5 for Lu$_2$NiMnO$_6$. Reflectance spectra indicate moderate reflectivity between 18-30$\%$ in the visible range, implying better transparency, whereas the absorption coefficient $\alpha(\omega)$ exceeds 10$^5$ cm$^{-1}$ above the gap, highlighting these materials' suitability for UV-photodetector and transparent-conductor applications.	
In the second set, absorption and reflectance were analyzed, where the inclusion of $f$-electrons brought significant changes to the material properties compared to the $f$-electron-excluded case. As the $f$-orbitals showed a higher density above the Fermi level in the partial density of states plots (Fig. \ref{fig:bs6_Re}), the corresponding absorption peaks appeared from 0$eV$up to 0.7 eV. This result closely resembled the findings of Azmat et al. \cite{Azmat24}. The order of absorption changed to perovskites containing Tm, Sm, Eu, Er, Pr, Yb, Nd, Pm, Ho, Ce, Tb, Gd, and Lu. The absence of free electrons in the stable hal$f$-filled and fully filled states of Gd and Lu, respectively, placed them at the bottom of the series. Due to its similar electronic configuration, Tb$_2$MnNiO$_6$ followed next in the sequence. Tm- and Sm-based compounds exhibited the highest absorbance in the visible region and also showed high reflectance at 0 eV. Reflectance remained below 30$\%$ for all compounds in the visible region and beyond. The spectral features in the visible range could be attributed to $d$–$d $ transitions between the orbitals of the rare-earth element and those of Mn and Ni, due to their close energy alignment. Therefore, RMNO could be harnessed as an optical material for optoelectronic applications, with properties tunable by the choice of the rare-earth element.

\begin{figure}[H]
\begin{center}
\includegraphics[height=5.5cm,width=7.5cm]{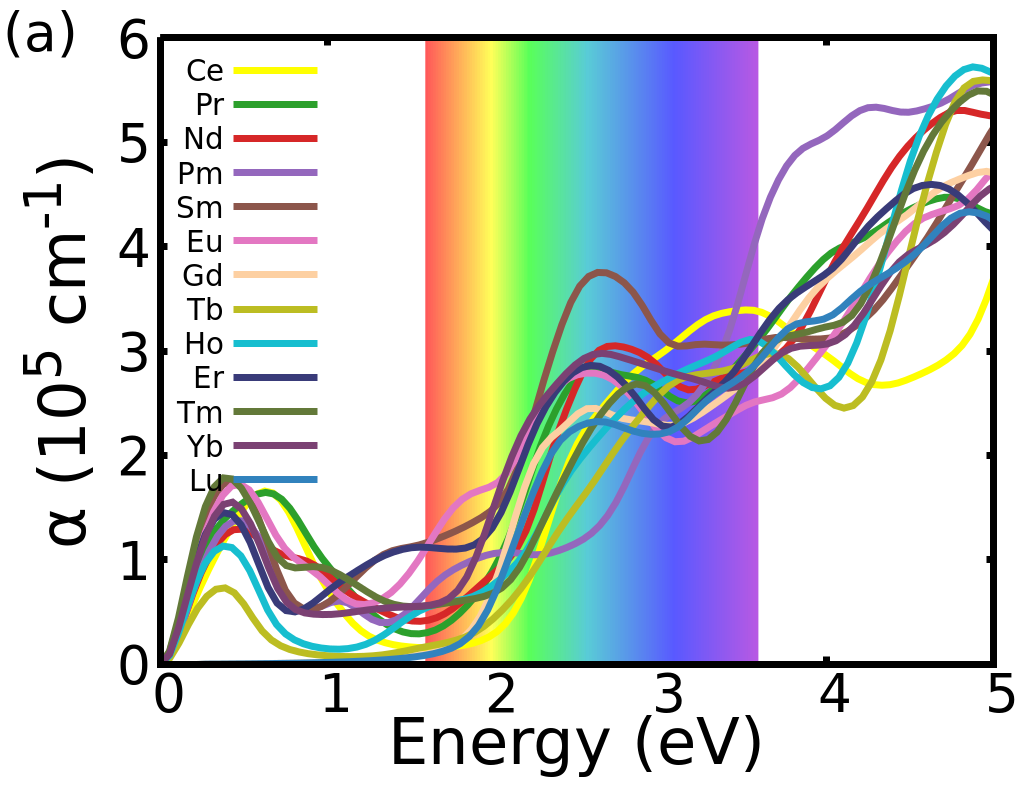}
\includegraphics[height=5.5cm,width=7.5cm]{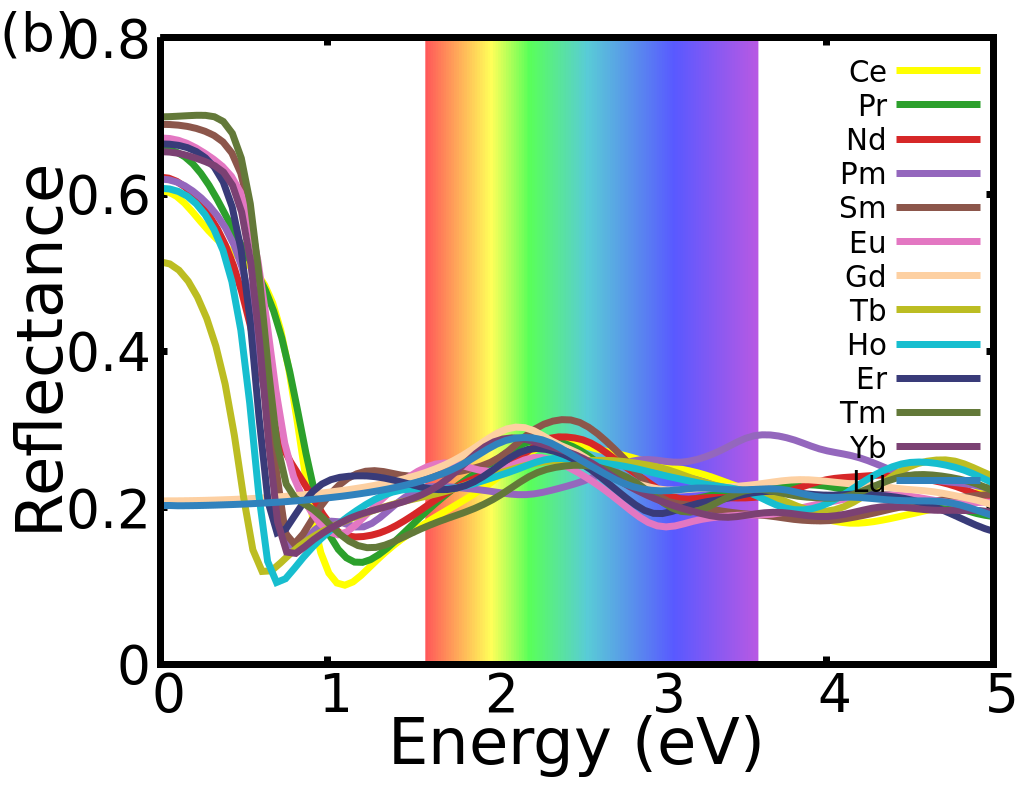}
\caption{(a) Absorption coefficient and (b) Reflectance for RMNO with $f$-orbitals.}
\label{fig:optical6_Re}
\end{center}
\end{figure}
	
\subsection{Thermal properties}
\begin{figure}[H]
\begin{center}
\includegraphics[height=5.5cm,width=7.5cm]{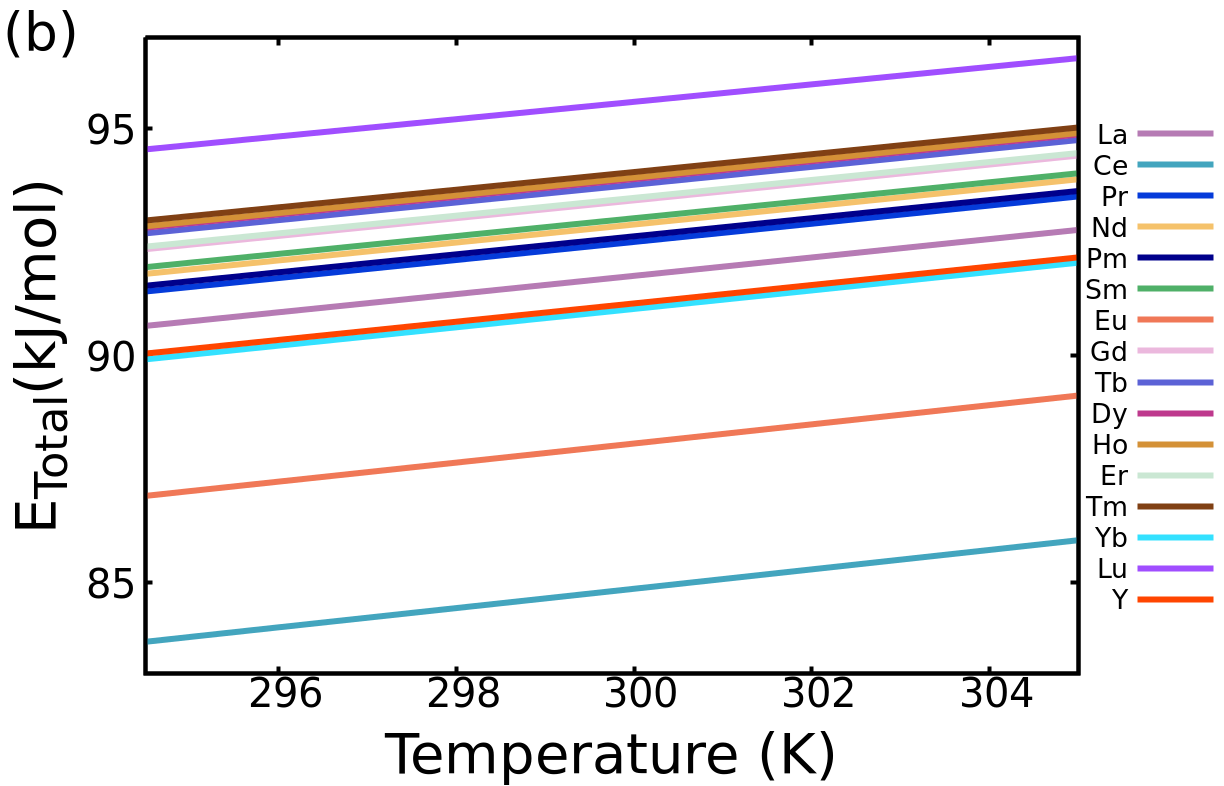}
\includegraphics[height=5.5cm,width=7.5cm]{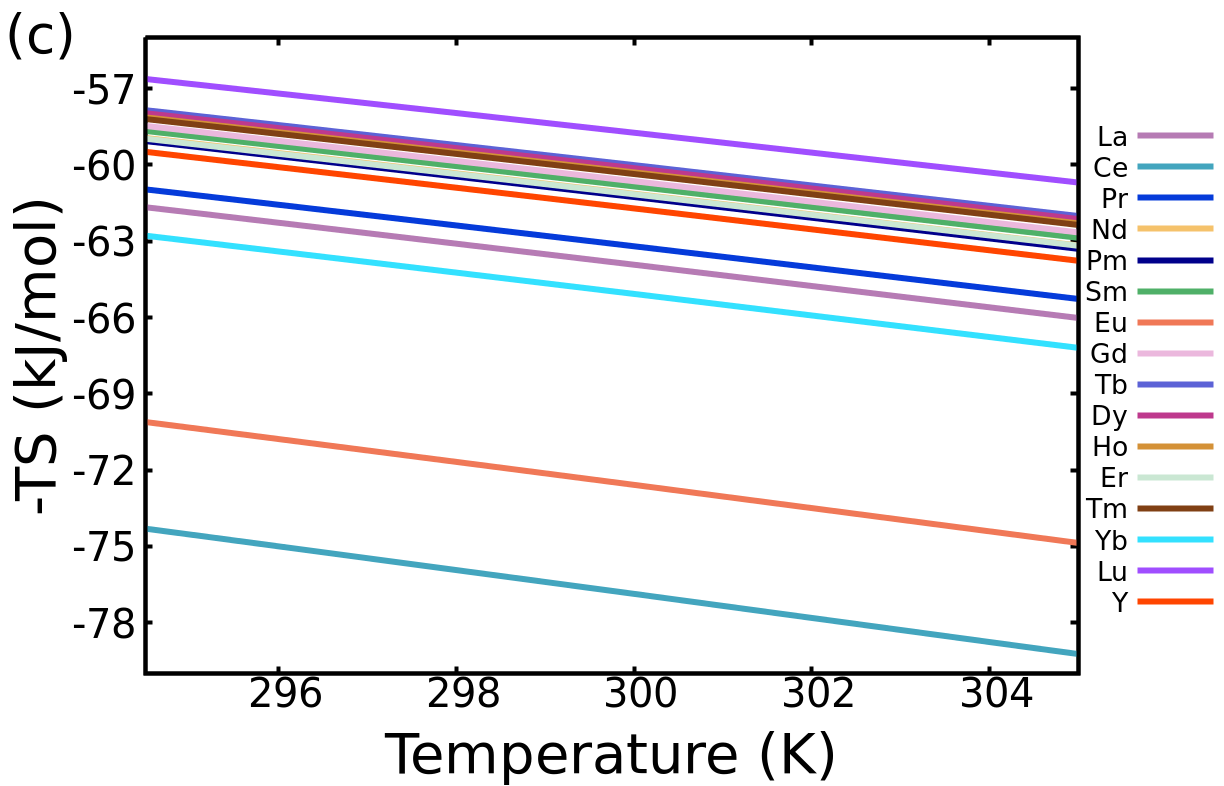}\\
\includegraphics[height=5.5cm,width=7.5cm]{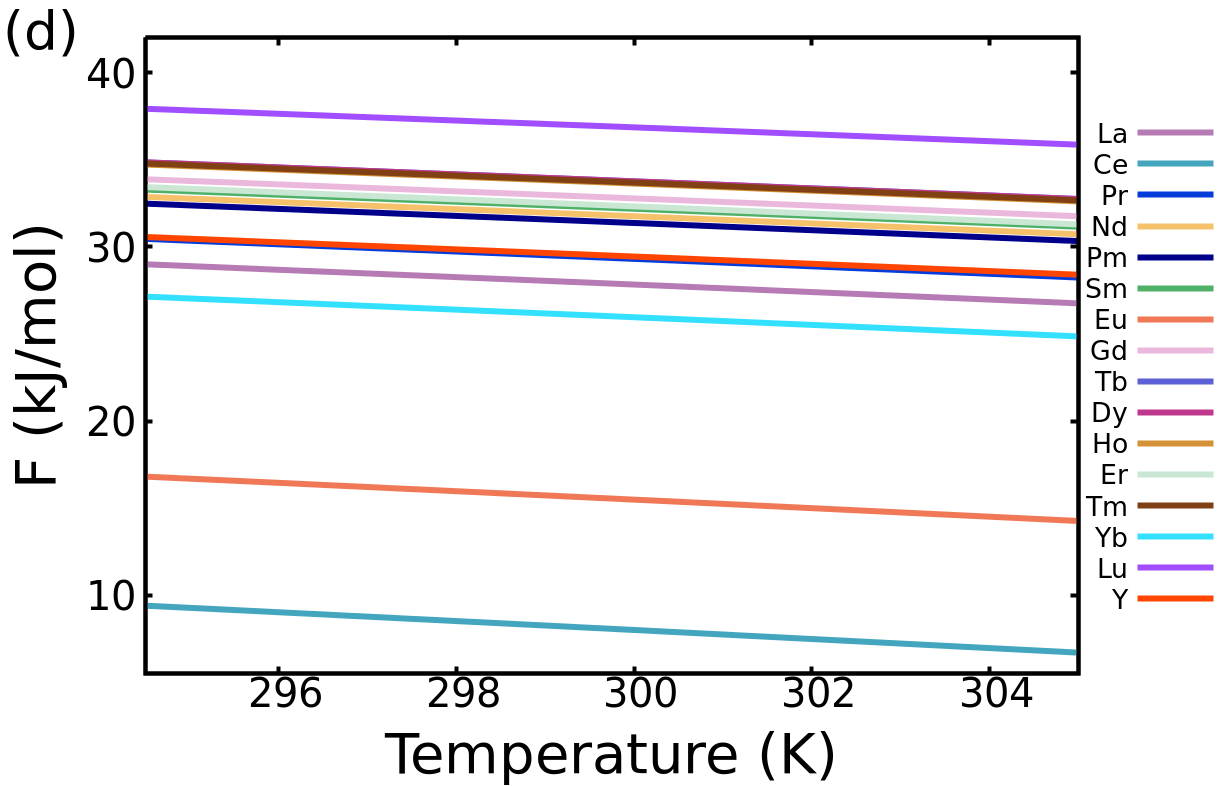}
\includegraphics[height=5.5cm,width=7.5cm]{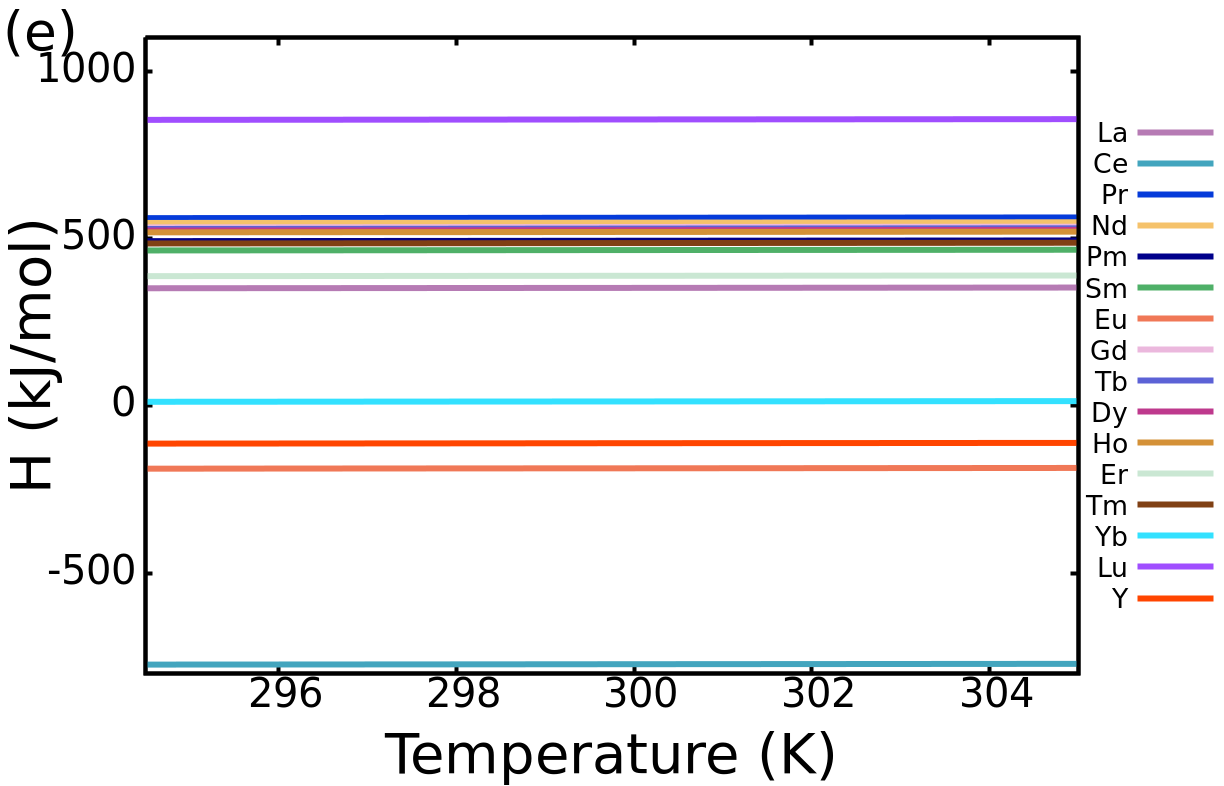}\\ 
\includegraphics[height=5.5cm,width=7.5cm]{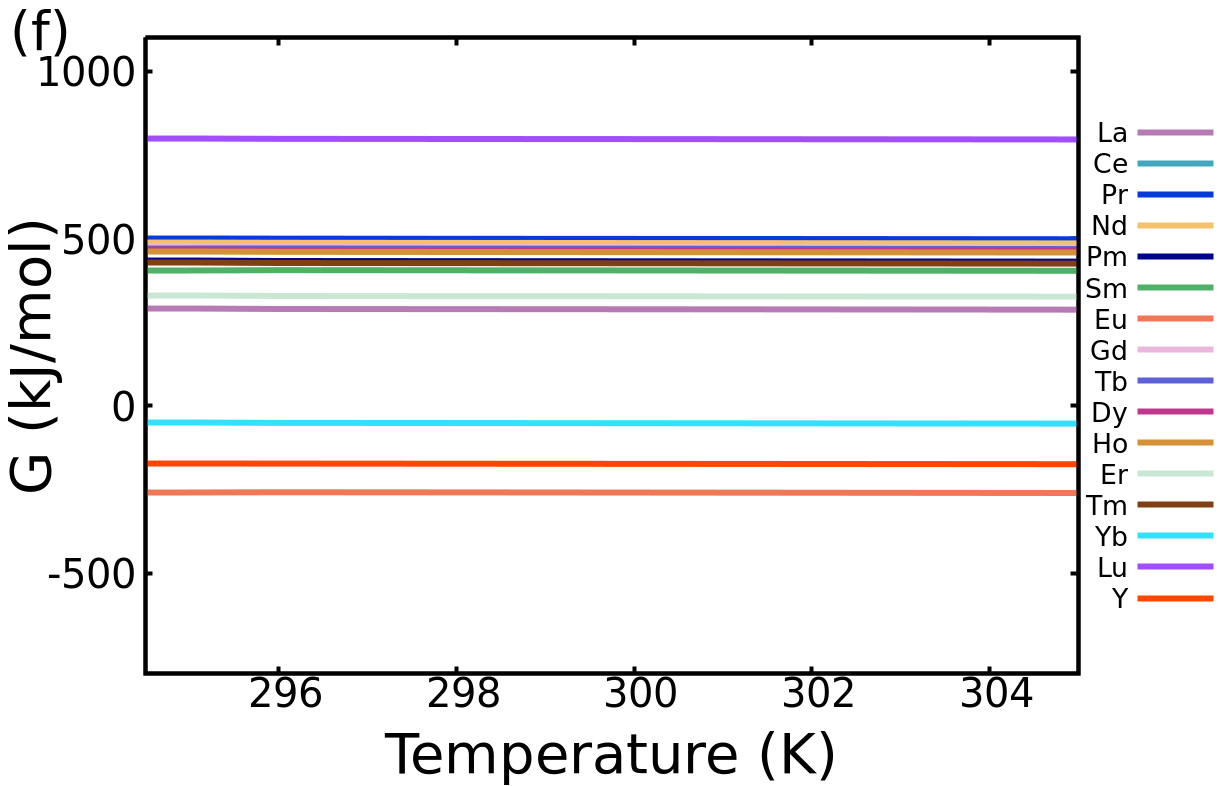}
\includegraphics[height=5.5cm,width=7.5cm]{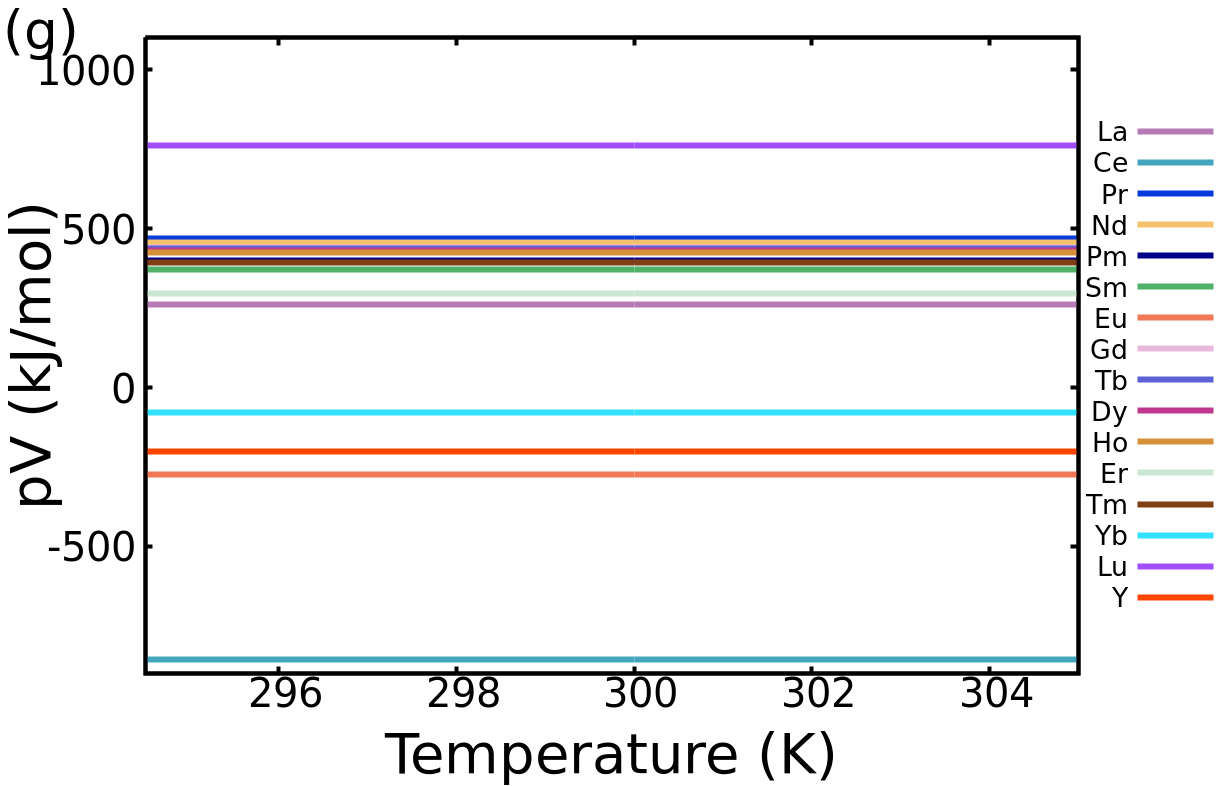}\\
\includegraphics[height=5.5cm,width=7.5cm]{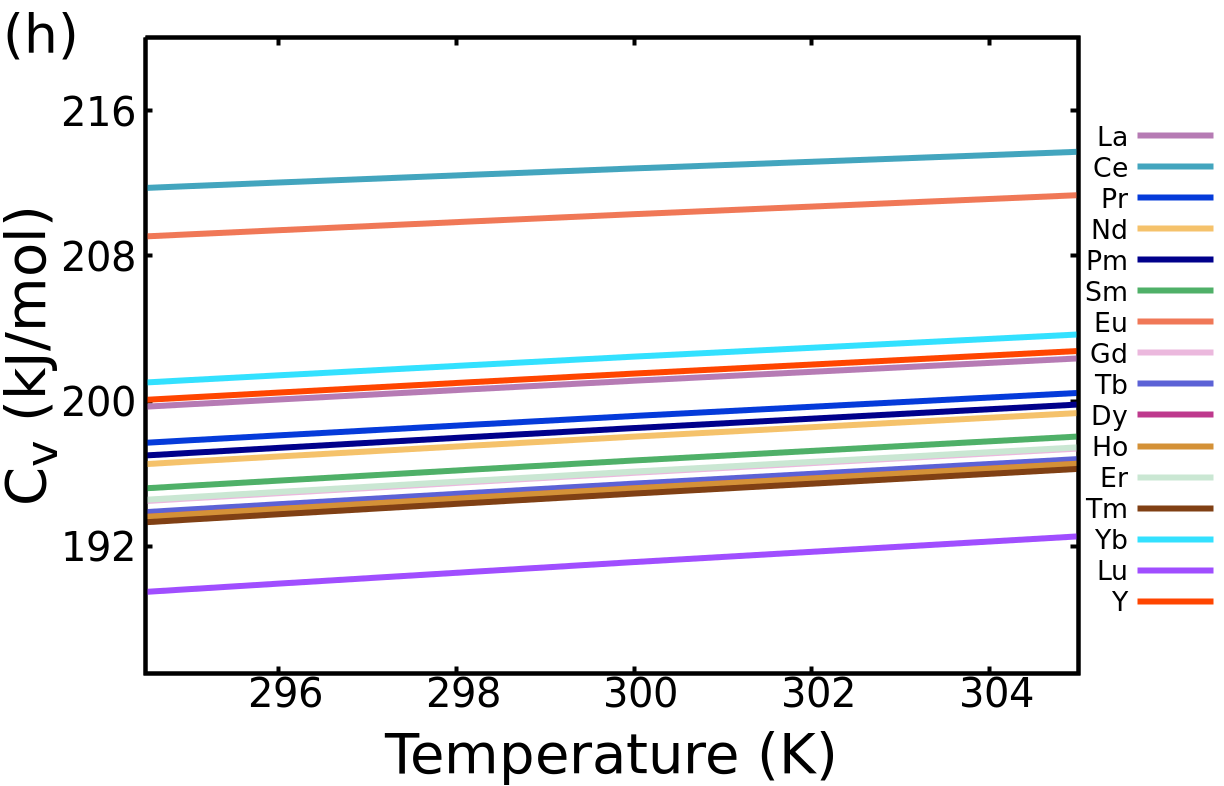}
\includegraphics[height=5.5cm,width=7.5cm]{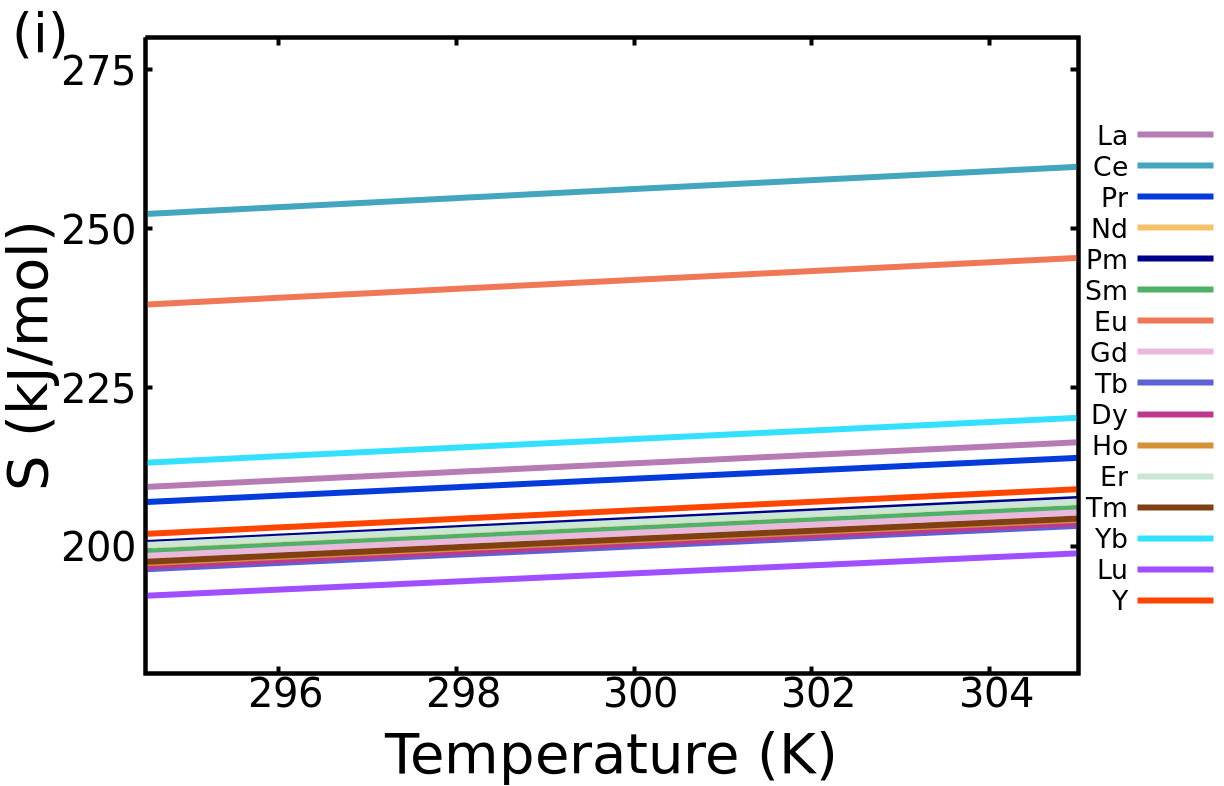}\\
\caption{(a) Thermal functions of RE$_2$NiMnO$_6$ and the magnified counter part at room temperature: (b) Total Energy, (c) -TS, (d)Helmholtz free energy, (e) Enthalpy, (f) Gibbs free Energy, (g) pV term, (h) Specific heat capacity and (i) Entropy.}
\label{fig:thermal6}
\end{center}
\end{figure}

Thermal properties were calculated and analyzed by using PHONOPY, which gathers force constants from both supercell and DFPT methods, for high accuracy within the limits of the quasi-harmonic approximation\cite{Togo23}. Derived harmonic phonon energy from canonical distribution of statistical mechanics is given in Sec. 4.4.6, along with other derived quantities \cite{Togo15}.
The calculated phonon thermal properties such as total energy, Helmholtz free energy, Enthalpy, Gibbs Energy and other thermodynamic functions are shown in Fig. \ref{fig:thermal6}(a) and a magnified version around room temperature, near 300 $K$ are shown in Fig. \ref{fig:thermal6}(b-i). 
The thermodynamic potentials and other derived properties of the materials were studied and analysed over the range 0-1500K. The data so obtained has been represented in the figures listed above. Overall, the thermodynamic behavior exhibited qualitatively similar trends across all materials, with only minor quantitative variations in magnitude and curvature. 
The internal energy, $U(T)$ increases smoothly and continuously with temperature, in direct agreement with the specific heat trends, reflecting steady thermal energy accumulation with no anomalous features. The total energy of the materials at 300 $K$ are shown in \ref{tab:thermal6}. The Specific heat capacity $C_v$, shows a general monotonic increase rising steeply at low temperatures, aligning with Debye behaviour, before gradually plateauing at higher temperatures. At high temperature range, $C_v$ asymptotically approaches 3$Nk_B$ in accordance with Dulong-Petit law which is a characteristic of crystalline solids. The absence of discontinuities or abrupt peaks in either $C_v$ or $U(T)$ suggests the absence of any structural, magnetic, or electronic phase transitions within the examined temperature range.
The Helmholtz free energy, F(T), which quantifies the maximum reversible work extractable from a system at constant volume and temperature, decreases monotonically with increasing temperature, consistent with thermodynamic expectations. Ce$_2$MnNiO$_6$ appears to be the most stable structure owing to its lowest Helmholtz free energy value at that temperature. 
\begin{table}[htbp]
\centering
\caption{Thermodynamic properties of lanthanide elements at T = 300$~K$.}
\begin{tabular}{lccccc}
\hline
Element & $E(T)$ & $C_v(T)$ & $F(T)$ & $S(T)$ \\
    & (meV) & (J/mol·K) & (meV) & (J/mol·K) \\
\hline
La & 91.76 & 201.12 & 27.84 & 213.08 \\
Ce & 84.87 & 212.81 & 8.00  & 256.22 \\
Pr & 92.51 & 199.19 & 29.31 & 210.66 \\
Nd & 92.89 & 198.05 & 31.76 & 203.78 \\
Pm & 92.63 & 198.52 & 31.37 & 204.20 \\
Sm & 93.03 & 196.74 & 32.19 & 202.81 \\
Eu & 88.07 & 210.30 & 15.49 & 241.93 \\
Gd & 93.42 & 196.05 & 32.78 & 202.13 \\
Tb & 93.77 & 195.46 & 33.76 & 200.02 \\
Dy & 93.89 & 195.20 & 33.76 & 200.42 \\
Ho & 93.92 & 195.20 & 33.63 & 200.97 \\
Er & 93.48 & 196.12 & 32.32 & 203.84 \\
Tm & 94.05 & 194.91 & 33.69 & 201.19 \\
Yb & 91.03 & 202.45 & 25.96 & 216.90 \\
Lu & 95.59 & 191.14 & 36.86 & 195.77 \\
\hline
\end{tabular}
\label{tab:thermal6}
\end{table}
\begin{figure}[H]
\begin{center}
\includegraphics[height=7.5cm,width=12cm]{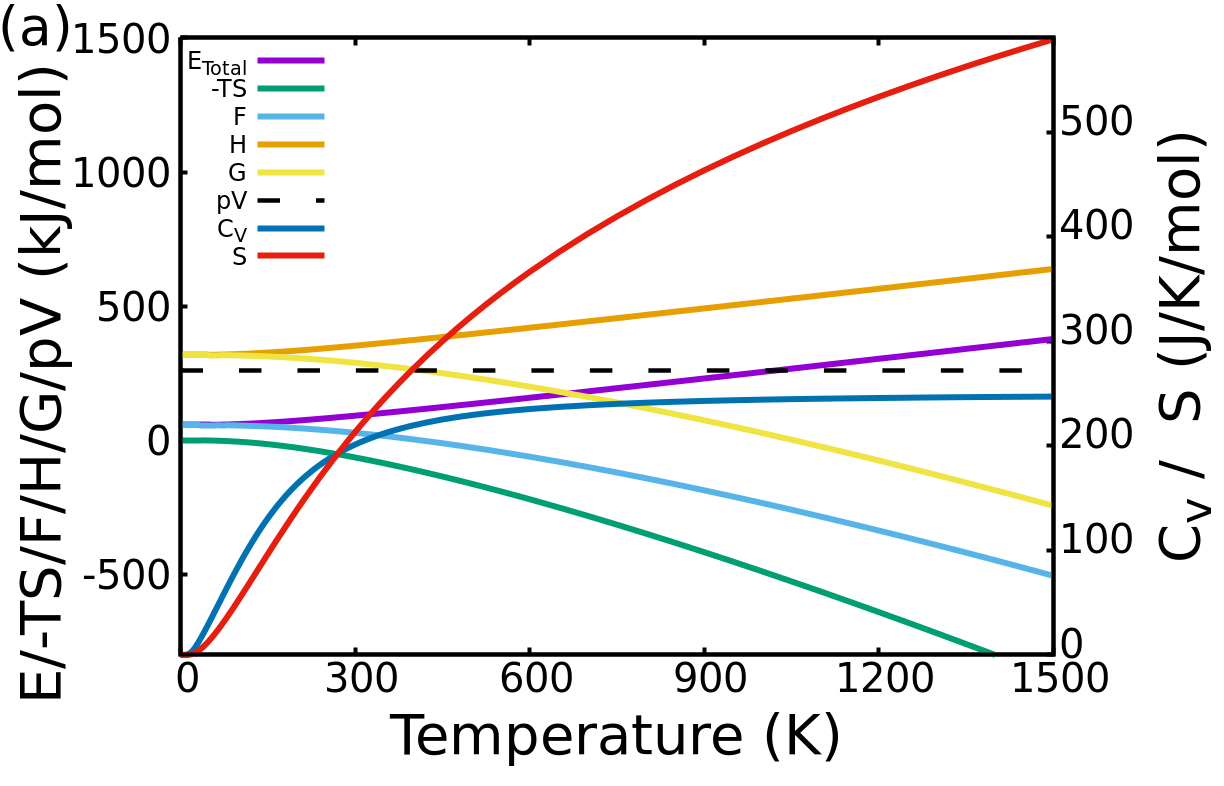}\\
\end{center}
\end{figure}
Entropies of the materials under study grow smoothly with temperature exhibiting general agreement with the second law of thermodynamics. The absence of sharp changes or curvature discontinuities in the entropy profiles further supports the conclusion of continuous thermodynamic evolution without phase instability. Similarly, Gibbs' Free energy too varies conforming to the variation of Helmholtz free energy with temperature. $pV$ remains nearly a constant throughout and explains the temperature variation of enthalpy.
\subsection{Magnetisation Density}
Magnetisation density for each atom is represented over its surface with yellow outer layer for positive magnetisation and green for negative. Following plots shows perspective view of 3D iso-surface of magnetisation density with iso-value $\pm$ 007 for better visualisation \cite{Bhuyan22}. Right side plots corresponds to positive and left side for negative magnetisation with their respective atoms. Positive magnetisation is seen for the metals in most of cases except Ce$_2$MnNIO$_6$, Pm$_2$MnNIO$_6$, Tb$_2$MnNIO$_6$, Ho$_2$MnNIO$_6$, Er$_2$MnNIO$_6$ and Tm$_2$MnNIO$_6$, where Ni atom shows negative density. 
\begin{figure}[H]
\begin{center}
\includegraphics[height=5.5cm,width=7.5cm]{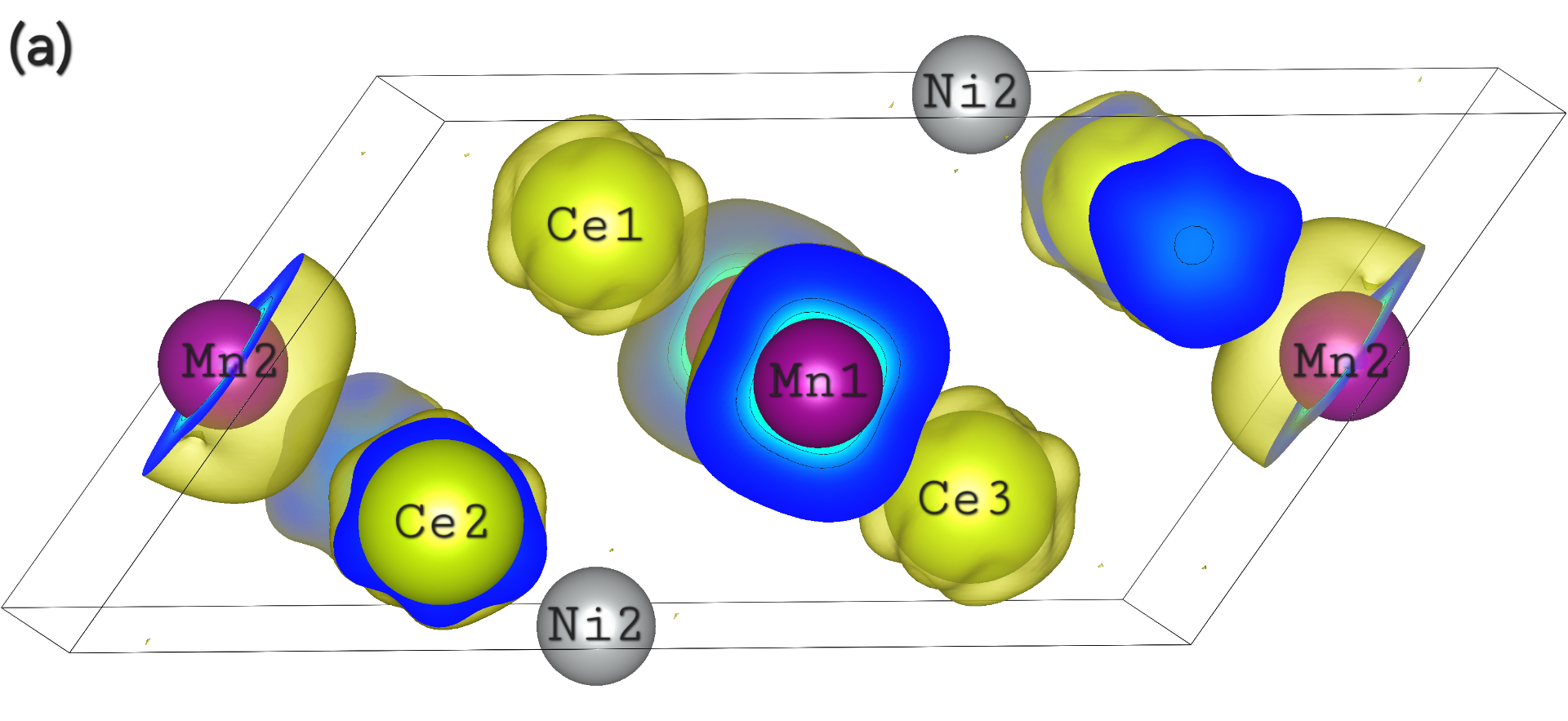}
\includegraphics[height=5.5cm,width=7.5cm]{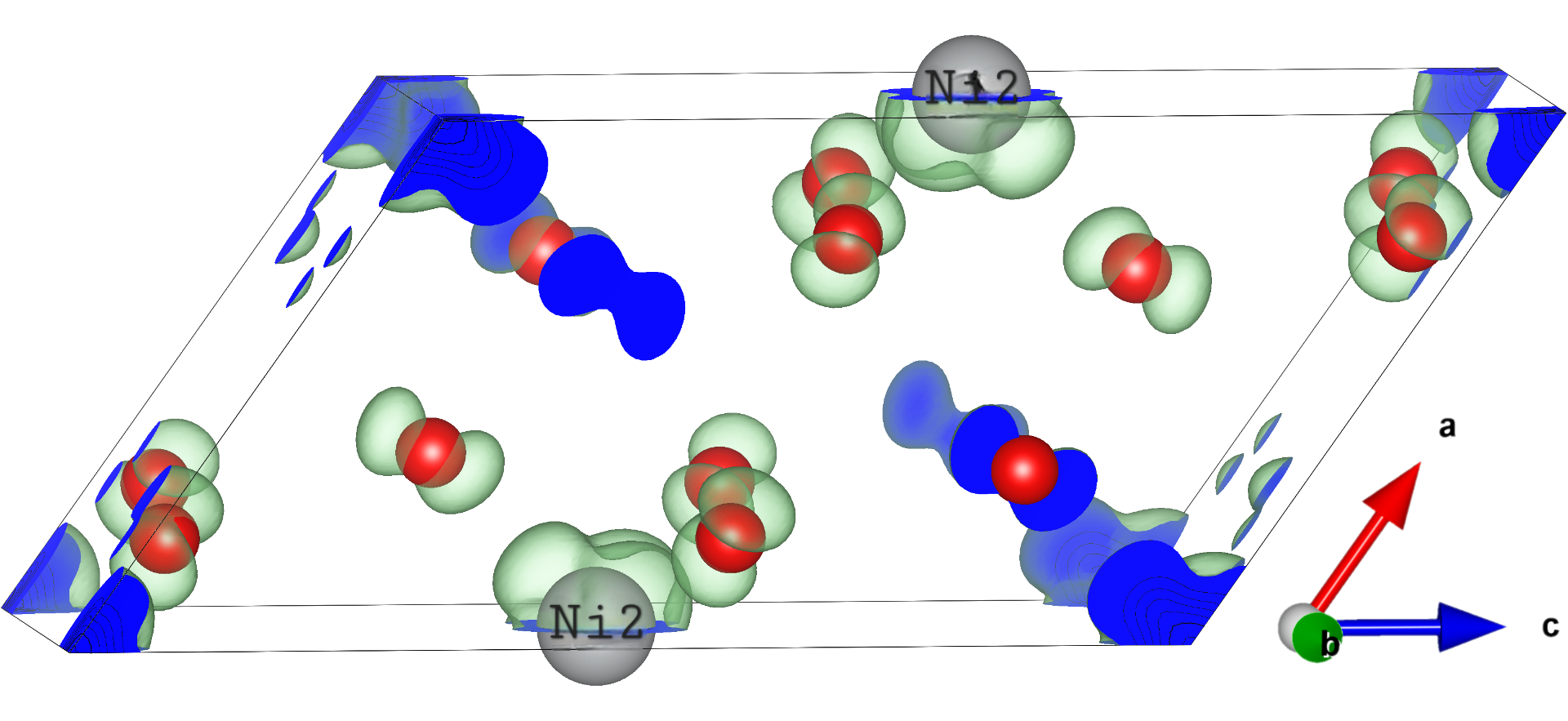}\\
\includegraphics[height=5.5cm,width=7.5cm]{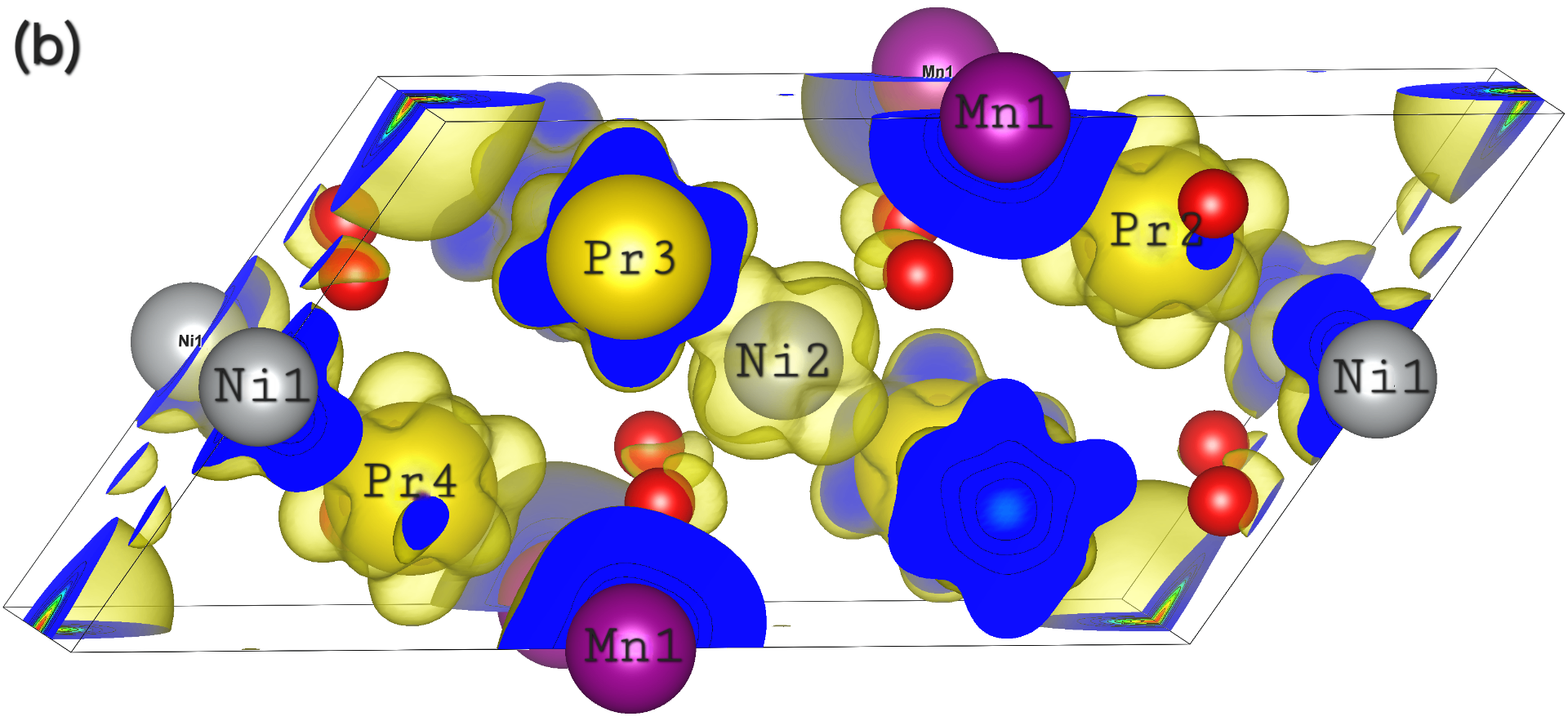}
\includegraphics[height=5.5cm,width=7.5cm]{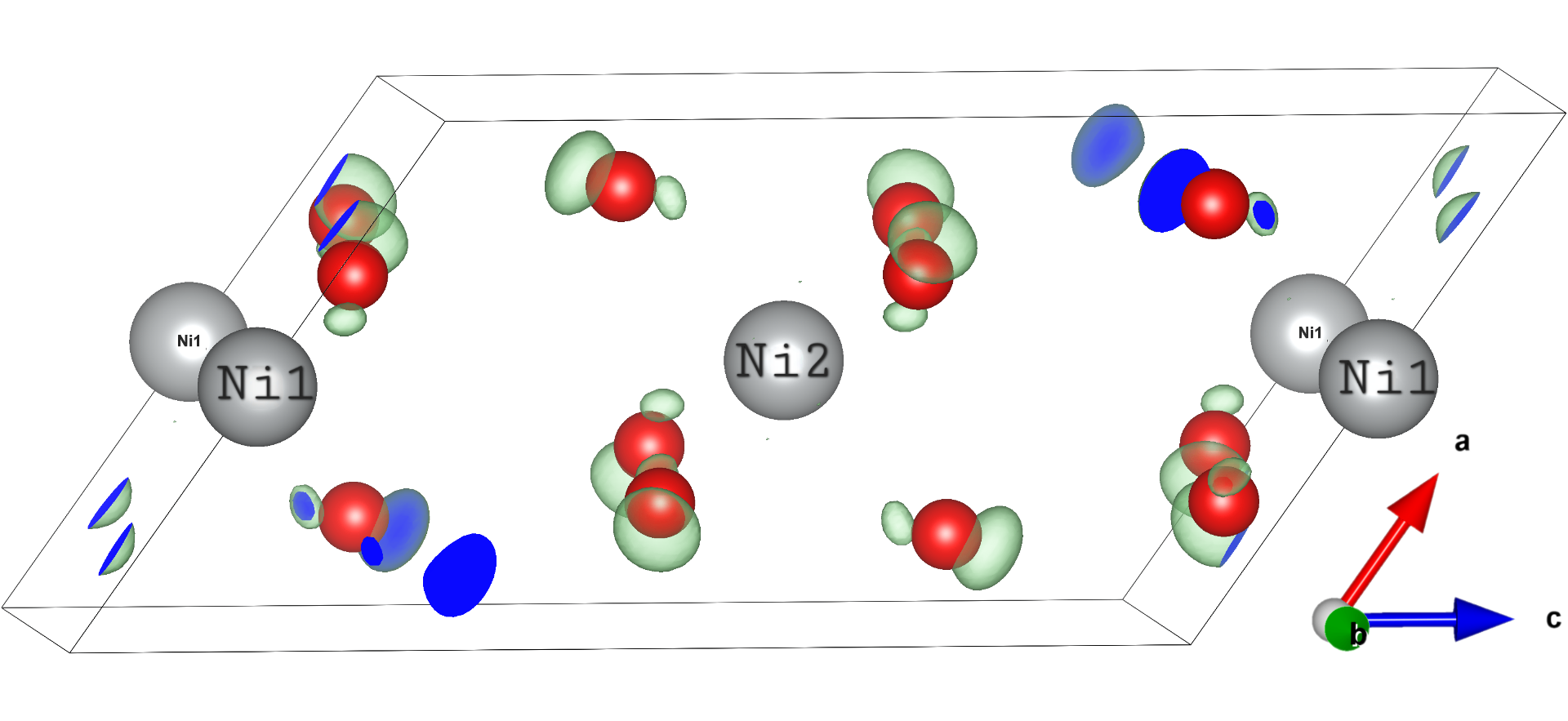}\\
\includegraphics[height=5.5cm,width=7.5cm]{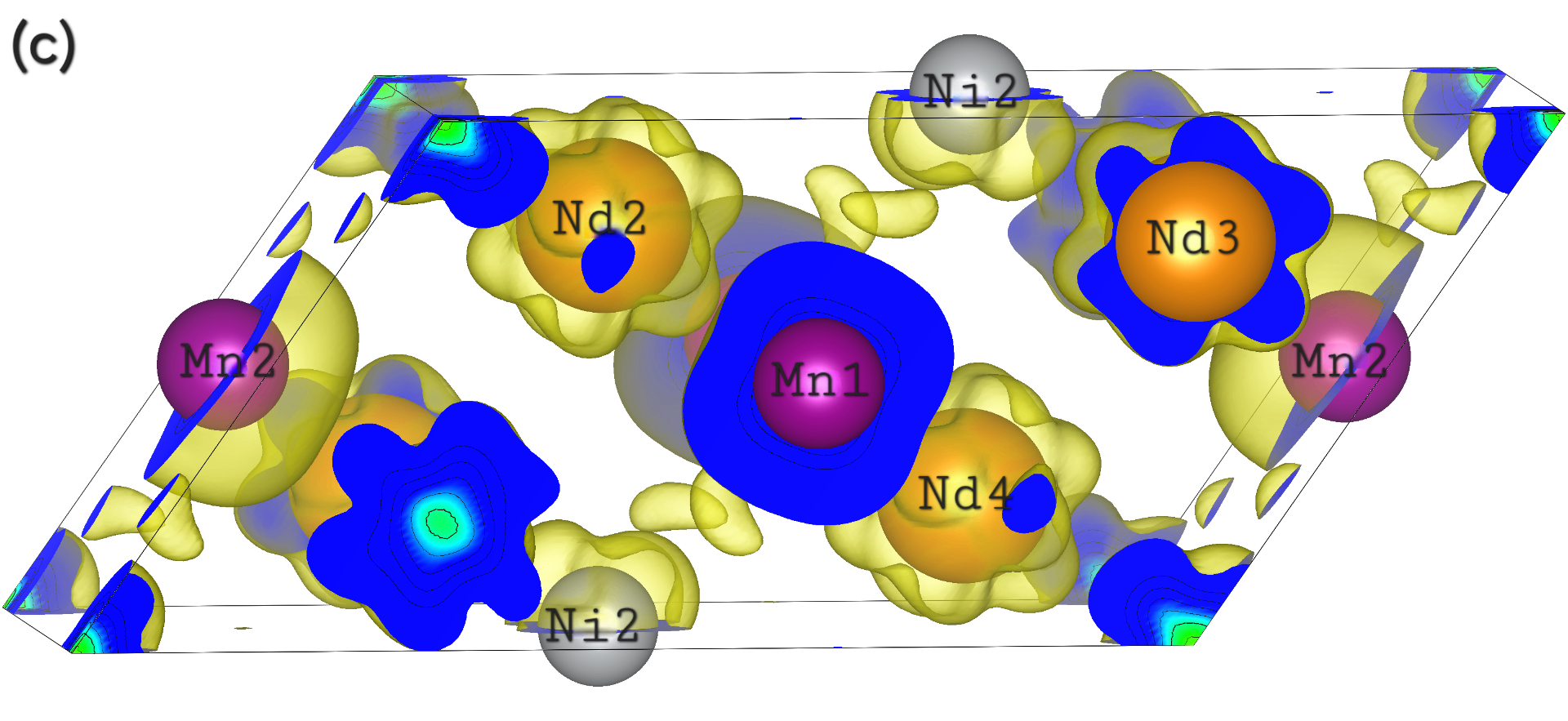}
\includegraphics[height=5.5cm,width=7.5cm]{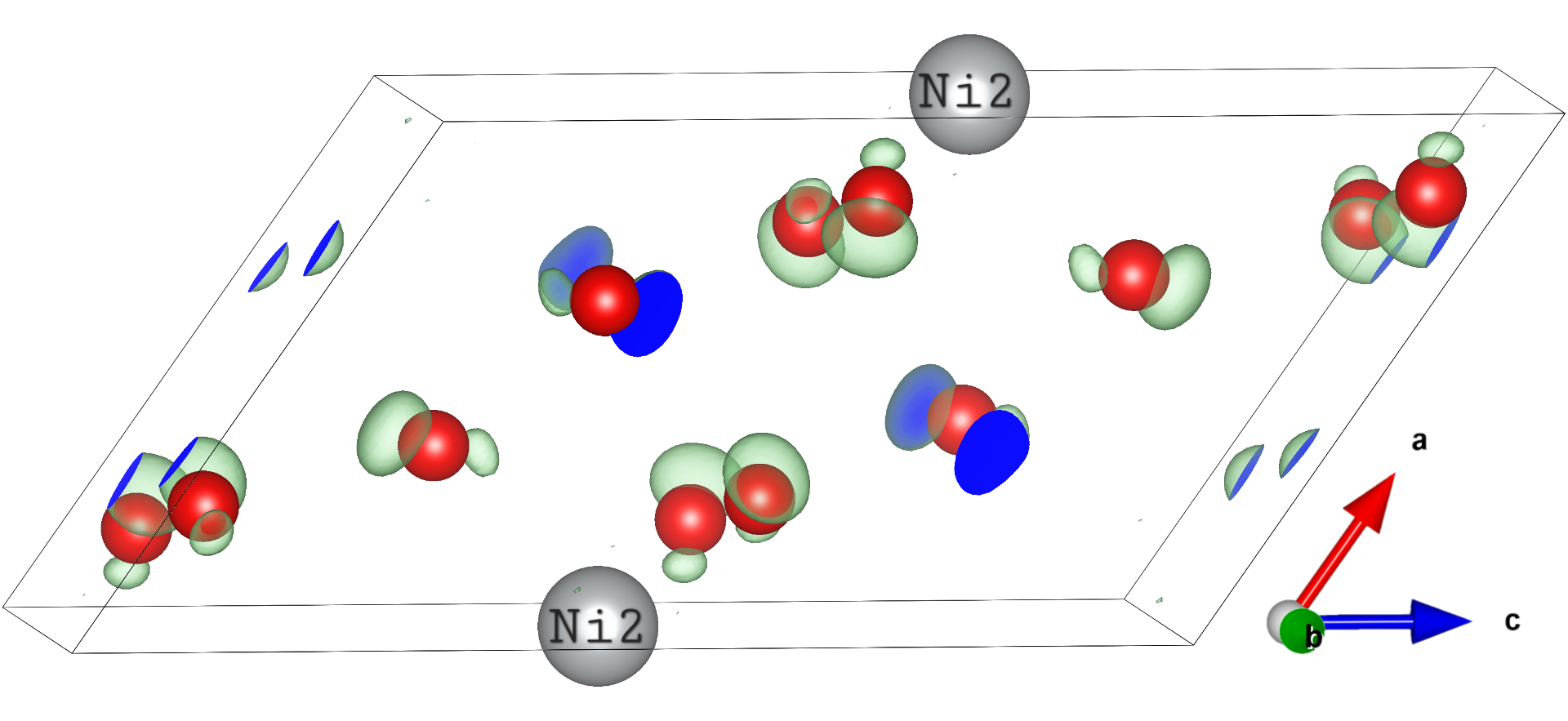}\\
\end{center}
\end{figure}
\begin{figure}[H]
\begin{center}
\includegraphics[height=5.5cm,width=7.5cm]{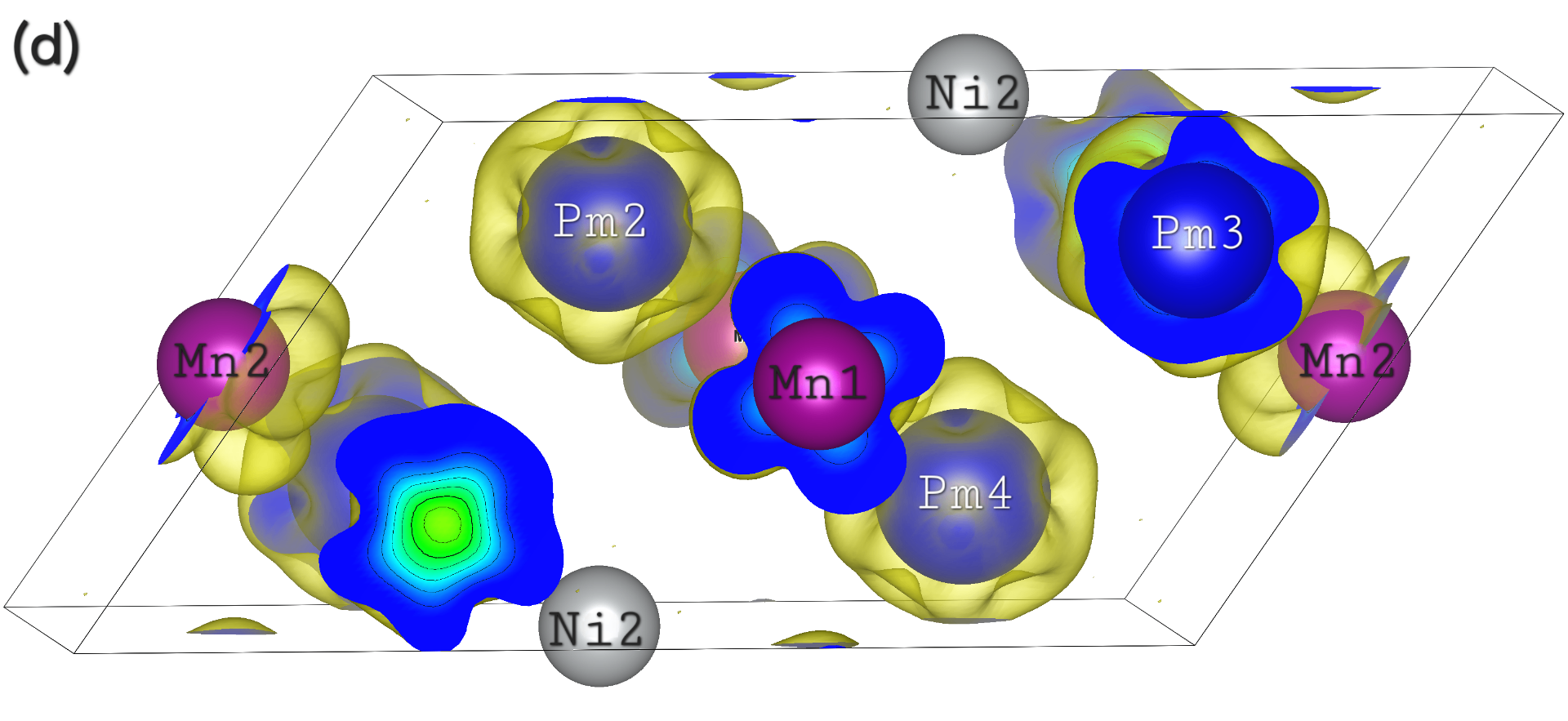}
\includegraphics[height=5.5cm,width=7.5cm]{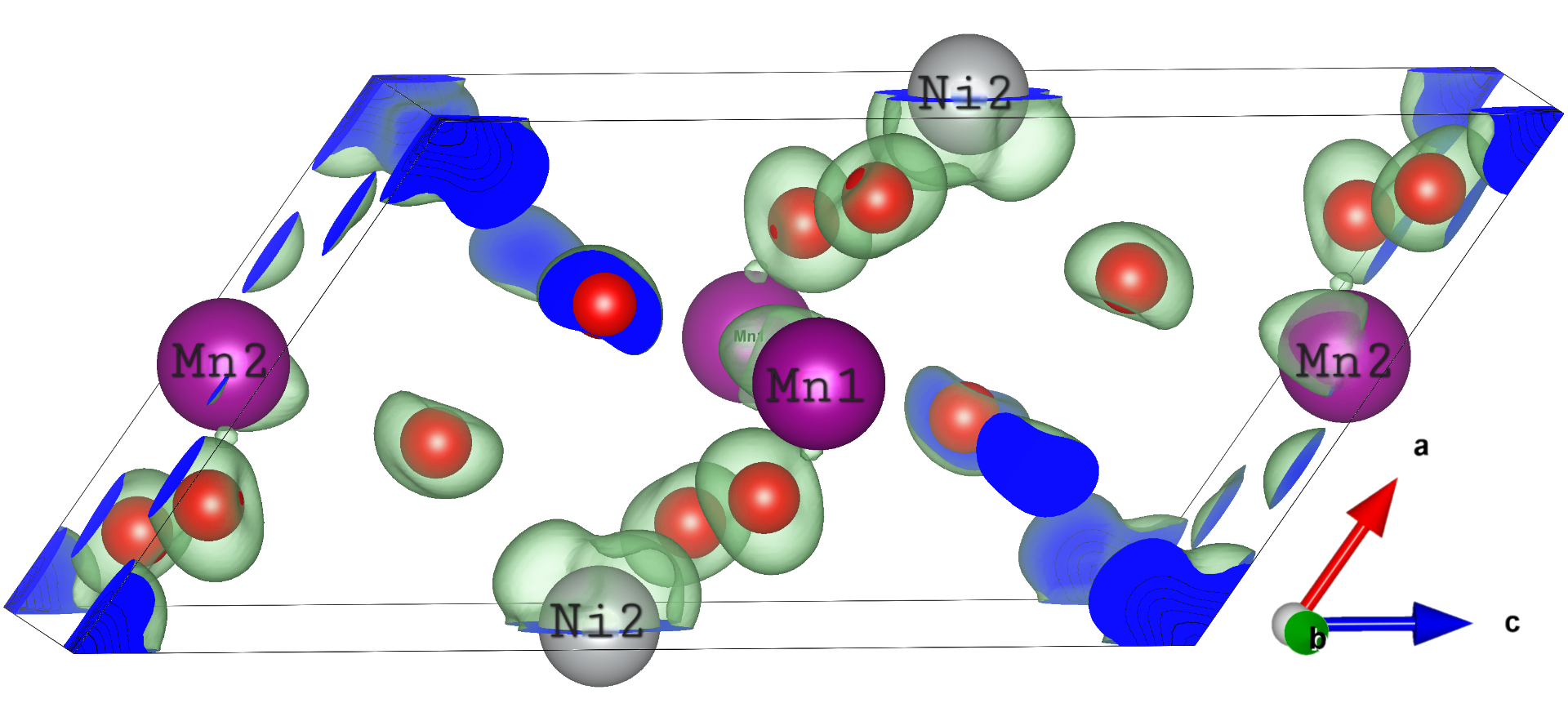}\\
\includegraphics[height=5.5cm,width=7.5cm]{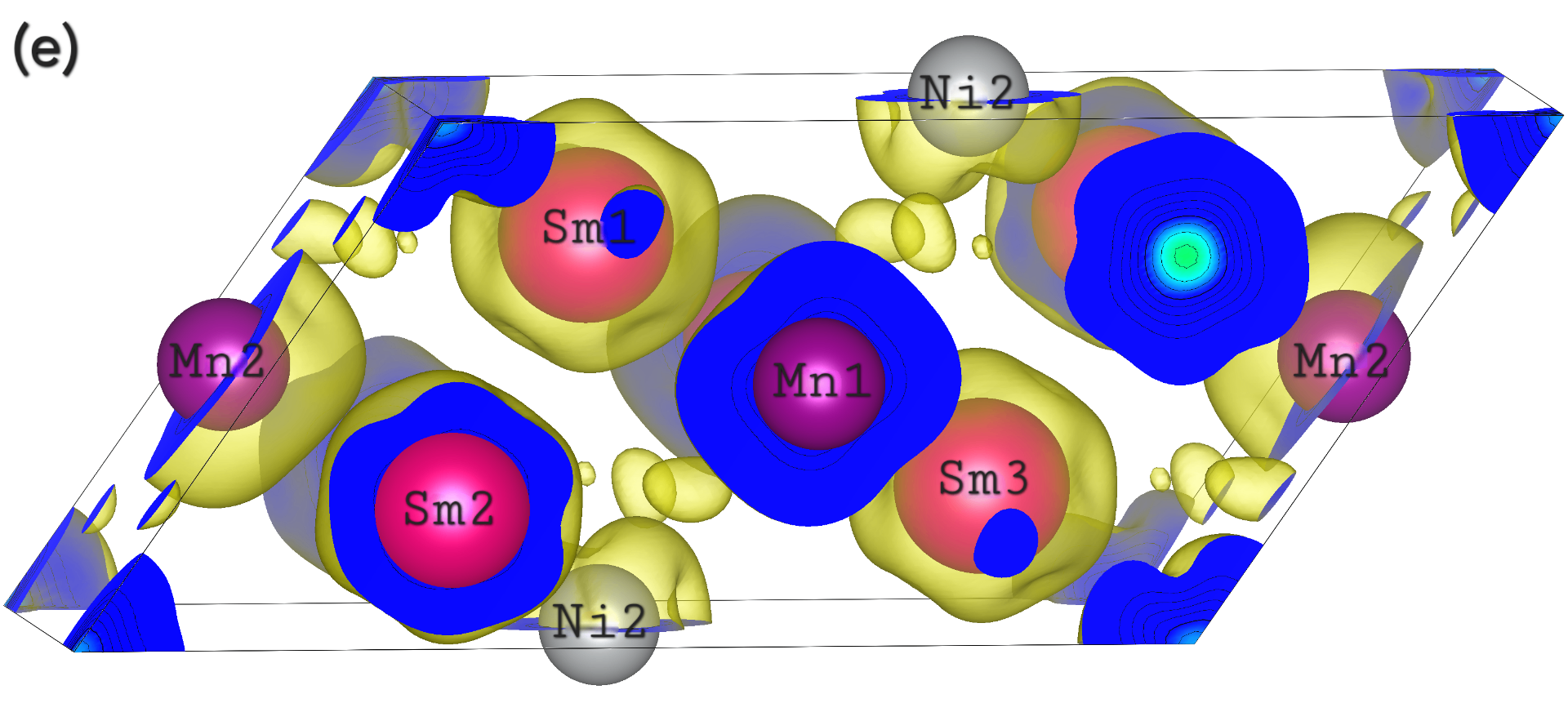}
\includegraphics[height=5.5cm,width=7.5cm]{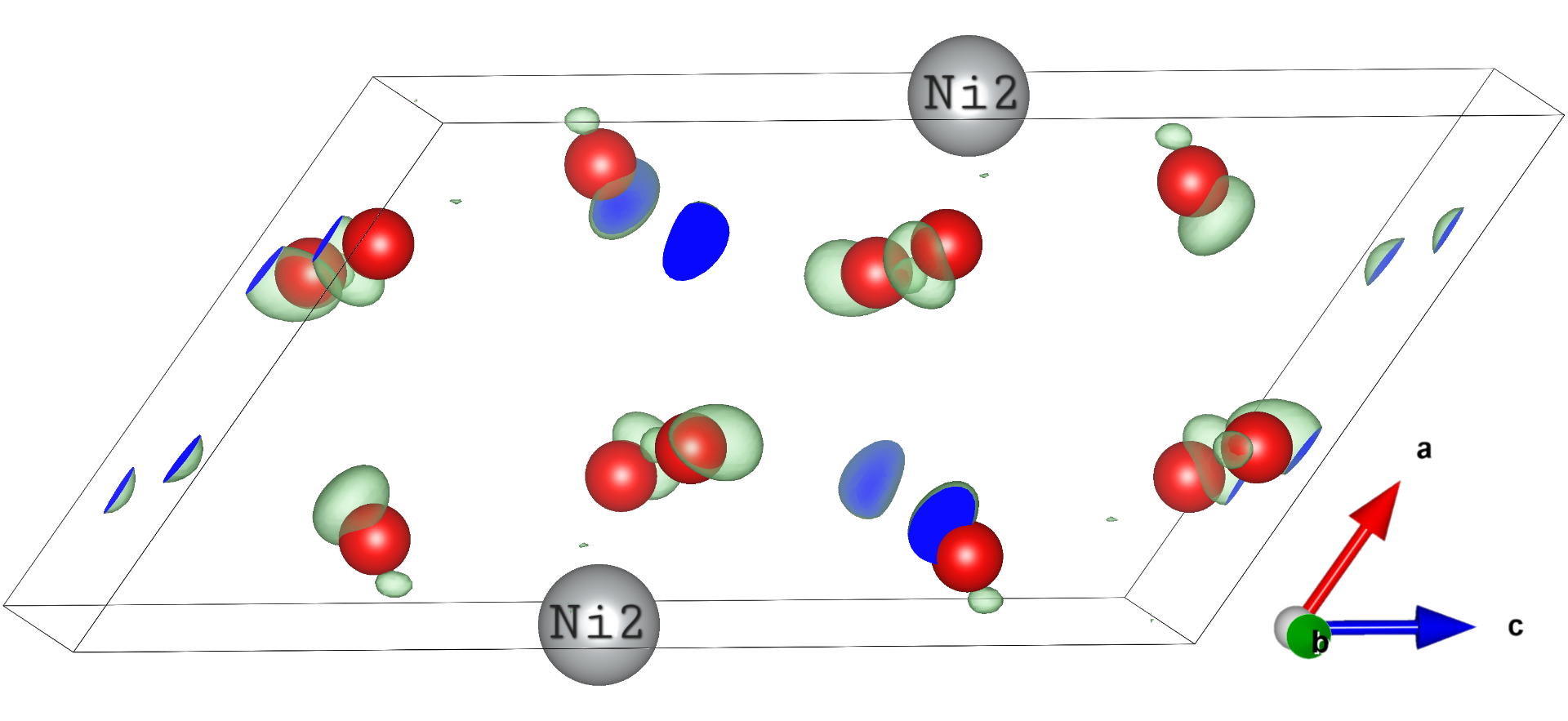}\\\includegraphics[height=5.5cm,width=7.5cm]{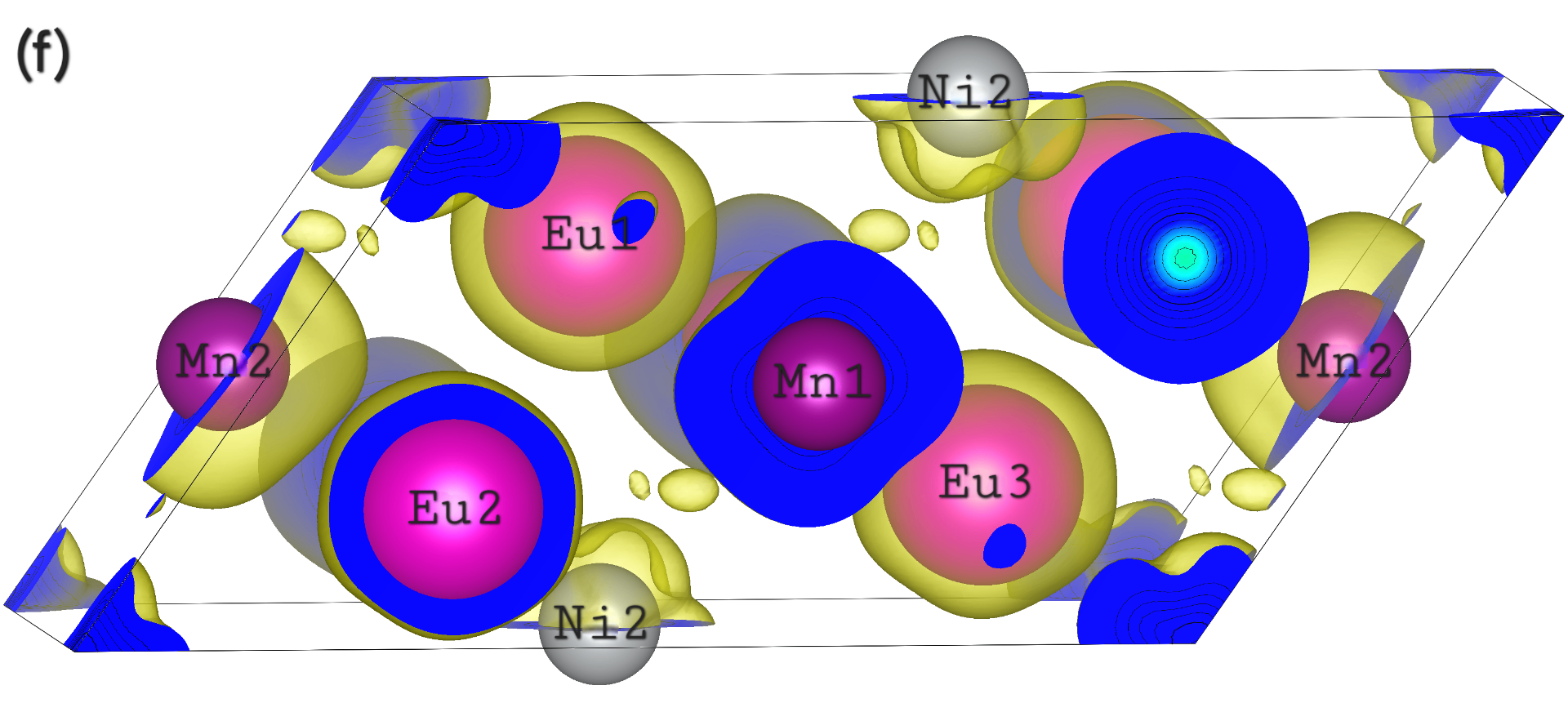}
\includegraphics[height=5.5cm,width=7.5cm]{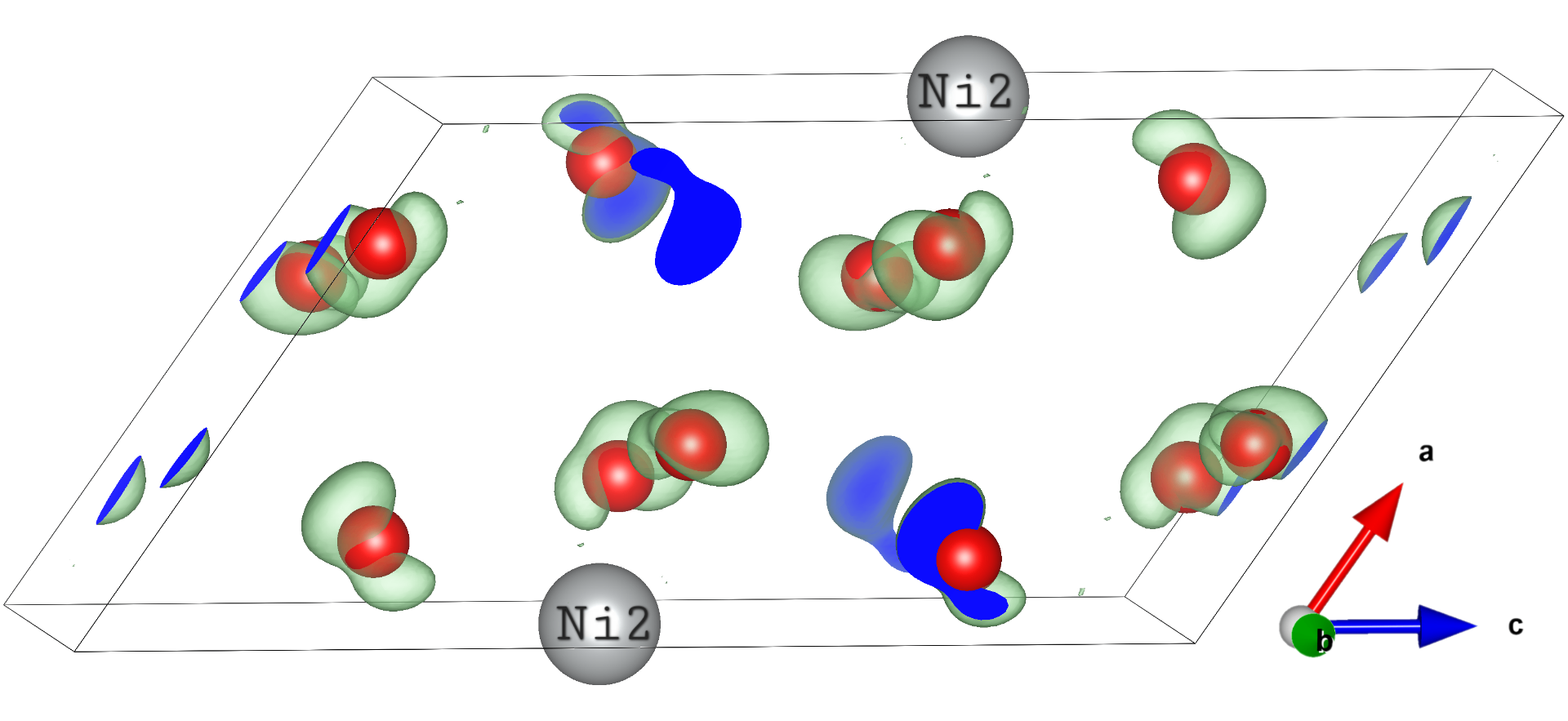}\\
\includegraphics[height=5.5cm,width=7.5cm]{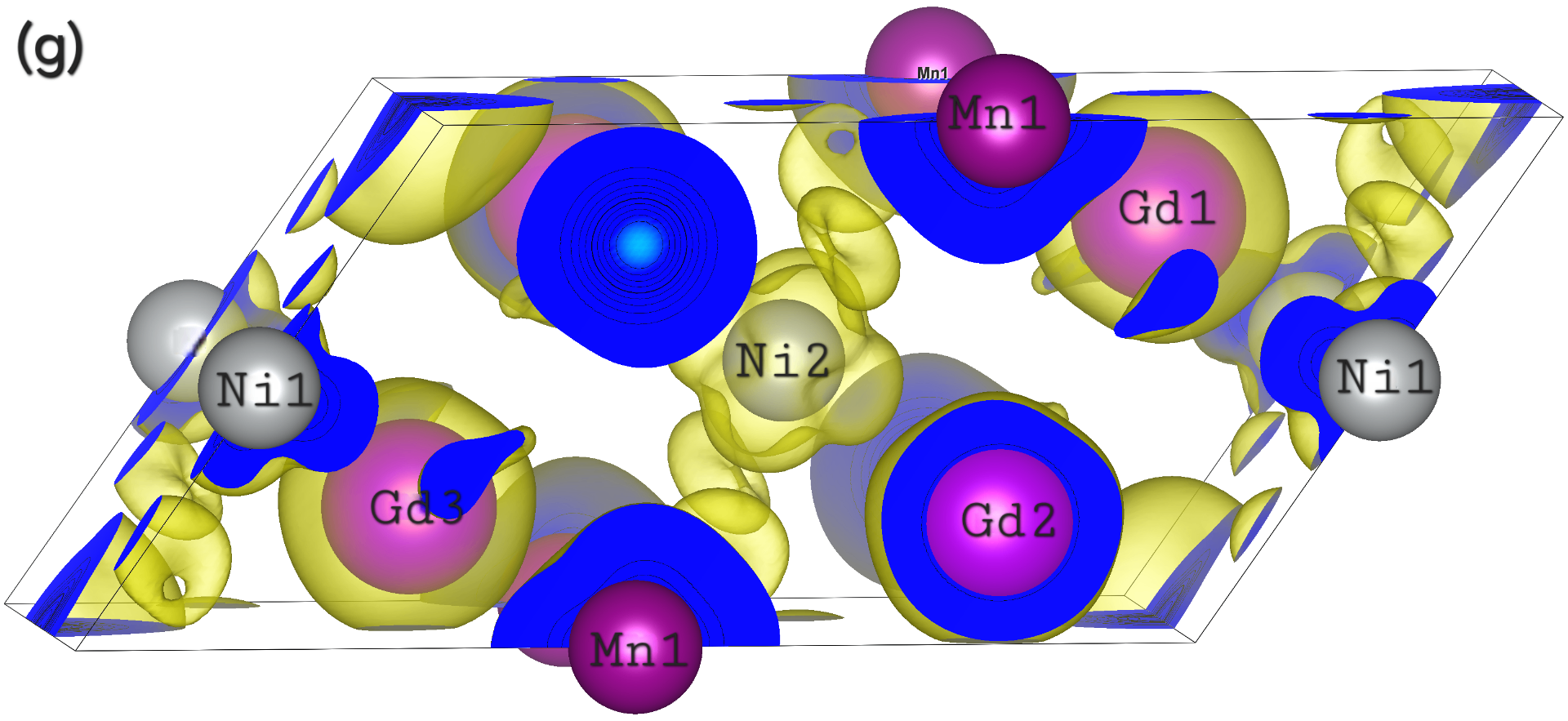}
\includegraphics[height=5.5cm,width=7.5cm]{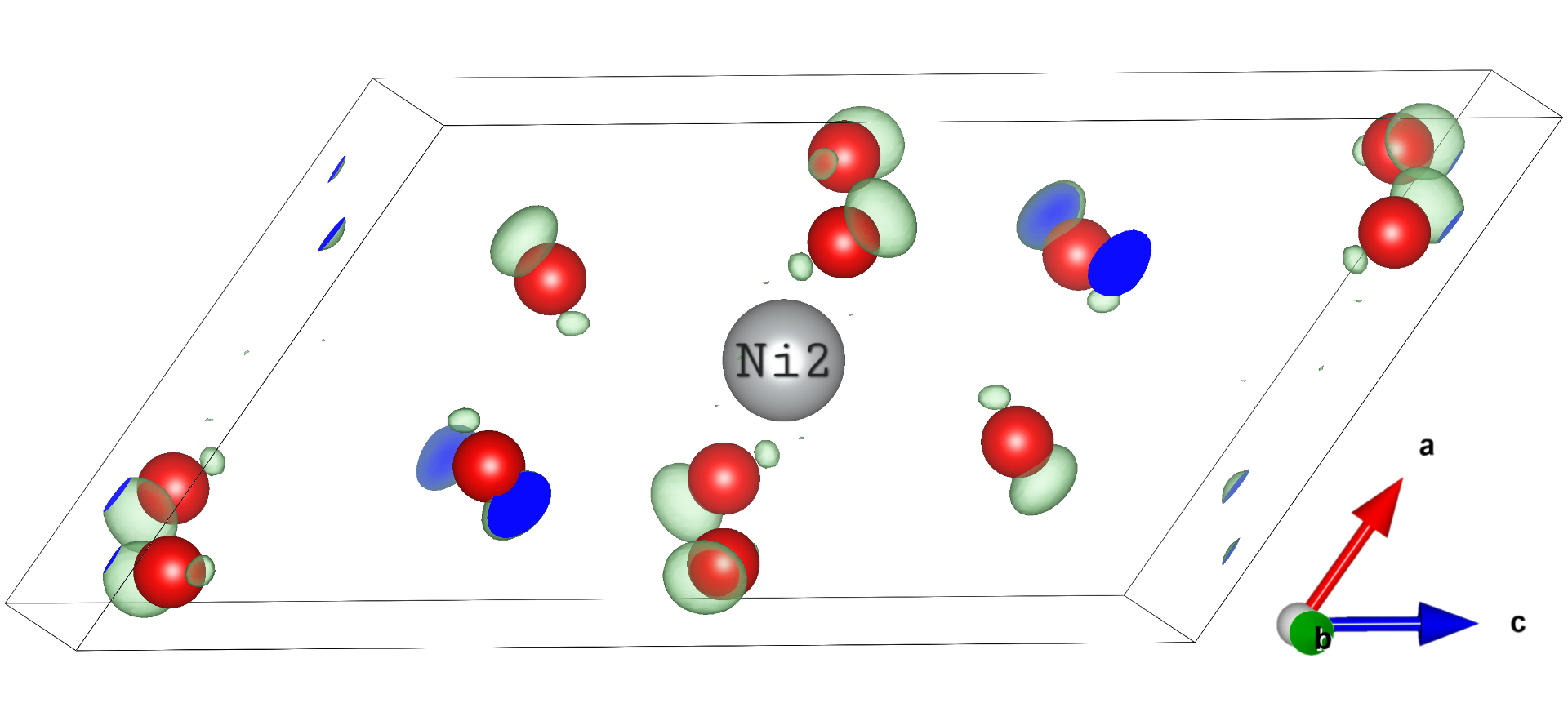}\\
\end{center}
\end{figure}
\begin{figure}[H]
\begin{center}
\includegraphics[height=5.5cm,width=7.5cm]{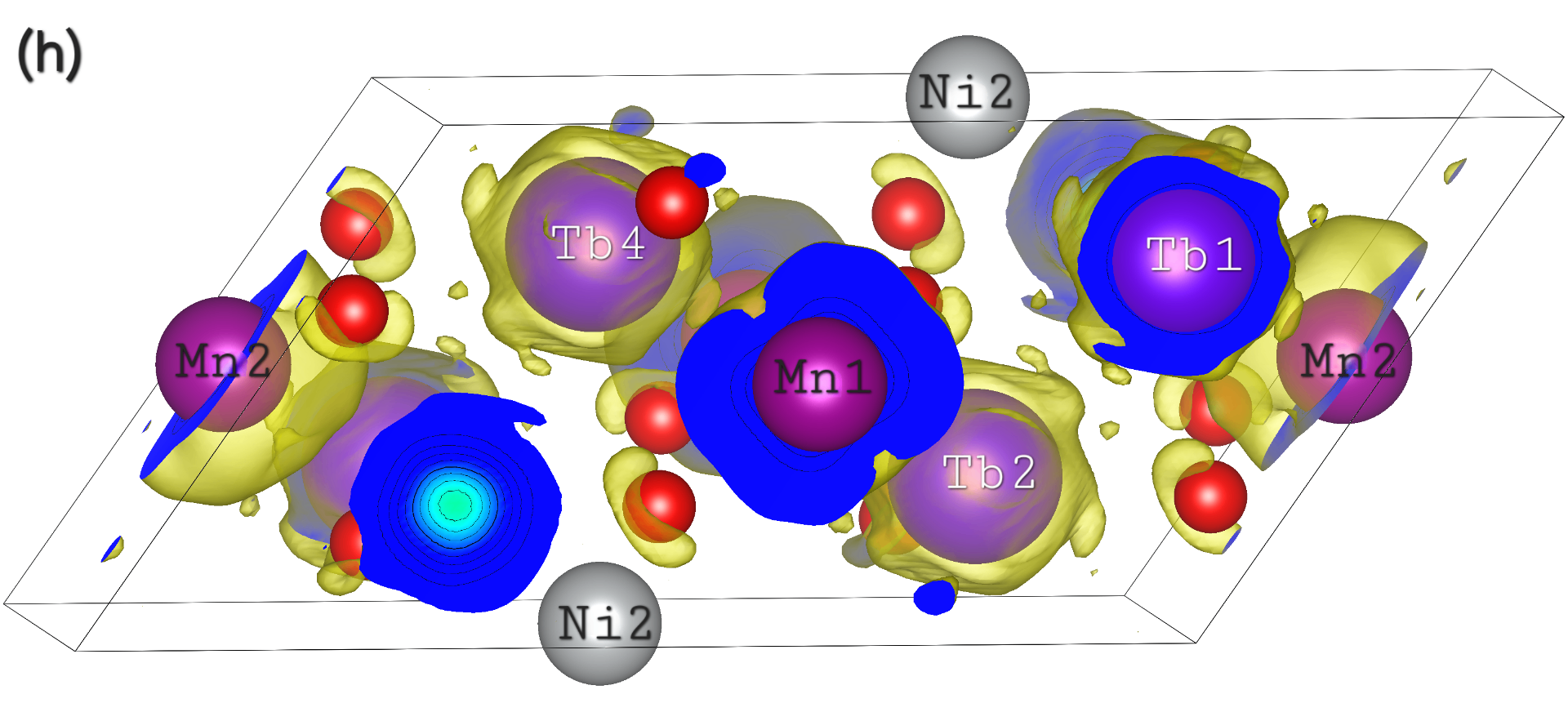}
\includegraphics[height=5.5cm,width=7.5cm]{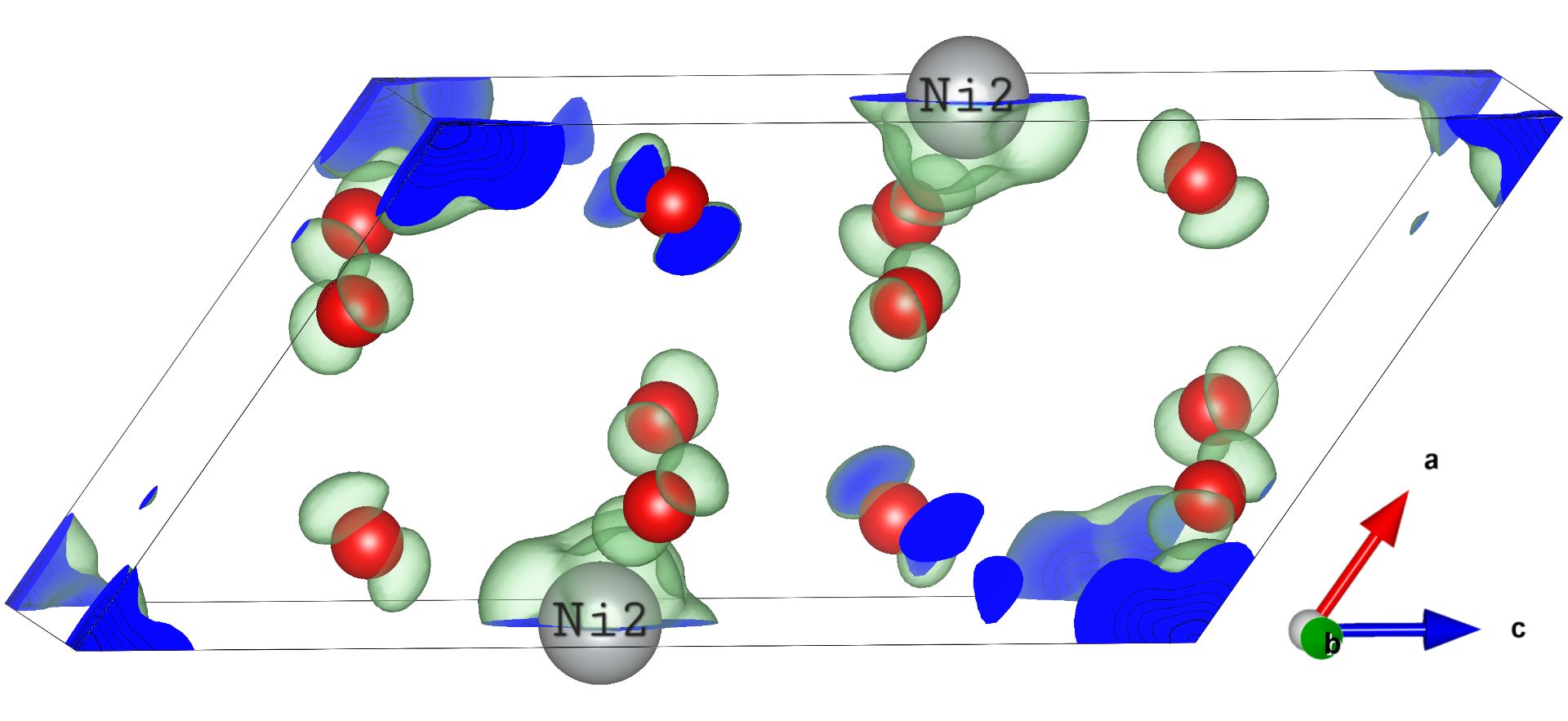}\\
\includegraphics[height=5.5cm,width=7.5cm]{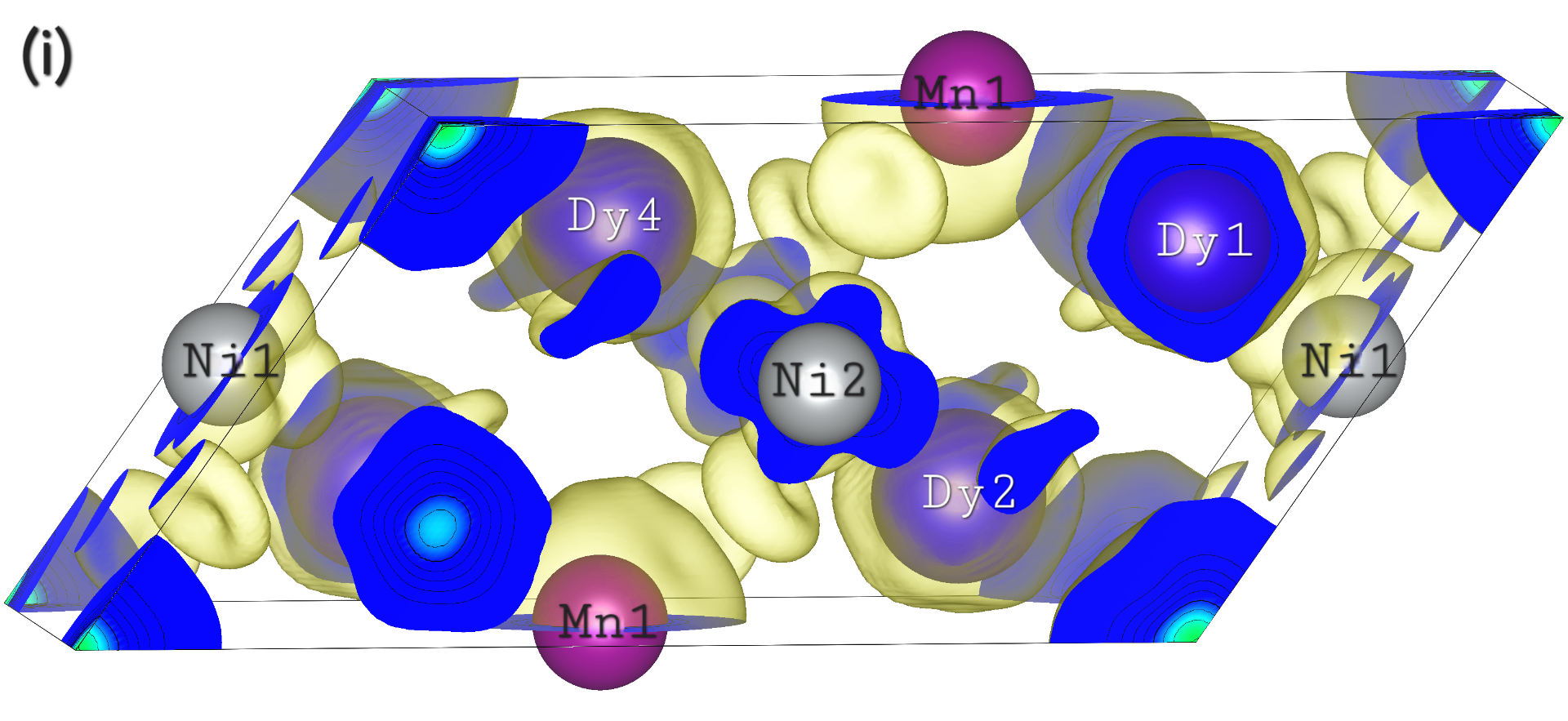}
\includegraphics[height=5.5cm,width=7.5cm]{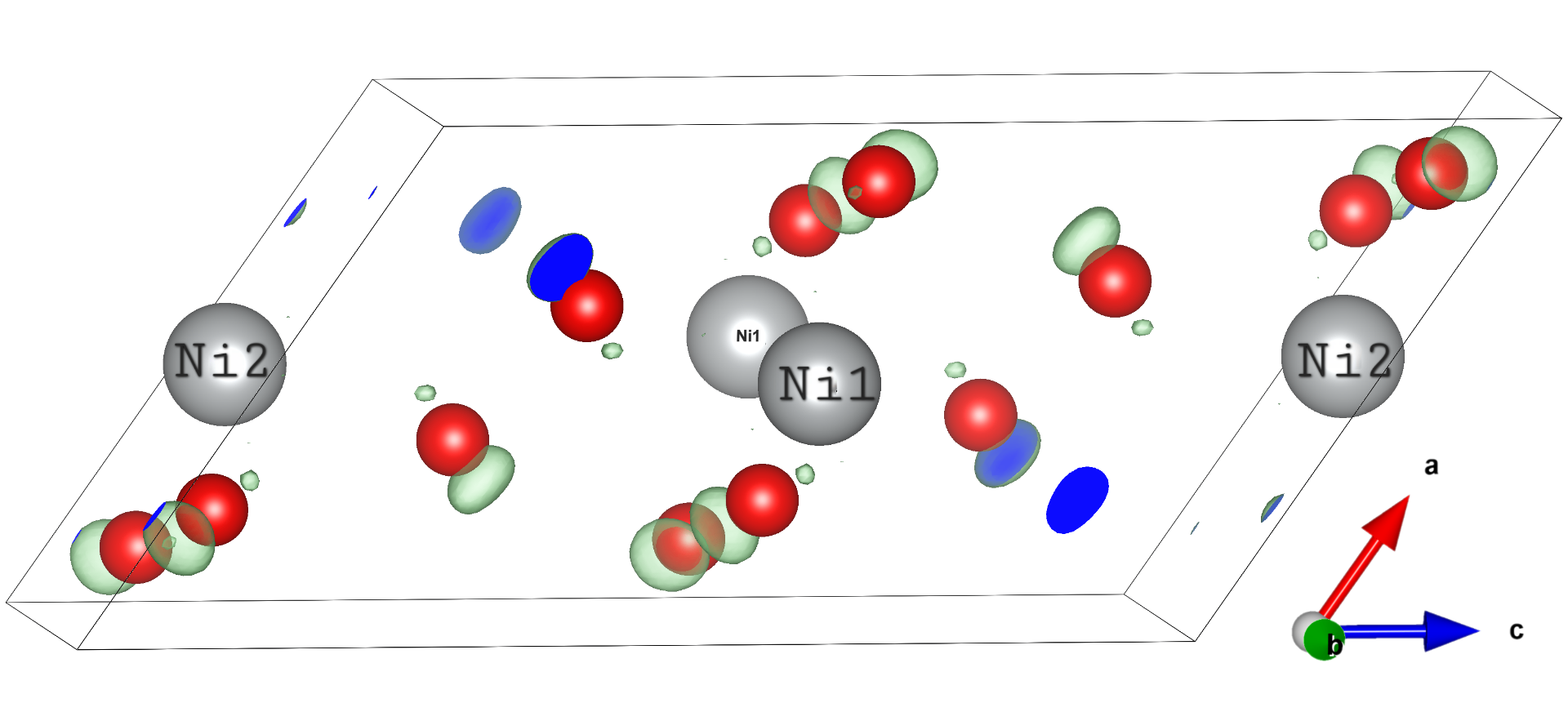}\\
\includegraphics[height=5.5cm,width=7.5cm]{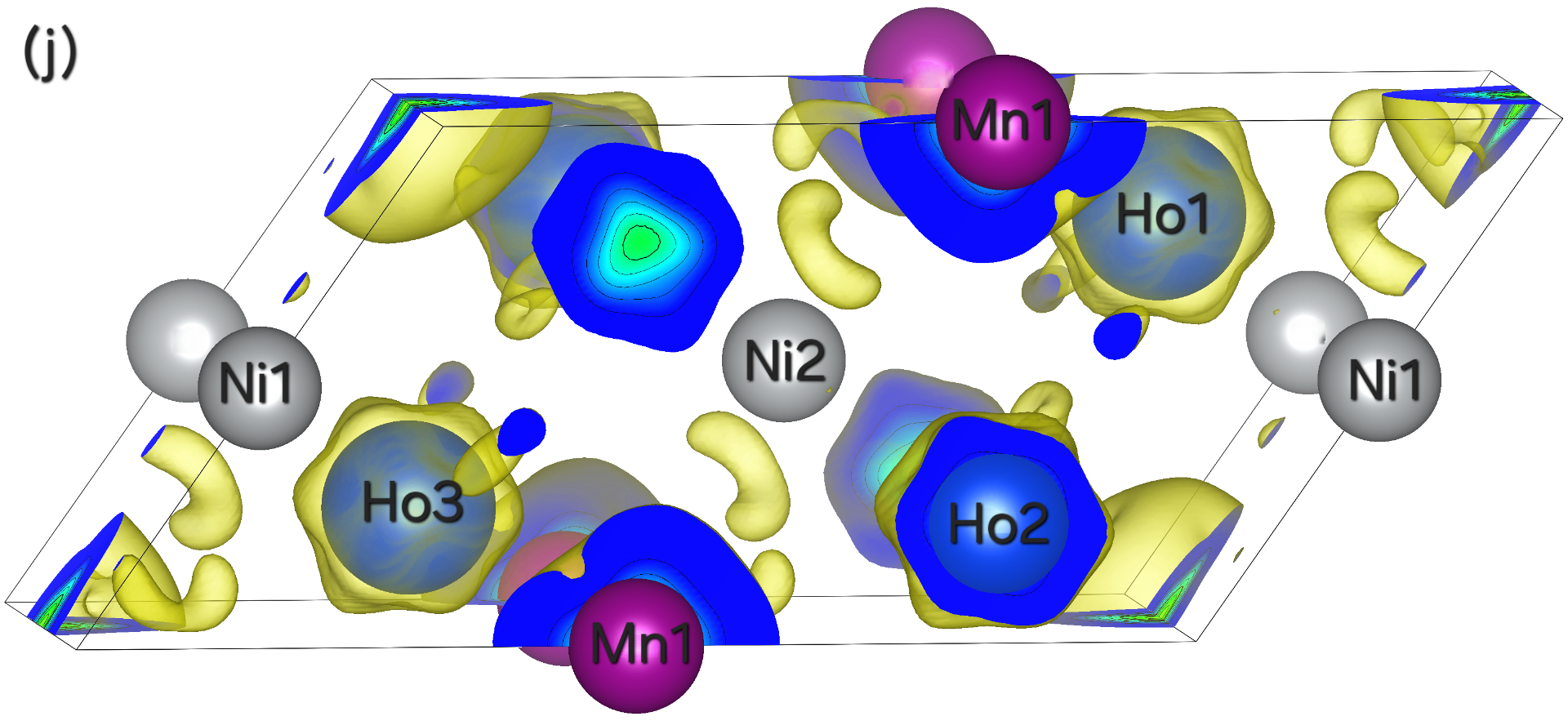}
\includegraphics[height=5.5cm,width=7.5cm]{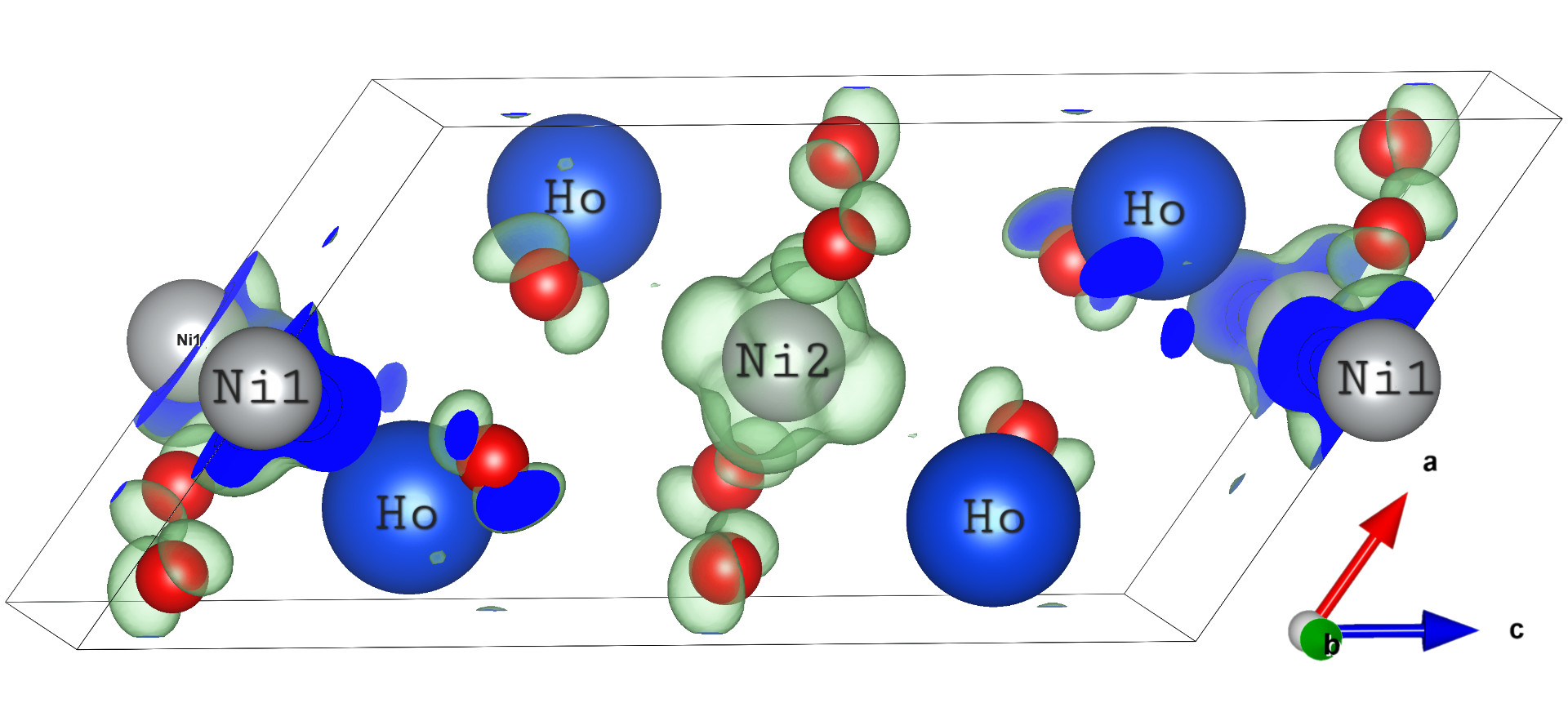}\\
\includegraphics[height=5.5cm,width=7.5cm]{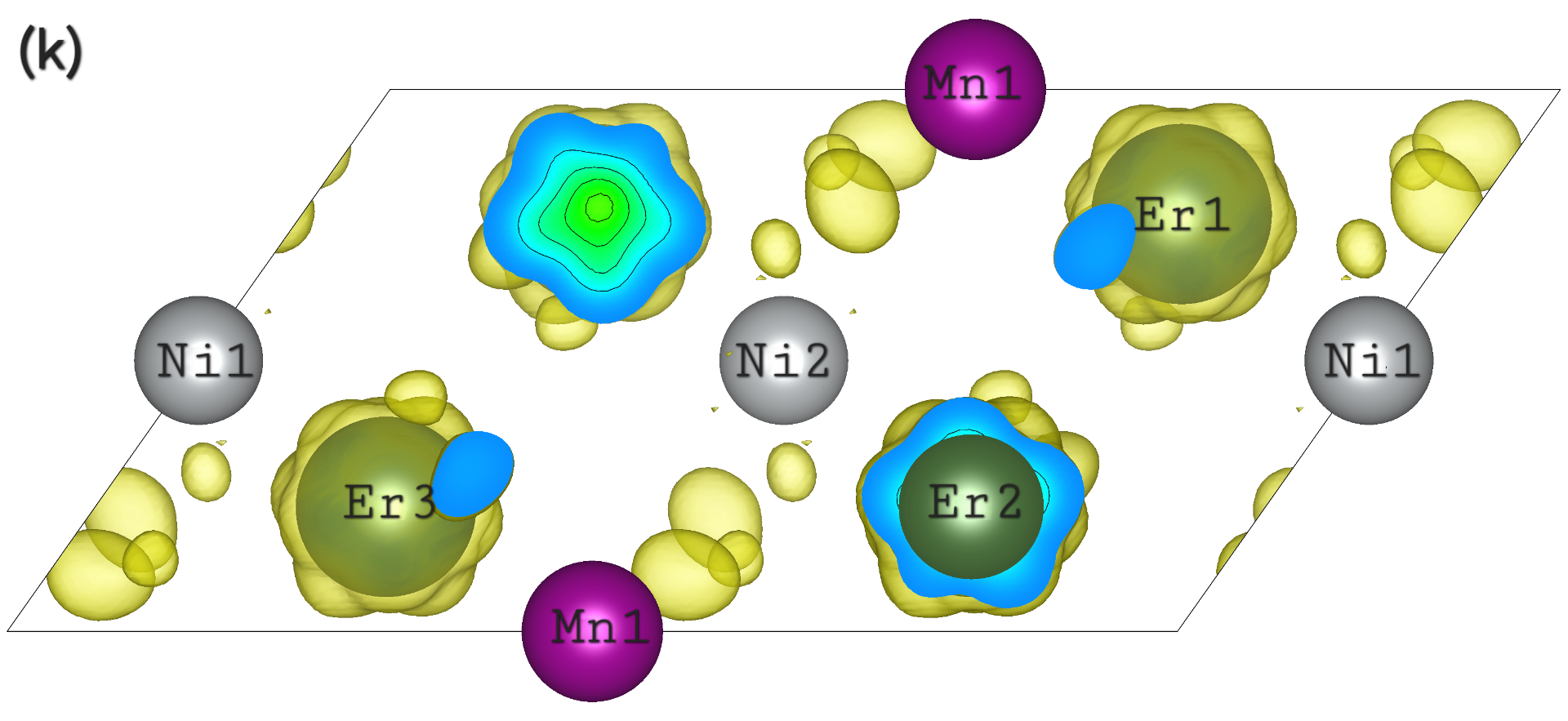}
\includegraphics[height=5.5cm,width=7.5cm]{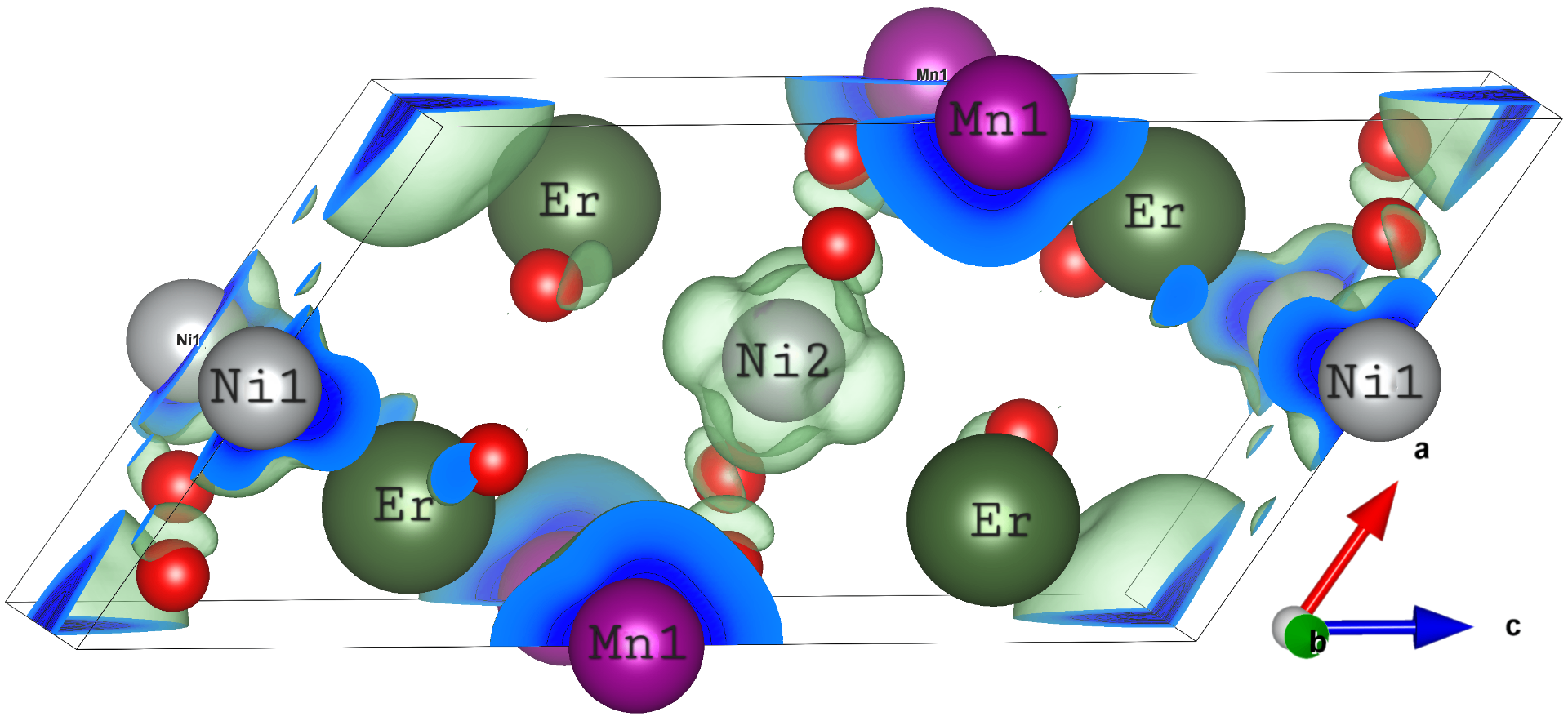}\\
\end{center}
\end{figure}
\begin{figure}[H]
\begin{center}
\includegraphics[height=5.5cm,width=7.5cm]{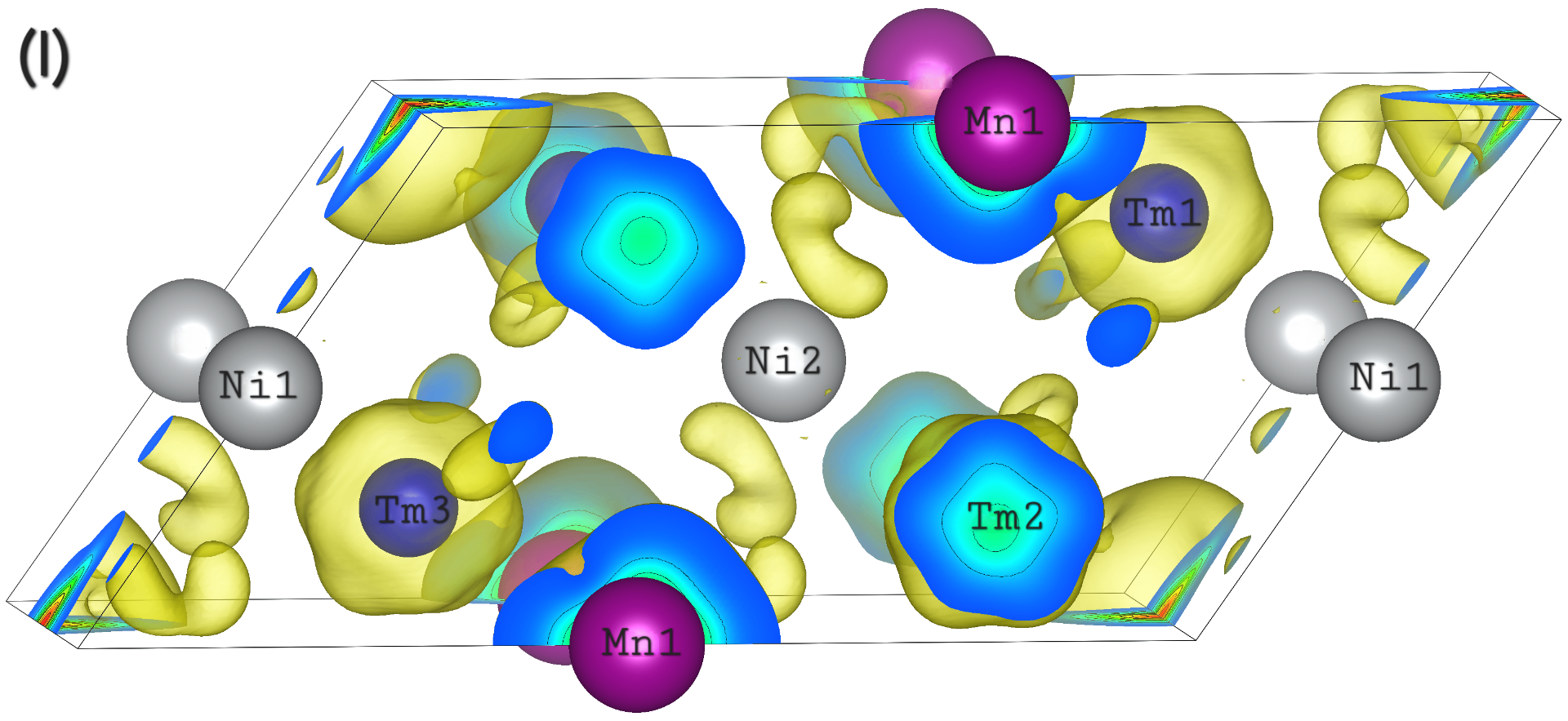}
\includegraphics[height=5.5cm,width=7.5cm]{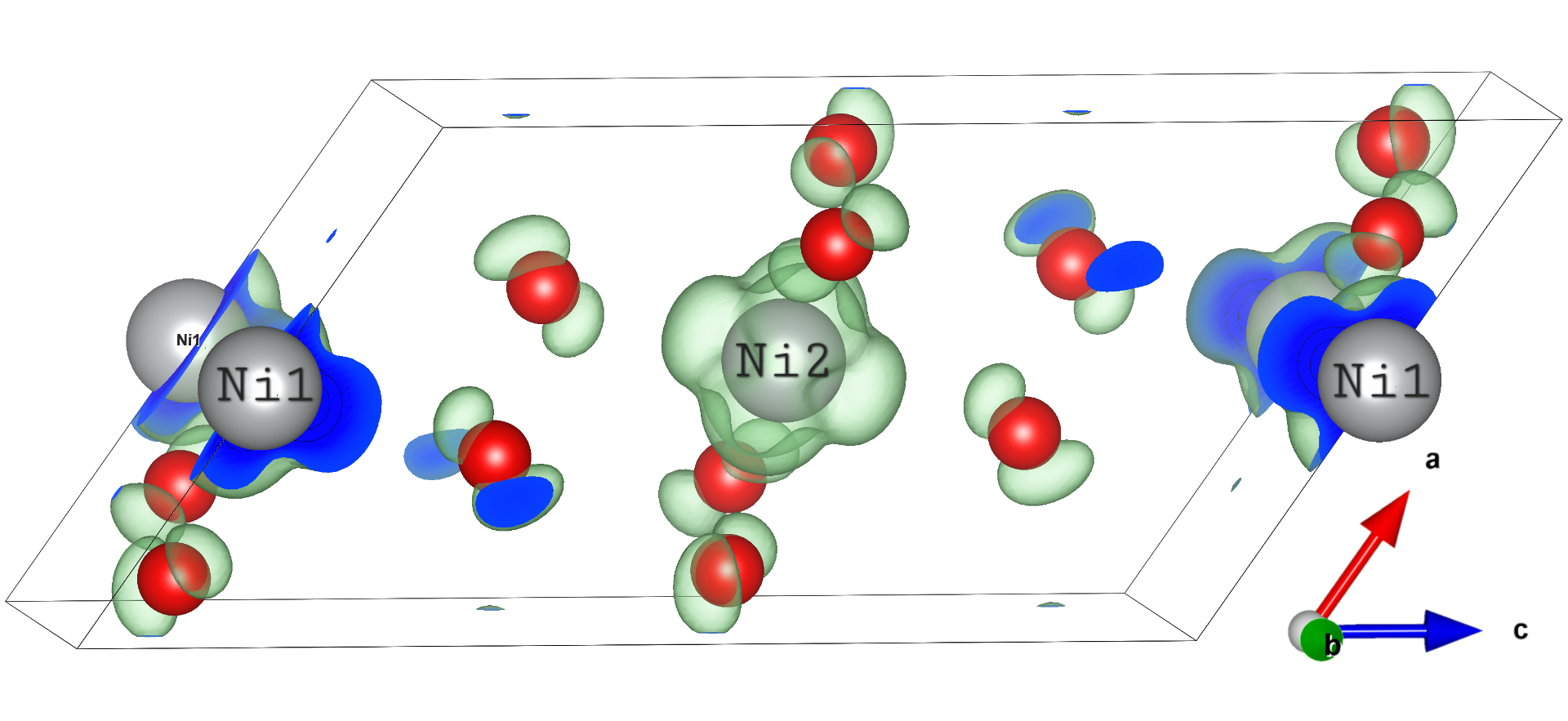}\\
\includegraphics[height=5.5cm,width=7.5cm]{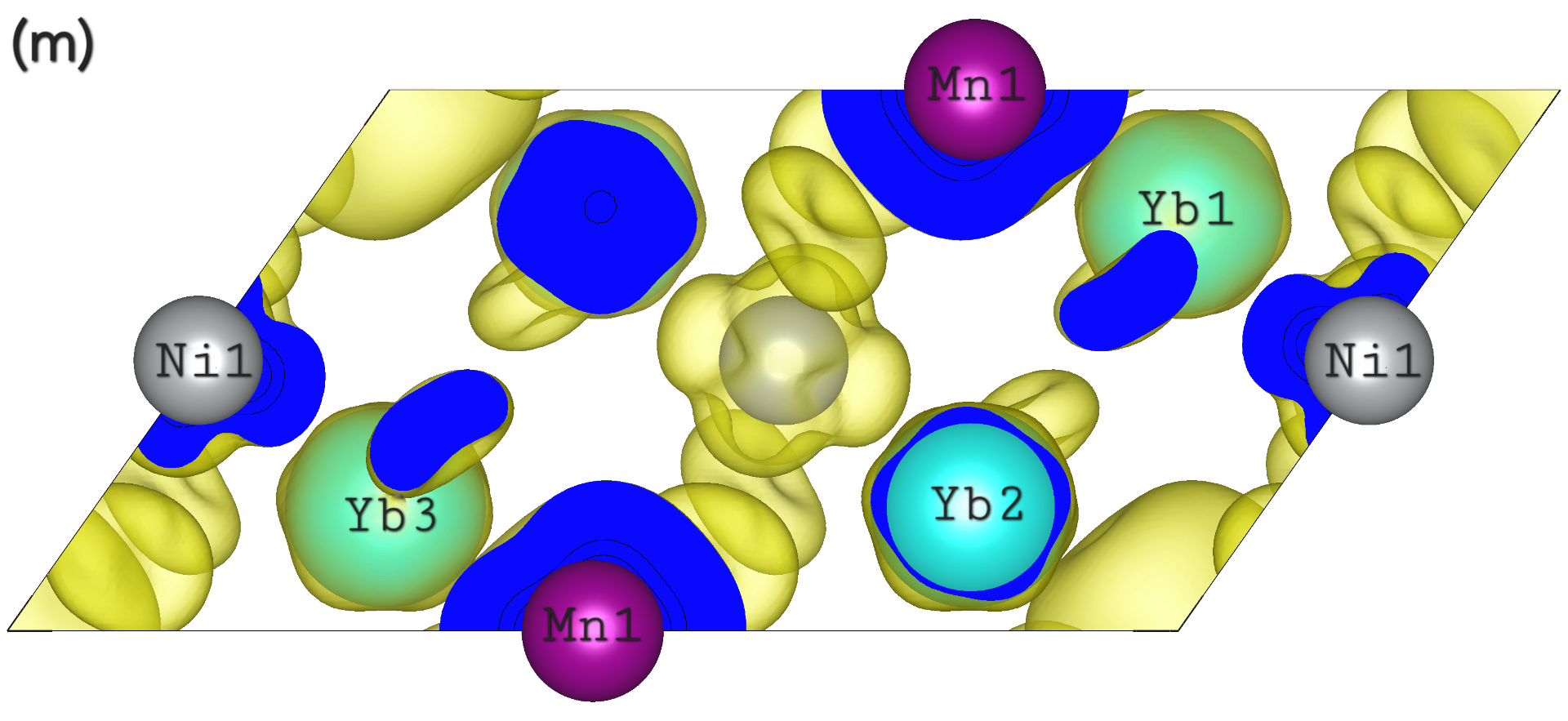}
\includegraphics[height=5.5cm,width=7.5cm]{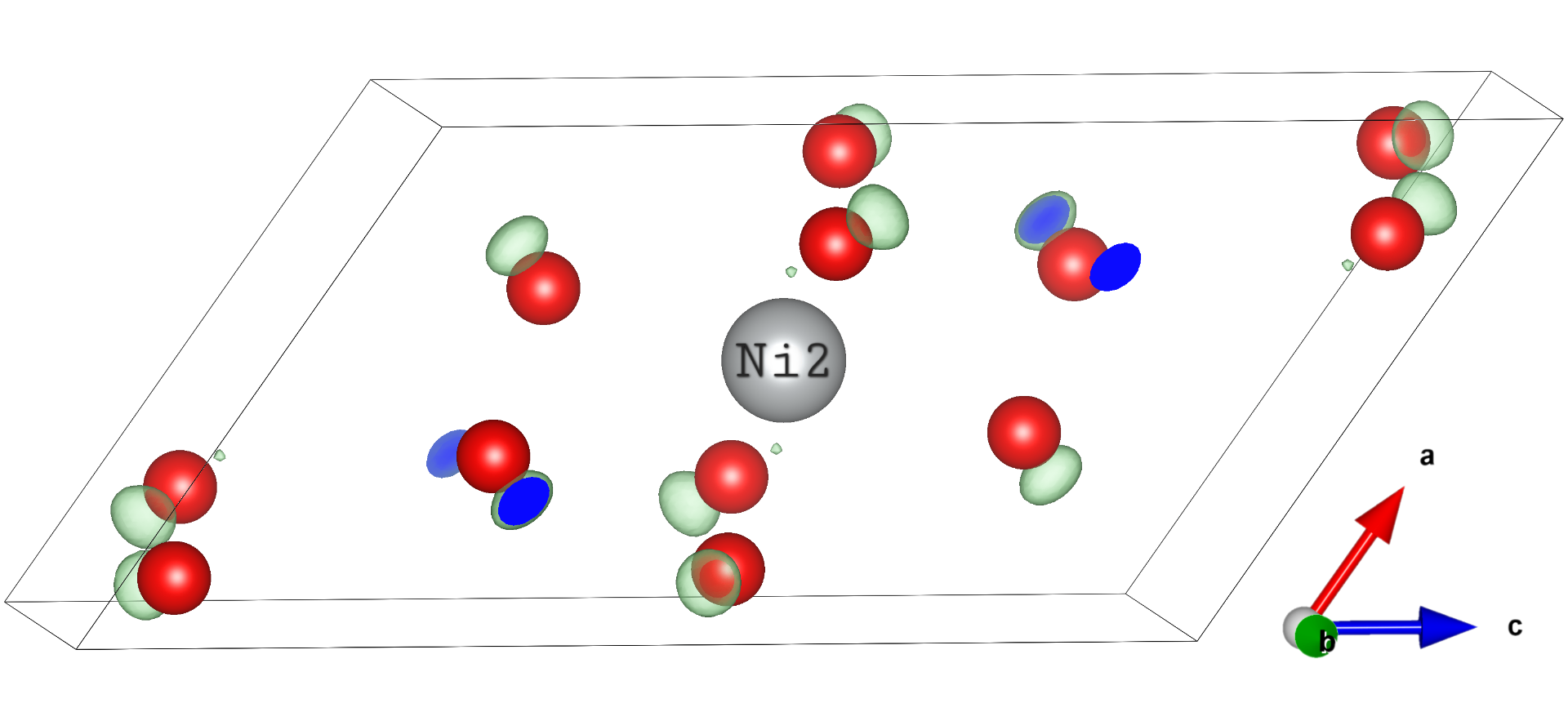}\\
\includegraphics[height=5.5cm,width=7.5cm]{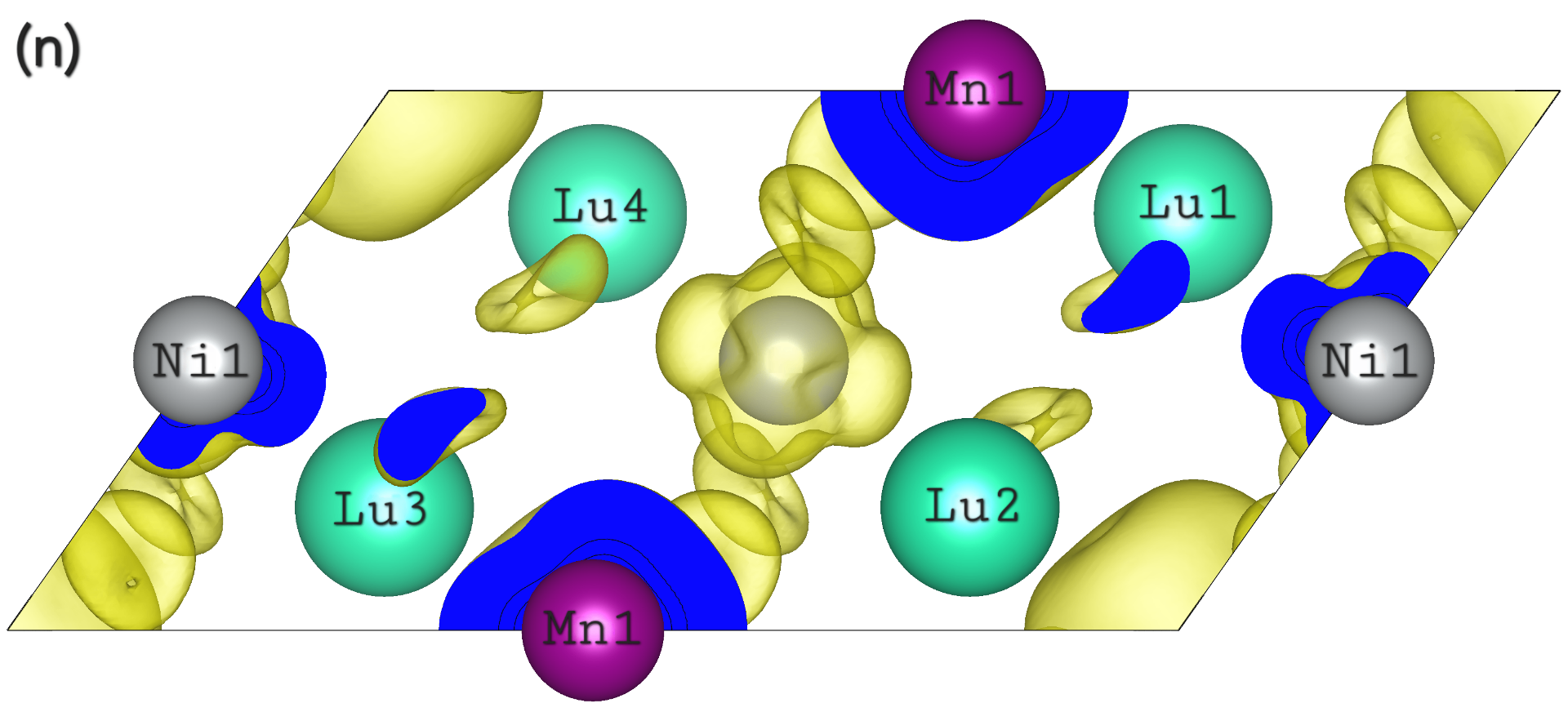}
\includegraphics[height=5.5cm,width=7.5cm]{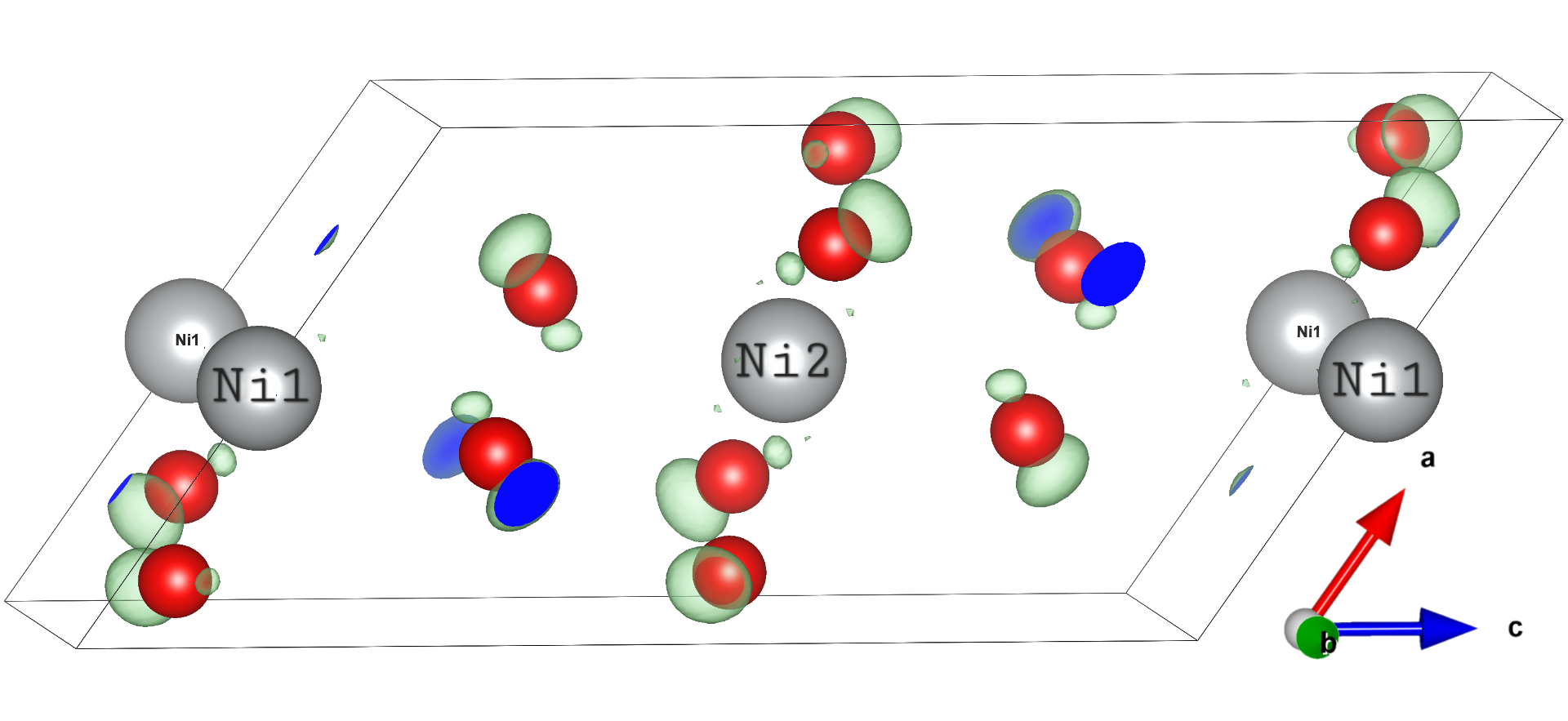}\\

\caption{Positive(right) and negative(left) magnetisation density in the order of Re$_2$MnNiO$_{6}$, Re as (a)Ce, (b)Pr, (c)Nd, (d)Pm, (e)Sm, (f)Eu, (g)Gd, (h)Tb, (i)Dy, (j)Ho, (k)Er, (k)Tm, (m)Yb and (n)Lu with $f$-electrons in valence band.}
\label{fig:mag6}
\end{center}
\end{figure}
Another exception for Er$_2$MnNIO$_6$, where Mn atoms also have negative magnetisation density. Ce$_2$MnNIO$_6$ and Yb$_2$MnNIO$_6$ have lowest positive magnetisation for Re atom. It is also observed that lanthanides and Mn atoms have more spherical density where as Ni atoms have distorted spherical density around it. Ni and O atoms density shapes are more dependent on neighbouring atoms. In this work rare earth composites with with Ce, Pm, Tb, Ho, Er and Tm shows ferrimagnetic nature because of their atomic contribution to both positive and negative density. Other structures shows ferromagnetic nature, where the highest total magnetic moment is found to be for Gd$_2$MnNIO$_6$, Eu$_2$MnNIO$_6$ and Sm$_2$MnNIO$_6$ with the value of 37.99, 33.99 and 30.81 $\mu_{B}$ (see Table. \ref{tab:magnet7})
\begin{table}[H]
\begin{center}
\begin{tabular}{|c|c|c|c|c|c|}
\hline 
RMNO & Re & Mn & Ni & O & Total \\ 
\hline 
Ce & 0.710 & 0.391 & -1.650 & -0.055 & 5.998 \\ 
\hline 
Pr & 1.805 & 3.434 & 1.655 & -0.16 & 17.998 \\ 
\hline 
Nd & 2.815 & 3.276 & 1.667 & -0.02 & 22.000 \\ 
\hline 
Pm & 4.002 & 1.026 & -1.692 & -0.1 & 14.000 \\ 
\hline 
Sm & 5.124 & 3.165 & 1.830 & -0.01 & 30.816 \\ 
\hline 
Eu & 6.048 & 3.218 & 1.626 & -0.059 & 33.993 \\ 
\hline 
Gd & 6.790 & 3.152 & 1.670 & 0.017 & 37.999 \\ 
\hline 
Tb & 5.876 & 2.985 & -1.664 & -0.02 & 26.589 \\ 
\hline 
Dy & 4.857 & 3.158 & 1.650 & 0.036 & 29.990 \\ 
\hline 
Ho & 3.825 & 2.894 & -1.684 & -0.01 & 18.032 \\ 
\hline 
Er & 2.759 & -3.183 & -1.631 & 0.045 & 2.000 \\ 
\hline 
Tm & 1.770 & 2.898 & -1.660 & 0.015 & 10.000 \\ 
\hline 
Yb & 0.802 & 3.132 & 1.680 & 0.075 & 14.000 \\ 
\hline 
Lu & 0.015 & 3.125 & 1.665 & 0.018 & 10.000 \\ 
\hline 
\end{tabular} 
\caption{Total and individual magnetic moment of RMNO (in Bohr magneton).}
\label{tab:magnet7}
\end{center}
\end{table}
\section{Conclusion}

Rare-earth elements, characterized by their partially filled 4$f$ orbitals, offer unique advantages over conventional transition metals due to their ability to exhibit multiple valence states. This feature facilitates enhanced electronic tunability and complex bonding behavior in rare-earth-based compounds, particularly in double perovskite oxides. However, accurate first-principles calculations involving rare-earth elements remain challenging. Standard density functional theory (DFT) functionals often struggle with the localized nature of $f$-electrons, leading to significant sel$f$-interaction errors. These inaccuracies compromise the reliability of exchange-correlation potentials and result in poorly predicted electronic structures. Actinide-containing systems are even more problematic in this regard, though lanthanides also present notable computational complexities.

In this study, we focus on lanthanide-based double perovskites of the form RMNO with $f$-electrons both in core and valence states. Structural optimization and phonon dispersion analyses confirm their dynamic stability of RMNO composites. While phonon modes at the $\Gamma$-point indicate stable configurations across all systems, certain compounds such as Ce$_2$MnNiO$_6$ and others exhibit phonon branches below zero beyond $\Gamma$ point, suggesting the possibility of phase transition. Electronic band structure calculations band gaps ranging from approximately 1.3$eV$to 2.0$eV$which can split indicating possibility of magnetic properties. When the initial magnetic moments of the rare-earth ions are explicitly considered, band splitting up to $\sim$1.5$eV$is observed, underscoring the sensitivity of the electronic structure to spin-polarized configurations. Thermodynamic parameters further affirm the structural and energetic stability of these materials. Optical absorption spectra, particularly for La$_2$MnNiO$_6$ and related compounds, exhibit pronounced features across the visible spectrum, with clear absorption edges near the onset of their respective band gaps.Spin density calculation separate the series in FiM and FM material containing 6 and 8 compound respectively with as high as $\sim$38 Bohr magneton. These findings underline the strong potential of rare-earth double perovskites in optoelectronic and energy-related industrial applications, warranting further comprehensive studies for their sustainable utilization.

\bibliographystyle{ieeetr}
\bibliography{RMNO}

\end{document}